\documentclass[%
superscriptaddress,
 amsmath,amssymb,
 aps,twocolumn,
floatfix,
]{revtex4-2}

\usepackage{graphicx}% Include figure files
\usepackage{fancyhdr}
\usepackage{dcolumn}% Align table columns on decimal point
\usepackage{bm}% bold math
\usepackage[version=4]{mhchem}
\usepackage[colorlinks=True,linkcolor=red,citecolor=blue,urlcolor=blue]{hyperref}% add hypertext capabilities
\newcommand{\cst}{\ce{CrSiTe3}}
\newcommand{\cgt}{\ce{CrGeTe3}}
\newcommand{\cri}{\ce{CrI3}}
\newcommand{\fgt}{\ce{Fe3GeTe2}}

\begin{document}

%\preprint{APS/123-QED}

\title{Pressure evolution of electronic structure and magnetism in the layered van der Waals ferromagnet {\cgt}}

\author{Han-Xiang Xu}
\affiliation{Research Institute for Interdisciplinary Science, Okayama University, Okayama 700-8530, Japan}

\author{Makoto Shimizu}
\altaffiliation[Present address: ]{Department of Physics, Graduate School of Science, Kyoto University, Kyoto 606-8502, Japan}
\affiliation{Department of Physics, Okayama University, Okayama 700-8530, Japan}

\author{Daniel Guterding}
\affiliation{Technische Hochschule Brandenburg, Magdeburger Straße 50,
14770 Brandenburg an der Havel, Germany}

\author{Junya Otsuki}
\affiliation{Research Institute for Interdisciplinary Science, Okayama University, Okayama 700-8530, Japan}

\author{Harald O. Jeschke}
\affiliation{Research Institute for Interdisciplinary Science, Okayama University, Okayama 700-8530, Japan}

\date{\today}

\begin{abstract}
Layered van der Waals ferromagnets, which preserve their magnetic properties down to exfoliated monolayers, are fueling an abundance of fundamental research and nanoscale device demonstration. {\cgt} is a prime example for this class of materials. Its temperature-pressure phase diagram features an insulator-to-metal transition and a significant increase of ferromagnetic Curie-Weiss temperatures upon entering the metallic state. We use density functional theory to understand the magnetic exchange interactions in {\cgt} at ambient and elevated pressure. We calculate Heisenberg exchange couplings, which provide the correct ferromagnetic ground state and explain the experimentally observed pressure dependence of magnetism in {\cgt}. Furthermore, we combine density functional theory with dynamical mean field theory to investigate the effects of electronic correlations and the nature of the high pressure metallic state in {\cgt}.
\end{abstract}

\maketitle

\section*{Introduction}
The chromium tellurides {\cst} and {\cgt} were initially investigated as bulk ferromagnetic semiconductors~\cite{Ouvrard1988,Marsh1988,Carteaux1991,Carteaux1995b,Carteaux1995} with ordering temperatures of $T_{\rm C}=32$\,K and $T_{\rm C}=61$\,K, respectively. Later it was discovered that the van der Waals bonded layers of {\cgt} can be exfoliated and even monolayers remain ferromagnetic~\cite{Gong2017}, a fact that had been predicted theoretically~\cite{Li2014}. Together with a few other layered ferromagnets like {\cri}~\cite{McGuire2015,Huang2017,Huang2018,Jiang2018} and {\fgt}~\cite{Deng2018}, these materials may facilitate a host of interesting applications~\cite{Burch2018}, which encompass fabrication of heterostructures~\cite{Gong2019}, e.g. with topological insulators~\cite{Ji2013, Mogi2018}, use as switchable resistive components in phase-change random access memory~\cite{Hatayama2018} or as thermoelectric materials~\cite{Yang2016, Lefevre2017}.

Following the successful exfoliation of ferromagnetic monolayers of {\cst}~\cite{Lin2016} and {\cgt}~\cite{Gong2017}, these materials came under intense scrutiny. {\cst} shows an insulating ferromagnetic ground state~\cite{Casto2015, Liu2016}. Applying pressure to this system leads to a insulator-to-metal transition~\cite{Cai2020}. At a pressure of $\sim 7.5~\textrm{GPa}$ a structural phase transition suppresses ferromagnetism and a superconducting state with a $T_{\rm c}$ of about $3~\mathrm{K}$ emerges~\cite{Cai2020}. At a pressure of around $15~\textrm{GPa}$ {\cst} undergoes a polymorphic transformation.~\cite{Xu2020} While ferromagnetism in bulk {\cst} has a Curie-temperature of $T_{\rm C} = 32~\mathrm{K}$, pressure dramatically enhances the Curie-temperature to $T_{\rm C} \sim 138~\mathrm{K}$.~\cite{Zhang2021} Various other studies have investigated the structural, electronic and magnetic properties of {\cst} and similar materials both experimentally~\cite{Williams2015, Achinuq2021, Niu2021, Zhang2019, Milosavljevic2018} and theoretically~\cite{Sivadas2015, Zhuang2015, Kang2019, Zhang2019, Chen2020, Milosavljevic2018}.

{\cgt} is particularly interesting, since it undergoes an insulator-to-metal transition at a pressure of about $5~\mathrm{GPa}$~\cite{Bhoi2021}, which is not coupled to a structural transition, in contrast to {\cst}~\cite{Cai2020}. Only at pressures of around $18~\textrm{GPa}$ a structural phase transition to an amorphous state is observed.~\cite{Yu2019} Besides application of pressure, the insulator-to-metal transition can also be triggered by intercalation of organic ions~\cite{Wang2019} into bulk crystals or ionic liquid gating to a field effect transistor~\cite{Wang2018, Verzhbitskiy2020}. This also applies to related materials~\cite{Huang2018, Deng2018, Jiang2018}. Both applying pressure and injecting charge carriers dramatically increases the Curie temperature of {\cgt} from $T_{\rm C} = 61~\mathrm{K}$ at ambient pressure to $T_{\rm C} \sim 200~\mathrm{K}$~\cite{Verzhbitskiy2020} or even higher~\cite{Bhoi2021}. {\cgt} has been characterized as a charge transfer insulator and its ferromagnetism has been attributed to a superexchange mechanism via the tellurium ions~\cite{Kang2019, Zhang2019, Watson2020, Bhoi2021}.

\begin{figure*}[t]
    \includegraphics[width=0.9\textwidth]{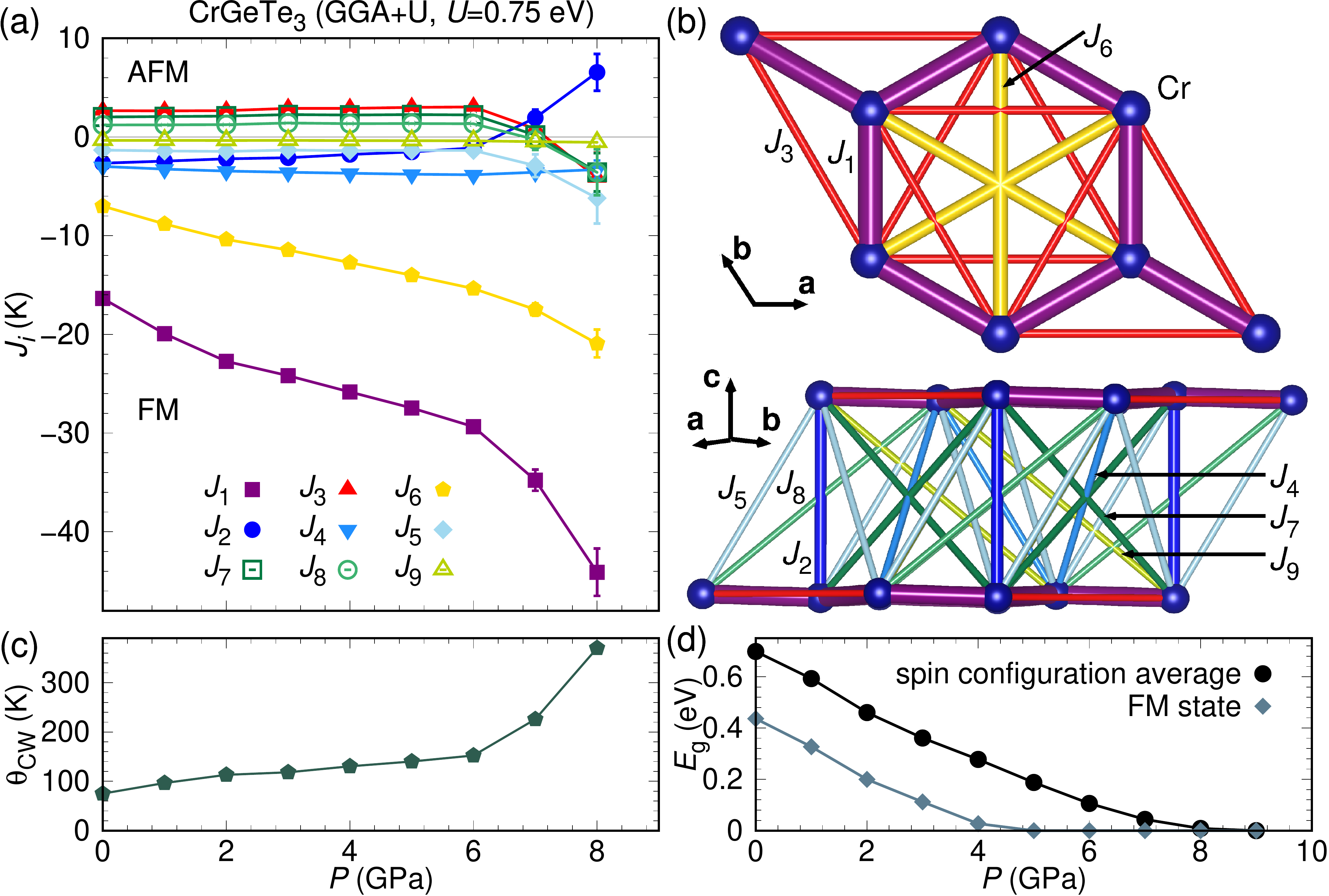}
    \caption{Result of DFT energy mapping for {\cgt} as function of pressure. Interpolated experimental crystal structures are used as explained in the text. (a) First nine exchange interactions in Kelvin, evaluated at fixed Hund's rule coupling $J_{\rm H}=0.72$\,eV and on-site interaction strength $U=0.75$\,eV. The exchange paths are visualized in (b). (c) Curie-Weiss temperature calculated via Eq.~\eqref{eq:tcw} from the Heisenberg interactions in (a). (d) Charge gaps obtained from the ferromagnetic state at $U=0.75$\,eV, $J_{\rm H}=0.72$\,eV (diamonds) and by averaging over the 70 spin configurations per pressure point from (a) (circles). }
    \label{fig:couplings}
\end{figure*}

It has been conjectured that the increased bandwidth under pressure closes the charge transfer gap, which in turn enhances the ferromagnetic superexchange and causes the greatly enhanced Curie temperature~\cite{Bhoi2021}. Various other experimental~\cite{Sakurai2021, Saito2021,Liu2017, Lin2017, Zeisner2019, Suzuki2019} and theoretical studies~\cite{Sivadas2015, Fang2018, Zeisner2019, Suzuki2019, Lee2020, Fumega2020, Tiwari2021} for {\cgt} have investigated further details of this material.

We perform theoretical calculations for {\cgt} at various pressures and focus on the issues of the enhanced Curie-temperature and the conjectured shrinking of the charge transfer gap. Our results confirm the experimentally observed pressure phase diagram~\cite{Bhoi2021}. We find that {\cgt} evolves from a charge transfer insulator into a correlated metal with significant pressure- and orbital-dependent electronic correlations, as the charge transfer gap closes with increasing applied pressure. At the same time, in-plane ferromagnetic exchange couplings are strongly enhanced. Inter-plane ferromagnetic exchange couplings also start to appear at the insulator-to-metal transition and contribute to the experimentally observed increase of the Curie temperature, which is explained by our calculations. Furthermore, we discuss similarities between the behavior of {\cgt} under pressure and with electron doping, which is crucial for applications of {\cgt} and other van der Waals ferromagnets.

\section*{Methods} 
We base our calculations on interpolated high pressure crystal structures of {\cgt} ($R\bar{3}$ space group, no. 148). For that purpose, we use the structural data from Ref.~\cite{Yu2019}. We use splines for the lattice parameters and a linear fit for the rather noisy fractional coordinates (see Appendix~\ref{app:interpolation}).
We study the electronic structure and magnetism using the full potential local orbital (FPLO) basis~\cite{Koepernik1999} and a generalized gradient approximation to the exchange and correlation functional~\cite{Perdew1996}. Our manybody calculations are based on a tight binding Hamiltonian obtained from projective Wannier functions~\cite{Eschrig2009,Koepernik2023}. We determine Heisenberg Hamiltonian parameters using an energy mapping method~\cite{Jeschke2013,Iqbal2017,Guterding2016} based on DFT+U
~\cite{Liechtenstein1995} total energies; we fix the value of the Hund's rule coupling $J_{\rm H}=0.72$\,eV~\cite{Mizokawa1996} and vary the onsite interaction $U$. We use energies of 70 distinct spin configurations to resolve 9 exchange interactions (see Appendix~\ref{app:energymapping}).
We perform DFT+DMFT calculations within DCore~\cite{Shinaoka2021} using a hybridization expansion continuous time quantum Monte Carlo (CT-QMC) method~\cite{Werner2006,Gull2011} as impurity solver (see Appendix~\ref{app:DMFT} for more details). We use the Pad{\'e} approximation for analytic continuation of the self energy. For Cr $3d$, we use interaction parameters $U=2.0$\,eV, $J_{\rm H}=0.72$\,eV. Based on those, Slater integrals and orbital dependent interaction matrices are calculated~\cite{Anisimov1993}.
For the DMFT calculations, we prepare 36 band tight binding models including 36 orbitals from two formula units of {\cgt}, in particular 10 Cr $3d$, 18 Te $5p$, 2 Ge $4s$ and 6 Ge $4p$ orbitals. Due to covalent bonding but lower prevalence in the composition, Ge orbital weights and DOS resemble Te weights and DOS but are much smaller; nevertheless, Ge orbitals are needed to improve the overall agreement between tight binding model and DFT band structure. In contrast, including Cr $4s$ orbitals does not improve the results.

\section*{Results} 
We first investigate the evolution of magnetism in {\cgt} under pressure and extract the Heisenberg Hamiltonian of the material via DFT energy mapping. This method has been shown to be very reliable for Cr$^{3+}$ based magnets~\cite{Ghosh2019}. We determine the parameters of the Heisenberg Hamiltonian written in the form
\begin{equation}
    H=\sum_{i<j} J_{ij} {\bf S}_i\cdot {\bf S}_j \,,
\label{eq:hamiltonian}\end{equation}
where ${\bf S}_i$ and ${\bf S}_j$ are spin operators and every bond is counted once.
Our results are summarized in Fig.~\ref{fig:couplings}. As {\cgt} is a small gap semiconductor at ambient pressure and becomes metallic under pressure, exchange interactions are expected to be rather long-ranged. Therefore, we resolve the first nine exchange interactions, extending up to 2.5 times the nearest neighbor distance. 

To estimate the appropriate on-site interaction $U$, we determine the Heisenberg parameters at ambient pressure, calculate the Curie-Weiss temperature $\theta_{\rm CW}$ from those parameters, and compare it to the experimental value~\cite{Bhoi2021}. Note that here, we do not focus on the ferromagnetic ordering temperature, because in a layered van der Waals magnet it depends logarithmically on both interlayer exchange and the magnetic anisotropy~\cite{Yasuda2005,Lado2017,Tiwari2021b}. We find that the on-site interaction $U$ has to be chosen as small as possible (see Appendix~\ref{app:energymapping} for further details). In Fig.~\ref{fig:couplings}\,(a), we show the evolution of exchange interactions with pressure for $U=0.75$\,eV, which is only slightly larger than the Hund's rule coupling $J_{\rm H}=0.72$\,eV. This means that the effective interaction $U-J_{\rm H}$ is close to zero; nevertheless, the result of this calculation is very different from the result of plain GGA which severely overestimates the strength of magnetic interactions. At ambient pressure, we find that the Hamiltonian of {\cgt} is dominated by two ferromagnetic interactions, $J_1$ and $J_6$.
The interaction paths are visualized in Fig.~\ref{fig:couplings}\,(b): $J_1$, $J_3$ and $J_6$ are first, second and third neighbors in the honeycomb lattice, respectively. The six other paths are interlayer exchange paths. 
We determine the Curie-Weiss temperature according to the mean field formula
\begin{equation}\begin{split}
       \theta_{\rm CW}=-\frac{1}{3}S(S+1)\big(& 3 J_1 + J_2 + 6 J_3 + 3 J_4 + 6 J_5 \\&+ 3 J_6 + 6 J_7 + 3 J_8 + 6 J_9\big) \,,
\label{eq:tcw}
\end{split}\end{equation}
where $S=3/2$ is the same spin eigenvalue that is used in the energy mapping procedure.
The pressure evolution of $\theta_{\rm CW}$ is shown in Fig.~\ref{fig:couplings}\,(c). The contribution to $\theta_{\rm CW}$ from the six interlayer couplings is almost constant at 8\,K between $P=0$ and $P=6$\,GPa and then increases strongly (see Fig.~\ref{fig:tcwparts}). The strong in-plane ferromagnetic exchange taken together with the initially small but always ferromagnetic effective interlayer exchange means that the Hamiltonian determined by DFT energy mapping clearly supports a ferromagnetic ground state of {\cgt} for all investigated pressures. Fig.~\ref{fig:couplings}\,(d) shows the charge gap of {\cgt}, obtained by averaging over the 70 distinct spin configurations that were calculated at all pressures. This measure places the transition pressure for the insulator-to-metal transition between $\sim 4.5$\,GPa and $\sim 7.5$\,GPa, based on either the ferromagnetic solution, which likely underestimates the transition pressure due to absence of spin fluctuations, and the spin configuration average, which is an upper limit due to the presence of many unfavorable magnetic configurations. While the Heisenberg Hamiltonian is usually applied to magnetic insulators, there are many cases where the magnetic properties of metallic magnets are also well described~\cite{Kim2017,Huang2023}, and the energy mapping approach used here works well for metallized {\cgt}. We would like to point out that in choosing a particular $U$ value for all pressures, we have taken a practical approach that gives us a good picture of the pressure evolution of the magnetic Hamiltonian. Nevertheless, we have convinced ourselves that somewhat larger choices of $U$ give qualitatively similar trends with pressure. 

The insulator-metal transition coincides with a significant strengthening of the dominant ferromagnetic exchange coupling $J_1$, and with couplings $J_3$, $J_7$ and $J_9$ turning ferromagnetic. This leads to a sharp upturn of the Curie-Weiss temperature (Fig.~\ref{fig:couplings}\,(c)). Our estimates for the charge transfer gap and the Curie-Weiss temperature as a function of pressure are in very good agreement with the experimental observations in Ref.~\cite{Bhoi2021}. In both experiment and our theoretical calculations, the Curie-Weiss temperature is nearly constant for a large pressure range and then increase rapidly near the insulator-to-metal transition. The prediction of the ferromagnetic ordering temperature $T_{\rm C}$ is complicated due to Mermin-Wagner physics that suppresses order in the pure Heisenberg 2D magnet. It is known that the ordering temperature is proportional to the in-plane exchange interactions but the order depends logarithmically on the relative strength of interlayer exchange~\cite{Irkhin1999,Yasuda2005} compared to the in-plane exchange or it has a power law dependence on the ratio between single-ion anisotropy~\cite{Costa2003} and in-plane exchange. Therefore, the in-plane and interlayer exchange interactions provide the qualitative trends for the Curie temperature, but a precise ordering temperature requires complex calculations that are beyond the scope of the present study. Notwithstanding some logarithmic or power law corrections, the substantial upturn of the Curie-Weiss temperature, which is also the mean field ordering temperature, indicates that the Hamiltonian we determined will lead to a significant increase of the Curie temperature $T_{\rm C}$ around $P=6$\,GPa.

\begin{figure}[t]
    \includegraphics[width=\columnwidth]{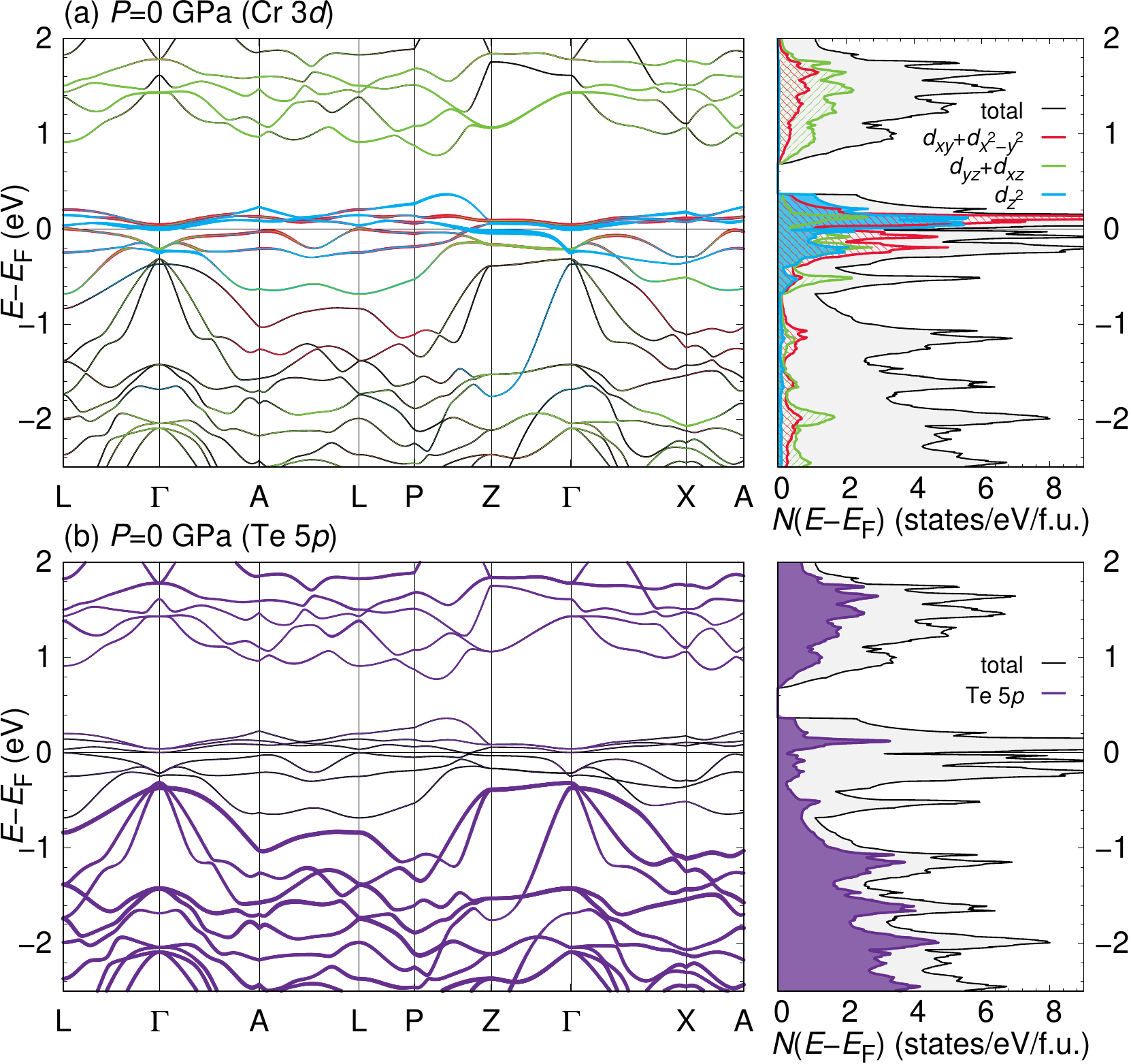}
    \caption{GGA band structure of {\cgt} at ambient pressure. (a) Band weights and partial densities of states for the three Cr $3d$ characters in trigonally distorted octahedral environment. (b) Band weight and partial density of states for summed up Te $5p$. High symmetry points of the rhombohedral space group are~\cite{Setyawan2010,Zhang2019} $L=(1/2,0,0)$, $A=(\nu/2,\nu/2,-\nu)$, $P=(\eta,\nu,\nu)$, $Z=(1/2,1/2,1/2)$, $X=(\nu,0,-\nu)$ where $\eta=(1+4\cos\alpha)/(2+4\cos\alpha)$, $\nu=3/4-\eta/2$. }
    \label{fig:ggabands}
\end{figure}

\begin{figure}[t]
    \includegraphics[width=\columnwidth]{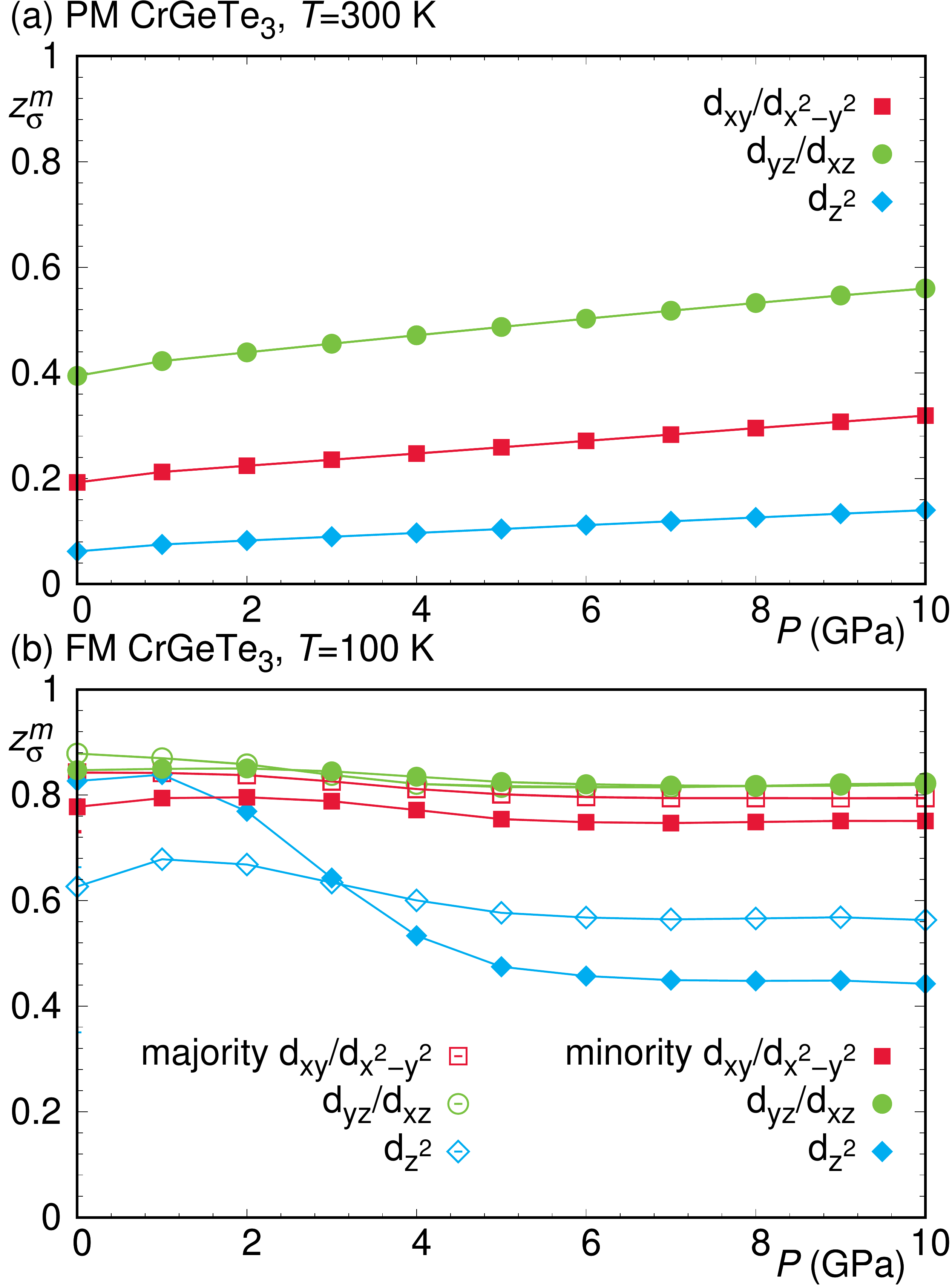}
    \caption{Pressure dependence of the renormalization factor (quasiparticle weight) $z^m_\sigma$ for different orbitals $m$ and spin directions $\sigma$. (a) Paramagnetic state at room temperature ($T=300$\,K). (b) Ferromagnetic state at $T=100$\,K. }
    \label{fig:renormalizationfactor}
\end{figure}

Now we investigate the importance of electronic correlations in {\cgt} under pressure. We use the same interpolated experimental structures (shown in Fig.~\ref{fig:interpolation}) as in our previous calculations. The nonmagnetic GGA band structure of {\cgt} has six Cr $3d$ bands (from two formula units) crossing the Fermi level. Cr is strongly hybridized with the spatially extended Te $5p$ orbitals (see Fig.~\ref{fig:ggabands}). In the trigonally distorted CrTe$_6$ octahedra, the crystal field splits the $3d$ orbitals into $a_{1g}$ ($d_{z^2}$), $e_g^\pi$ ($d_{xy}$, $d_{x^2-y^2}$), and $e_g^\sigma$ ($d_{xz}$, $d_{yz}$) manifolds. 

We first analyze the quasiparticle weight
\begin{equation}
    z_\sigma^m = \bigg(1-\frac{{\rm Im}\Sigma_\sigma^m(\omega_0)}{\omega_0}\bigg)^{-1} \,,
\end{equation}
where $\Sigma_\sigma^m(\omega_n)$ is the self energy for spin $\sigma$ and orbital $m$ and $\omega_0$ is the lowest positive Matsubara frequency~\cite{Arsenault2012}. This approximation is reasonable as the imaginary parts of the self energies of \ce{CrGeTe3} are linear to a good approximation for low frequencies. The quasiparticle weight is a measure for the strength of electronic correlations. While an uncorrelated system has quasiparticle weights equal to $z_\sigma^m = 1$, decreasing quasiparticle weights $z_\sigma^m$ signal an increase in electronic correlations. The inverse of the quasi-particle weight is the mass enhancement  $$\frac{1}{z_\sigma^m} =\frac{m^*}{m_{\rm DFT}}\,,$$ which indicates the increase of the effective mass $m^*$ compared to the band mass $m_{\rm DFT}$ due to electronic correlations. In case correlations open a gap, the quasiparticle weight indicates the mass renormalization in the bands closest to the Fermi level.
Figure~\ref{fig:renormalizationfactor}\,(a) shows the quasiparticle weight at room temperature in the paramagnetic state. At all pressures, the $d_{z^2}$ orbital shows the strongest renormalization, followed by $d_{xy}$ and $d_{x^2-y^2}$. As a function of pressure, the quasiparticle weight slightly increases. With constant interaction parameters, this is to be expected under pressure, because the band width increases and screening improves as the material is compressed.

This picture changes in the ferromagnetic state at $T=100$\,K (Fig.~\ref{fig:renormalizationfactor}\,(b)). In the magnetic insulator near ambient pressure, we find quasiparticle weights between 0.6 and 0.9, indicating only weak correlation effects. The correlation strength increases at the insulator-metal transition, which happens near $P=3$\,GPa in our DFT+DMFT calculations. This pressure value is somewhat lower than what is observed in experiment. At the insulator-metal-transition, in particular the $d_{z^2}$ orbital becomes substantially more strongly correlated.

\begin{figure*}[t]
    \includegraphics[width=\textwidth]{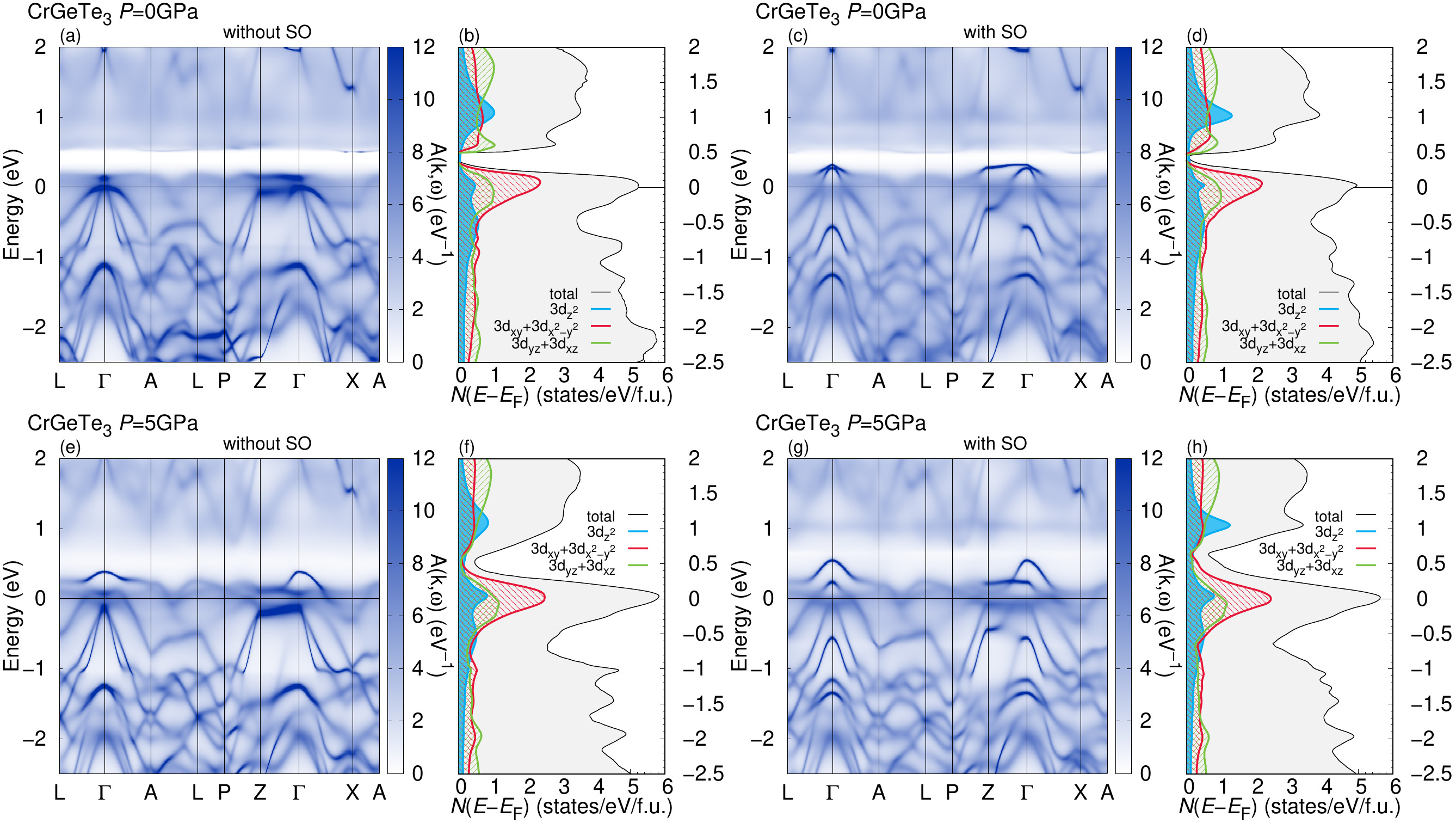}
    \caption{Momentum resolved and integrated spectral functions for {\cgt} in the paramagnetic state at room temperature ($T=300$\,K). (a)-(b) Ambient pressure without spin-orbit coupling, (c)-(d) including spin-orbit coupling in the DFT Hamiltonian. (e)-(f) Elevated pressure of $P=5$\,GPa without effects of spin-orbit coupling, and (g)-(h) $P=5$\,GPa including spin-orbit coupling.  Interaction parameters for Cr $3d$ are $U=2.0$\,eV, $J_{\rm H}=0.72$\,eV. The double counting correction is reduced by 4{\%}.}
    \label{fig:PMspectralfunction}
\end{figure*}

We now study the DFT+DMFT spectral function for {\cgt}. At ambient pressure, our calculations predict the outcome of future angle-resolved photoemission experiments. Such experiments may not be feasible under higher pressures, but our calculations under pressure may still give relevant insights. Fig.~\ref{fig:PMspectralfunction} presents both angle-resolved and angle-integrated spectral functions at $T=300$\,K; the paramagnetic solution shown here is the ground state at room temperature. Fig.~\ref{fig:PMspectralfunction}\,(a) and (c) compare the ambient pressure spectral function $A(\omega)$ without and with spin-orbit coupling, respectively. 

The effects of spin-orbit coupling are large for the heavy Te atoms, but not for the lighter Cr atoms. Therefore, we include the effects of spin-orbit coupling at the level of the DFT calculation from which we obtain the non-interacting Hamiltonian. However, we use a projection of the fully relativistic bands on $lm\sigma$ quantum numbers and solve the dynamical mean field problem for Cr $3d$ without explicitly accounting for spin-orbit coupling in the interactions. 

Comparing Figs.~\ref{fig:PMspectralfunction}\,(a) and (c) with the GGA orbital weights in Fig.~\ref{fig:ggabands}, we see significant effects of spin-orbit coupling for example in the occupied bands around the $\Gamma$ point which mostly carry Te orbital character. We consider Fig.~\ref{fig:PMspectralfunction}\,(c) more relevant for comparison with experiment than Fig.~\ref{fig:PMspectralfunction}\,(a).
While our paramagnetic DFT+DMFT results show significant correlation effects as seen from the quasiparticle weights, these are not strong enough to open a gap. Experimental resistivity, while significantly reduced at room temperature compared to low temperatures, in contrast still shows semiconducting behavior~\cite{Bhoi2021}. We speculate that charge self-consistency or better treatment of spin orbit coupling may be needed to explain the small gap at room temperature. This needs to be investigated in a future study.

\begin{figure*}[t]
    \includegraphics[width=\textwidth]{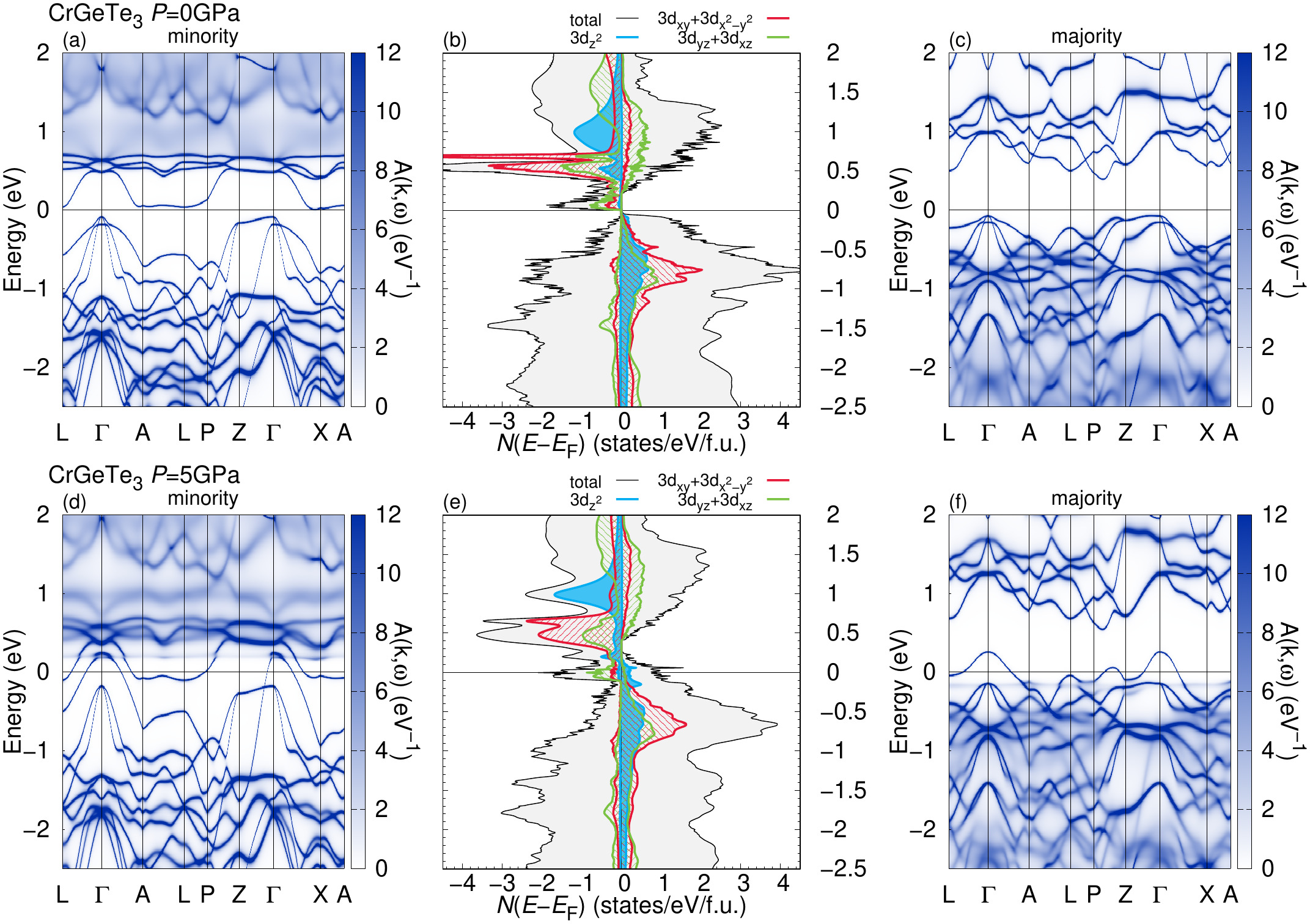}
    \caption{DFT+DMFT spectral function and density of states of {\cgt} at $T=100$\,K in the ferromagnetic state. (a)-(c) Ambient pressure $P=0$\,GPa. (d)-(f) Elevated pressure $P=5$\,GPa. Interaction parameters for Cr $3d$ are $U=2.0$\,eV, $J_{\rm H}=0.72$\,eV. The double counting correction is reduced by 4{\%}.}
    \label{fig:FMspectralfunction}
\end{figure*}

We now continue the analysis at low temperatures. Figure~\ref{fig:FMspectralfunction} shows angle resolved and integrated spectral functions of {\cgt} at $T=100$\,K in the ferromagnetic state. While the experimentally observed ordering temperature of {\cgt} is somewhat lower at $T_{\rm C}=61$\,K, it is computationally challenging to reach good convergence for lower temperatures, since more Matsubara frequencies are needed to cover the same energy range. Nevertheless, the spectral function does not change dramatically between $T=100$\,K and $T=50$\,K. At ambient pressure (Figs.~\ref{fig:FMspectralfunction}\,(a) to (c)), we find a ferromagnetic (Slater) insulator in agreement with the experiments of Bhoi {\it et al.}~\cite{Bhoi2021}. The spectral functions at $P=5$\,GPa show a ferromagnetic metallic state. We clearly observe an insulator-to-metal transition, which is predominantly induced by the widening of Te bands under pressure, i.e. the transition is band width controlled.

\section*{Discussion}
In our DFT calculations for the charge transfer gap, we find gap values and a pressure dependence (see Fig.~\ref{fig:couplings}(d)), which are consistent with experimental~\cite{Ji2013} and other theoretical~\cite{Fumega2020, Kang2019, Lee2020} estimates. Some studies have estimated band gaps, which are orders of magnitude smaller, probably due to differences in methodology.~\cite{Fang2018, Zeisner2019}

Our DFT calculations for the magnetic interactions of {\cgt} under pressure show a dramatic increase in the strength of ferromagnetic exchange couplings in the vicinity of the insulator-to-metal transition; some longer range antiferromagnetic couplings even turn ferromagnetic. These results are in good agreement with experimental findings in Ref.~\cite{Bhoi2021}. In particular, we find that smooth changes in the structural parameters produce the observed strong increase of the Curie-Weiss temperature. Therefore, this increase is of purely electronic origin and not connected to a structural transition, in agreement with the conjecture in Ref.~\cite{Bhoi2021}.

Previous studies have focused entirely on the intra-layer ferromagnetic exchange, while estimating inter-layer exchange to be antiferromagnetic~\cite{Bhoi2021, Gong2017, Zhang2019}. We find that, at least in {\cgt}, several inter-layer antiferromagnetic exchange terms also turn ferromagnetic at the insulator-to-metal transition (see Fig.~\ref{fig:couplings}). Therefore, also the inter-layer exchange contributes significantly to the observed increase in the Curie-Weiss temperature (see Fig.~\ref{fig:tcwparts}). We conclude that pressure does not only enhance the intra-layer ferromagnetic exchange and close the charge gap, as reported in other studies. Using pressure to pass through the insulator-to-metal transition makes magnetic interactions in {\cgt} also significantly more three-dimensional, with many of the longer range exchange terms assuming larger absolute values. To our knowledge, this fact has not been reported elsewhere.

Importantly, our results show that the inter-layer contribution to the Curie-Weiss temperature only becomes significantly ferromagnetic at the insulator-to-metal transition (see Fig.~\ref{fig:tcwparts}), while theoretical results by Fumega {\it et al.}~\cite{Fumega2020} predict a strong continuous increase of the inter-layer contribution already in the insulating state. The latter study uses a less detailed Hamiltonian, which corresponds to only taking into account the couplings $J_1$ and $J_2$ of our model. Although the sign of exchange couplings agrees with our study, their computational approach averages out the effects of longer-range couplings, which we are able to resolve in detail, and yields a different pressure dependence of the Heisenberg Hamiltonian parameters. Furthermore, Fumega {\it et al.} did not determine the behavior of exchange couplings beyond the insulator-to-metal transition. Hence, their study misses the peculiar sharp increase of the Curie temperature at and beyond the transition.

The exchange Hamiltonian at ambient pressure has also been determined by Sivadas {\it et al.}~\cite{Sivadas2015}, taking into account only the terms $J_1$, $J_2$ and $J_3$ from our model. They obtained a roughly similar ferromagnetic $J_1$, but predicted $J_2$ and $J_3$ to be antiferromagnetic at ambient pressure. We find $J_2$ to be ferromagnetic at ambient pressure. Again, the results probably differ from ours because this study did not take into account important longer-range exchange paths like $J_6$ and investigated fewer spin configurations.

A DFT+U study at ambient pressure has also been performed by Suzuki {\it et al.}~\cite{Suzuki2019}, who find $J_1$ and $J_2$ to be ferromagnetic, but do not resolve further exchange couplings. This study also compares DFT+U calculations for {\cgt} at ambient pressure to momentum-integrated photoemission experiments and concludes that the effective interaction $U_\text{eff} = U-J_{\rm H}$ should be chosen close to $U_\text{eff} = 1.1$~eV. Comparing DFT+U calculations to photoemission spectra may not give a fully consistent picture of dynamical correlations in {\cgt}, especially since we find the majority spin $d_{z^2}$ orbital to be strongly correlated already at ambient pressure. Qualitative differences in DFT+U compared to DFT+DMFT spectral functions and optical conductivity have also been reported for the related material {\cri}.~\cite{Kvashnin2022} In contrast, we choose $U_\text{eff} \approx 0$ in order to reproduce the experimentally observed Curie-Weiss temperature. Choosing a larger Coulomb repulsion $U$ would lead to severely overestimated exchange couplings and Curie-Weiss temperature (see Fig.~\ref{fig:0GPacouplings} and Table~\ref{tab:0GPacouplings}). This is also observable in the study by Fang {\it et al.}~\cite{Fang2018}, which overestimates the exchange couplings of a simplified Hamiltonian by using a large effective interaction.

For the Cr magnetic moments, we find a small decrease with increasing pressure, both in our DFT and DFT+DMFT calculations (see Fig.~\ref{fig:moments}). While experiments seem to observe a significant decrease in moments only in the pressure range from $P=3$\,GPa to $P=5$\,GPa~\cite{Bhoi2021} and more or less constant behaviour outside of this range, our calculations rather show a smooth decrease of the magnetic moment across the entire pressure range. While the origin of this minor difference is currently unclear, the overall size of the magnetic moment at ambient pressure is consistent with literature results.~\cite{Kvashnin2022, Kang2019, Zhang2019}

The reduction of magnetic moments with increasing pressure seems to be related to moderate changes in the electronic filling of Cr $3d$ states, which increases slightly (see Fig.~\ref{fig:occupations}\,(a)). The excess electrons populate the minority spin states and, therefore, decrease the observed magnetic moment. In addition, the filling of majority spin states decreases (see Fig.~\ref{fig:occupations}\,(b)), which further decreases the magnetic moments. This can be understood from Fig.~\ref{fig:FMspectralfunction}\,(d) where one can see partial occupation of minority $3d$ states and partial depletion of majority $3d$ states at moderate pressure. This suggests a connection to experiments, which observed a strong increase of the Curie temperature in {\cgt} upon electron doping.~\cite{Wang2019, Verzhbitskiy2020} This increase has been attributed to strong double-exchange upon electron doping~\cite{Wang2019} in addition to ferromagnetic superexchange in the undoped material~\cite{Bhoi2021, Watson2020}. Similar to the doping study of Wang {\it et al.}~\cite{Wang2019}, we find that previously unoccupied $e_g^\sigma$ ($d_{xz}$, $d_{yz}$) orbitals become occupied under pressure (see Fig.~\ref{fig:FMspectralfunction}(e) and Fig.~\ref{fig:occupations}). These orbitals may mediate a ferromagnetic double-exchange~\cite{Anderson1955} in addition to the ferromagnetic superexchange via the $a_{1g}$ ($d_{z^2}$) orbitals, which is already present at ambient pressure.~\cite{Wang2019} Based on the available data we can, however, not decide which exchange mechanism is dominant under pressure. Nevertheless, the apparent similarity between applying pressure~\cite{Bhoi2021, Fumega2020} and direct charge carrier injection~\cite{Wang2019, Verzhbitskiy2020, Burch2018, Cai2020, Deng2018, Huang2018, Jiang2018, Wang2018} certainly deserves further investigation for the whole family of van der Waals ferromagnets.

The importance of electronic correlations for layered van der Waals ferromagnets has been highlighted for {\cri}~\cite{Kvashnin2022}, {\cst}~\cite{Zhang2019} and other materials in this family~\cite{Lee2020}. We also find that correlations play a significant role. In the ferromagnetic state we find that especially the $d_{z^2}$ orbital is strongly correlated (see Fig.~\ref{fig:renormalizationfactor}(b)). For other orbitals the strength of electronic correlations increases only slightly with pressure (i.e. the quasiparticle weight decreases). The majority spin $d_{z^2}$ orbital is the most correlated orbital in {\cgt} at zero pressure. This changes at a pressure around $P \sim 3$\,GPa, where the minority spin $d_{z^2}$ electrons rapidly change their nature from being relatively uncorrelated to being the most correlated electrons in the material. At this pressure of $P \sim 3$\,GPa, the insulator-to-metal transition happens in our DFT+DMFT calculations. As the minority $d_{z^2}$ orbital occupation changes from essentially empty to slightly occupied near $E_{\rm F}$, the few $d_{z^2}$ carriers are substantially affected by electronic correlations. In general, we find the minority carriers to be more heavy than the majority carriers, in agreement with both experiment and theory for the related ferromagnet CoS$_2$~\cite{Fujiwara2022}.

Since {\cgt} is isostructural to {\cst} and Ge is chemically similar to Si, we can compare our DFT+DMFT spectral functions to ARPES experimental data and DFT+DMFT calculations for {\cst}~\cite{Zhang2019}. Our results for the $\Gamma$-A and P-Z paths agree reasonably well, in particular with spin-orbit coupling. For the path $\Gamma$-$X$ we can compare our DFT+DMFT spectral function in the ferromagnetic state at ambient pressure to ARPES results by Suzuki {\it et al.}~\cite{Suzuki2019}. Even though the experiment is carried out at $T=150$~K, the spectral function agrees well with out results at $T=100$~K (see Fig.~\ref{fig:FMspectralfunction}(a) and (c)). To our knowledge, quasiparticle weights and mass enhancements have not yet been determined in other studies on this class of materials.

\section*{Conclusions} 
We have performed DFT and DFT+DMFT calculations for the layered van der Waals ferromagnet {\cgt} under pressure. We estimated the Heisenberg exchange parameters for the material as a function of pressure. We showed that ferromagnetic exchange couplings increase rapidly, in absence of a structural phase transition, as the charge transfer gap closes. Since our Hamiltonian also contains long-range couplings, we were able to resolve both intra- and inter-layer exchange parameters. We showed that both intra- and inter-layer ferromagnetic couplings contribute to the observed rapid increase of the Curie-Weiss temperature under pressure, which is fully explained by our calculations.

Our estimates for the correlation strength of Cr orbitals show that the majority spin $d_{z^2}$ electrons are most strongly correlated at ambient pressure. The minority spin electrons start out relatively uncorrelated at low pressure and rapidly become more correlated at the insulator-to-metal transition. We observed that application of pressure also leads to the partial occupation of previously unoccupied Cr $e_g^\sigma$ ($d_{xz}$, $d_{yz}$) states, similar to what has been observed in electron doped crystals of {\cgt}.

The obvious similarity between the influence of pressure, which we investigated here, and the effect of charge doping reported in the literature deserves further attention, since the possibility to fabricate electronic devices based on {\cgt} crucially depends on combining the switchable resistivity and unusual magnetic properties of the material.

The spectral functions we calculated are in reasonable agreement with experimental and theoretical results for {\cgt} and the related material {\cst}. Our results may serve as a reference for comparison to future ARPES experiments on {\cgt}.

\acknowledgements
The computation in this work has been done using the facilities of the Supercomputer Center, the Institute for Solid State Physics, the University of Tokyo.
M.S.~was supported by JSPS KAKENHI grants No.~22H01181.
J.O.~was supported by JSPS KAKENHI grants No.~20K20522, No.~21H01003, and No.~21H01041.

\appendix

\section{Interpolation of crystal structures}
\label{app:interpolation}

\begin{figure}[htb]
    \includegraphics[width=\columnwidth]{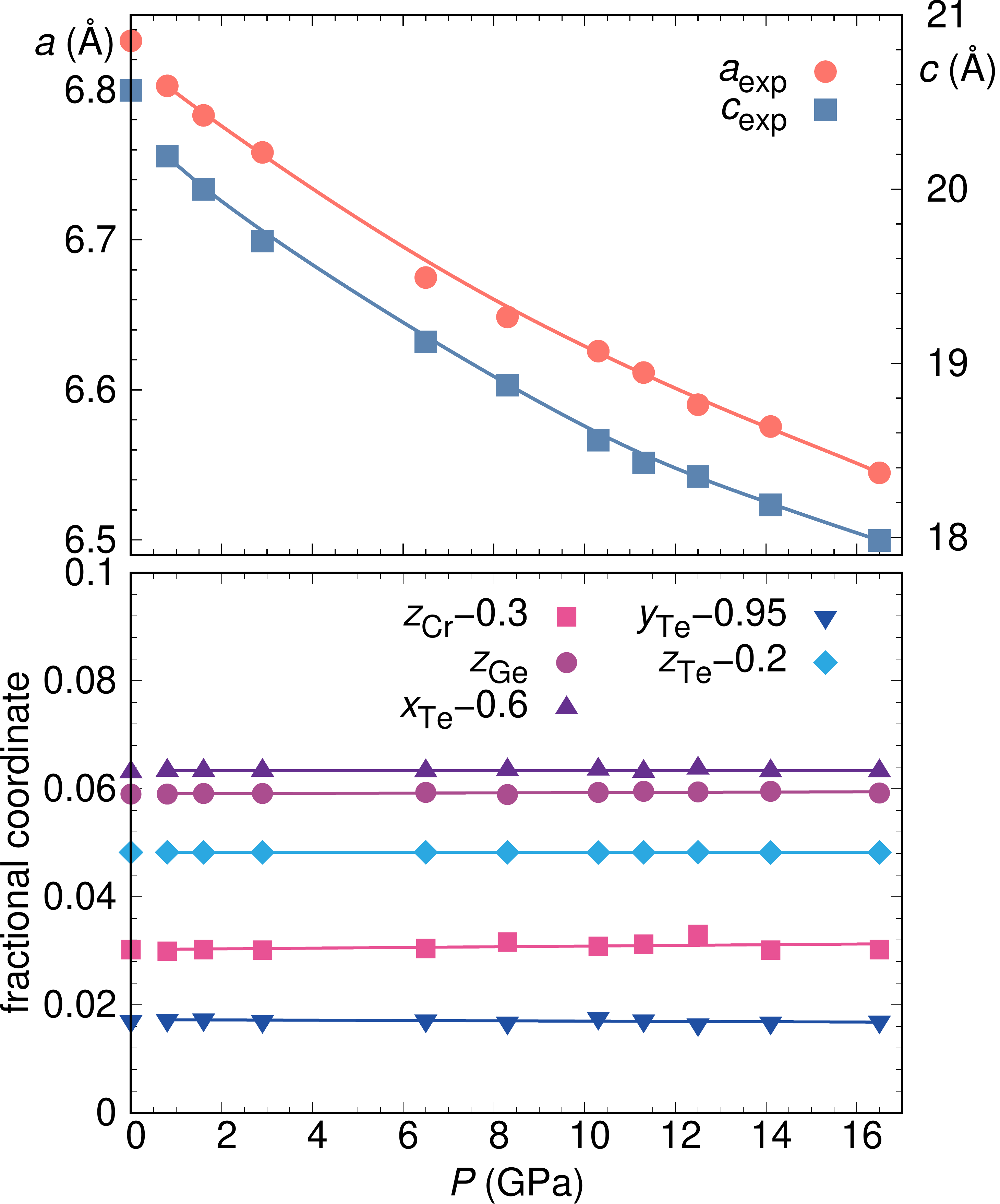}
    \caption{Experimental structural data of {\cgt} as function of pressure, shown together with interpolation. Data points are from Ref.~\cite{Yu2019}.}
    \label{fig:interpolation}
\end{figure}

We base our high pressure calculations for {\cgt} on the interpolation of the experimental crystal structures determined by Yu {\it et al.}, Ref.~\onlinecite{Yu2019}. In Figure~\ref{fig:interpolation}, we show the experimental data points together with our interpolation. This allows us to perform calculations on an equidistant set of high pressure structures. For the internal coordinates, we use only linear interpolation as the data points are rather noisy. 

\section{Ambient pressure energy mapping for {\cgt}}
\label{app:energymapping}

\begin{figure}[htb]
    \includegraphics[width=\columnwidth]{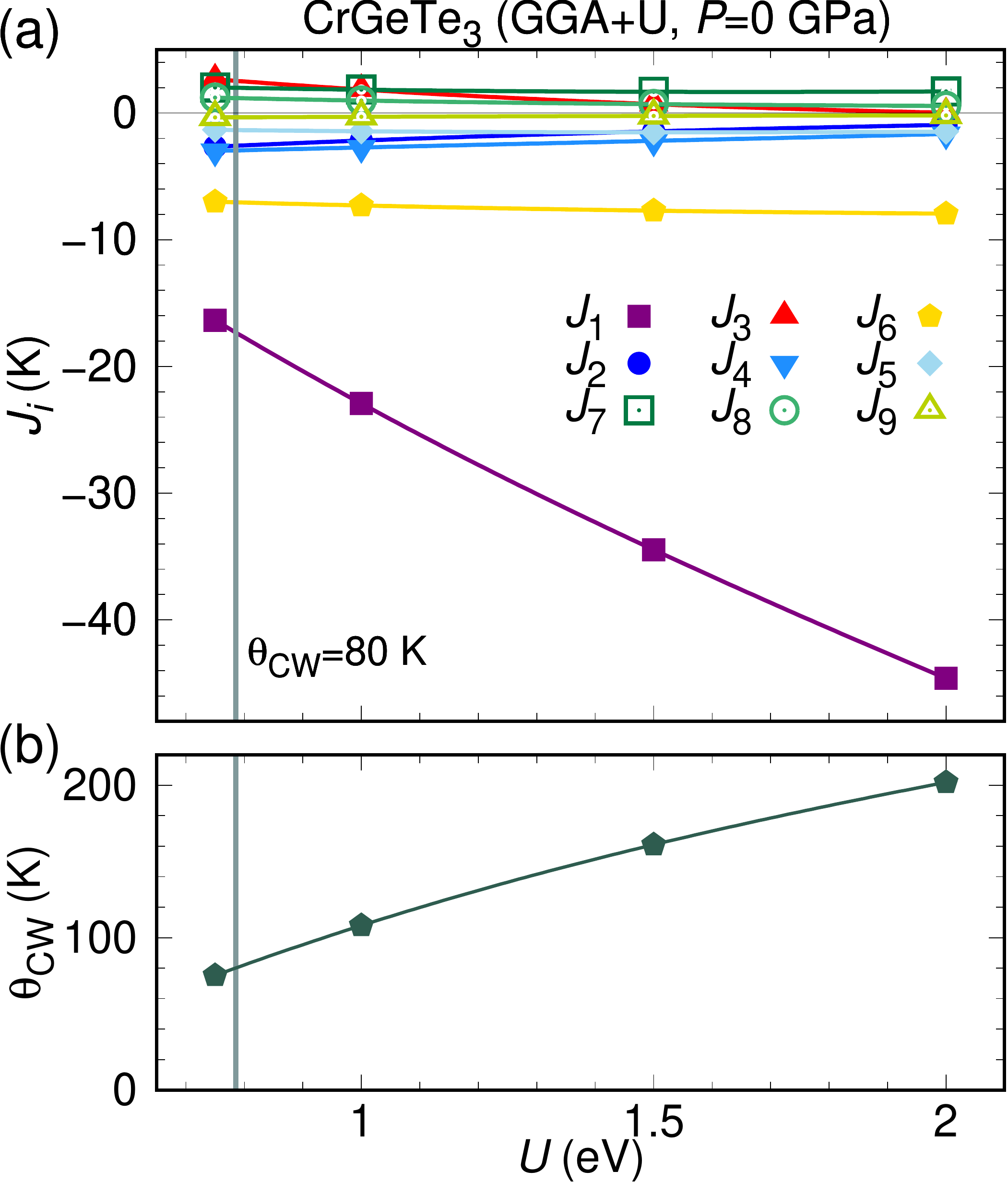}
    \caption{Exchange couplings of {\cgt} at ambient pressure as function of on-site interaction strength $U$, calculated by DFT energy mapping. The experimental value of the Curie-Weiss temperature $\theta_{CW}=80$\,K (Ref.~\onlinecite{Ji2013}) is matched by a low $U$ value of 0.79\,eV. }
    \label{fig:0GPacouplings}
\end{figure}

\begin{figure}[htb]
    \includegraphics[width=\columnwidth]{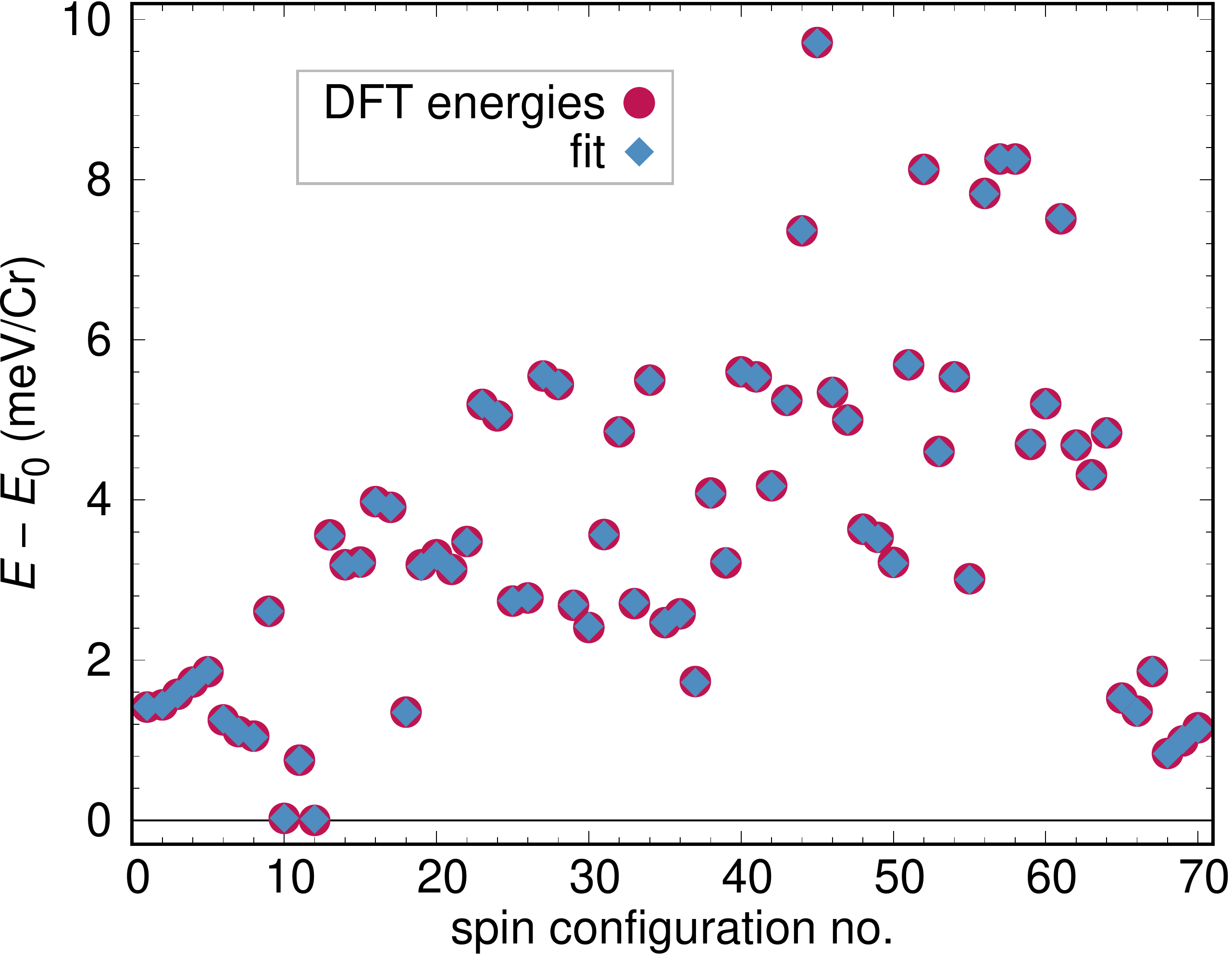}
    \caption{Quality of the energy mapping for {\cgt} at $P=0$\,GPa at an on-site interaction strength of $U=0.75$\,eV. The energies are plotted against the spin configurational space. The fit of the 70 DFT energies by nine exchange interactions is excellent.}
    \label{fig:0GPaenergies}
\end{figure}

The Heisenberg Hamiltonian parameters for {\cgt} at ambient pressure have been determined by several groups that use a mapping of DFT total energies at varying levels of sophistication. In Fig.~\ref{fig:0GPacouplings}, we present our calculation for the $P=0$\,GPa structure of Ref.~\onlinecite{Yu2019}. We use a 6-fold supercell of the primitive rhombohedral cell of {\cgt}, obtained using the linear transformation 
$$P=\begin{pmatrix}2&1&0\\-1&1&0\\0&0&2\end{pmatrix}\,.$$ Energies of 70 distinct spin configurations have been mapped to the nine exchange paths shown in Fig.~\ref{fig:couplings}\,(b). An example of the very good quality of the fit is shown in Fig.~\ref{fig:0GPaenergies}, indicating that the energy mapping procedure works very well for {\cgt}. Table~\ref{tab:0GPacouplings} shows the four calculated sets of couplings we determined. The last column gives the Curie-Weiss temperature calculated with Eq.~\ref{eq:tcw}. The line in bold face is interpolated to yield the experimental value of $\theta_{CW}=80$\,K determined in Ref.~\cite{Ji2013}. Note that a slightly different value of $\theta_{CW}=92$\,K was determined in Ref.~\cite{Zeisner2019}. We expect that the value of $U$, which is necessary to describe magnetism in {\cgt} correctly, does not vary significantly with pressure. Therefore, we use the value $U=0.75$\,eV, which is valid at ambient pressure, across the entire pressure range.

\begin{table*}[htb]
    \begin{tabular}{c|c|c|c|c|c|c|c|c|c|c}
    $U$\,(eV)&$J_1$\,(K)&$J_3$\,(K)&$J_2$\,(K)&$J_4$\,(K)&$J_6$\,(K)&$J_5$\,(K)&$J_7$\,(K)&$J_8$\,(K)&$J_9$\,(K)&$\theta_{CW}$\,(K)\\\hline
0.75 & -16.36(7) & 2.66(8) & -2.66(6) & -2.99(3) & -7.01(4) & -1.33(8) & 2.02(6) & 1.21(7) & -0.35(1) & 75 \\%# CrGeTe3, U=0.75, 4x4x4
{\bf 0.785} & {\bf -17.32(7)} & {\bf 2.54(8)} & {\bf -2.59(6)} & {\bf -2.95(3)} & {\bf -7.05(4)} & {\bf -1.35(8)} & {\bf 1.99(6)} & {\bf 1.17(7)} & {\bf -0.34(1)} & {\bf 80}\\
1 & -22.92(8) & 1.85(8) & -2.17(6) & -2.72(3) & -7.30(5) & -1.44(8) & 1.84(6) & 0.98(7) & -0.30(1) & 108 \\%# CrGeTe3, U=1, 4x4x4
1.5 & -34.47(7) & 0.70(8) & -1.45(6) & -2.20(3) & -7.71(4) & -1.52(8) & 1.67(6) & 0.70(7) & -0.23(1) & 161 \\%# CrGeTe3, U=1.5, 4x4x4
2 & -44.61(6) & 0.02(6) & -0.94(5) & -1.65(3) & -7.95(4) & -1.47(6) & 1.70(5) & 0.55(6) & -0.17(1) & 202 %# CrGeTe3, U=2, 4x4x4
    \end{tabular}
    \caption{Exchange interactions of {\cgt} at ambient pressure, calculated for four different values of the on-site interaction strength $U$. The line in bold face is interpolated to yield the experimental value of the Curie-Weiss temperature (Ref.~\cite{Ji2013}), using Eq.~\eqref{eq:tcw}. Statistical errors from the fitting procedure are given.}
    \label{tab:0GPacouplings}
\end{table*}

\section{Additional energy mapping results}

\begin{table*}[htb]
    \begin{tabular}{c|c|c|c|c|c|c|c|c|c|c}
    $P$\,(GPa)&$J_1$\,(K)&$J_3$\,(K)&$J_2$\,(K)&$J_4$\,(K)&$J_6$\,(K)&$J_5$\,(K)&$J_7$\,(K)&$J_8$\,(K)&$J_9$\,(K)&$\theta_{CW}$\,(K)\\\hline
0 & -16.36(7) & -2.66(6) & 2.66(8) & -2.99(3) & -1.33(8) & -7.01(4) & 2.02(6) & 1.21(7) & -0.35(1) & 75.1 \\%# CrGeTe3, U=0.75, 4x4x4
1 & -19.9(1) & -2.4(1) & 2.6(1) & -3.3(1) & -1.4(1) & -8.8(1) & 2.1(1) & 1.2(1) & -0.4(1) & 96.6 \\%# CrGeTe3, U=0.75, 4x4x4
2 & -22.7(2) & -2.2(1) & 2.7(2) & -3.5(1) & -1.5(2) & -10.4(1) & 2.1(1) & 1.2(1) & -0.4(1) & 113.2 \\%# CrGeTe3, U=0.75, 4x4x4
3 & -24.2(2) & -2.1(1) & 2.9(1) & -3.6(1) & -1.3(2) & -11.5(1) & 2.3(1) & 1.4(1) & -0.3(1) & 118.3 \\%# CrGeTe3, U=0.75, 4x4x4
4 & -25.8(2) & -1.8(1) & 2.9(2) & -3.7(1) & -1.4(2) & -12.7(1) & 2.2(2) & 1.3(2) & -0.4(1) & 130.4 \\%# CrGeTe3, U=0.75, 4x4x4
5 & -27.5(2) & -1.5(2) & 3.0(2) & -3.8(1) & -1.4(2) & -14.0(1) & 2.3(2) & 1.4(2) & -0.4(1) & 140.1 \\%# CrGeTe3, U=0.75, 4x4x4
6 & -29.4(2) & -1.1(2) & 3.0(2) & -3.8(1) & -1.4(2) & -15.4(2) & 2.2(2) & 1.3(2) & -0.4(1) & 152.2 \\%# CrGeTe3, U=0.75, 4x4x4
7 & -34.8(1.1) & 1.9(9) & 0.8(1.2) & -3.6(5) & -2.9(1.2) & -17.5(7) & 0.2(9) & -0.3(1.1) & -0.5(2) & 226. \\%# CrGeTe3, U=0.75, 4x4x4
8 & -44.1(2.4) & 6.6(1.9) & -4.1(2.6) & -3.3(1.0) & -6.2(2.6) & -20.9(1.5) & -3.6(2.0) & -3.6(2.4) & -0.5(3) & 370.0   %# CrGeTe3, U=0.75, 4x4x4
    \end{tabular}
    \caption{Exchange interactions of {\cgt} as function of pressure, calculated for the on-site interaction strength $U=0.75$\,eV. Statistical errors from the fitting procedure are given.}
    \label{tab:couplings}
\end{table*}

\begin{figure}[htb]
    \includegraphics[width=\columnwidth]{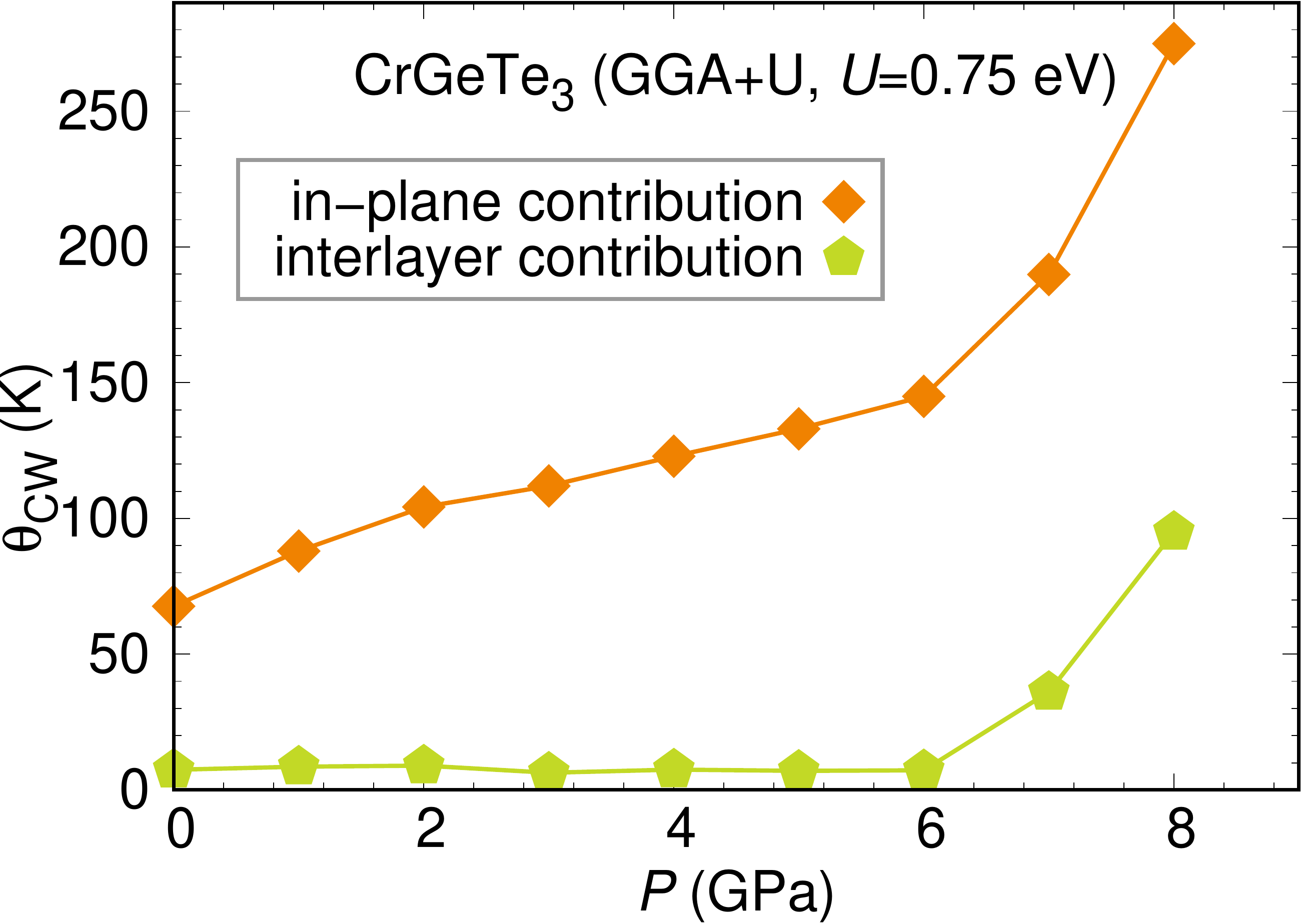}
    \caption{In-plane and interlayer contributions to the predicted Curie-Weiss temperature of {\cgt} as function of pressure, determined from Eq.~\eqref{eq:tcw}. }
    \label{fig:tcwparts}
\end{figure}

\begin{figure}[htb]
    \includegraphics[width=\columnwidth]{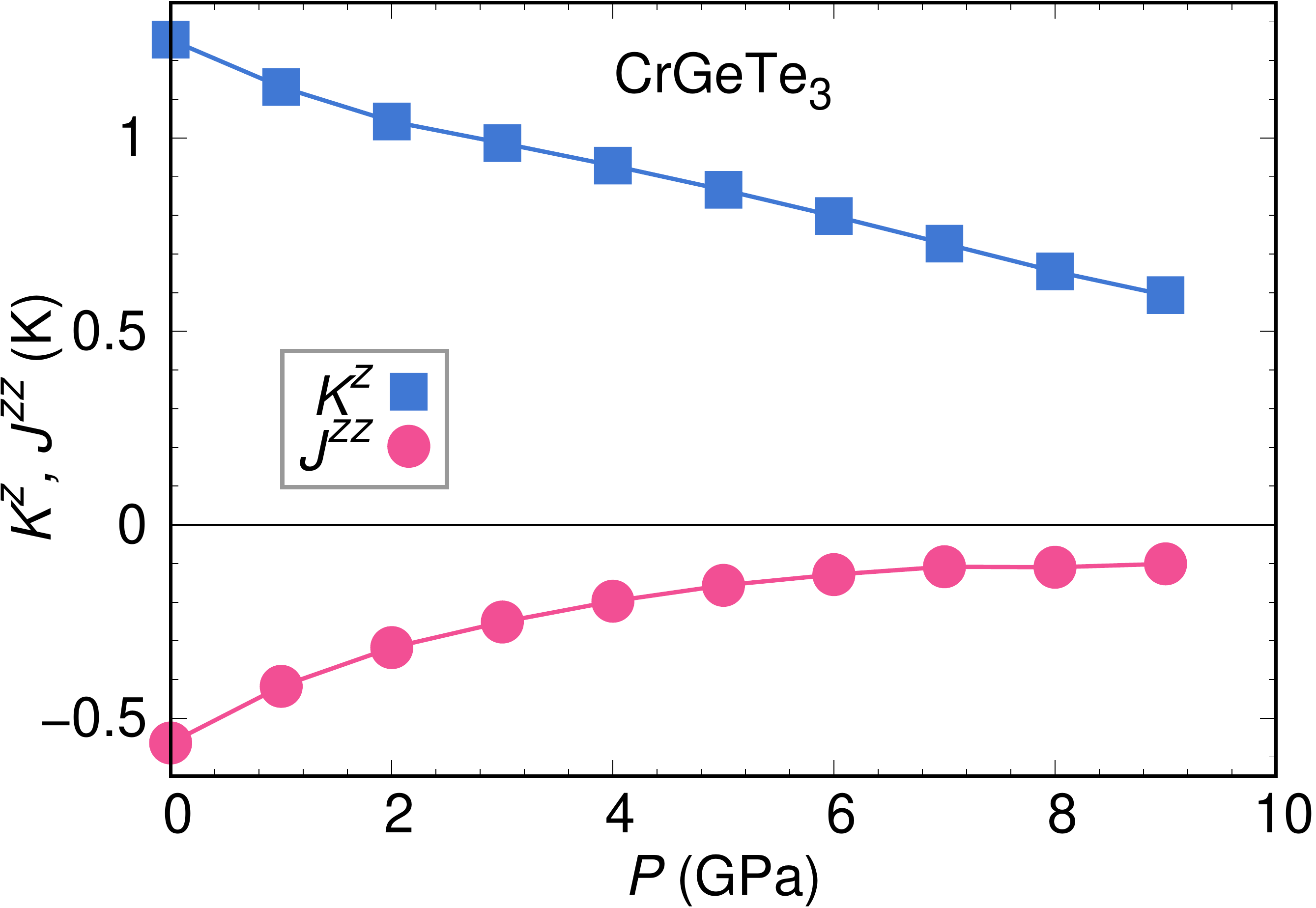}
    \caption{Single ion anisotropy $K^z$ and exchange anisotropy $J^{zz}$ of {\cgt} as function of pressure, calculated for the on-site interaction strength $U=0.75$\,eV. }
    \label{fig:anisotropy}
\end{figure}

We present the detailed results of the DFT energy mapping for {\cgt} as function of pressure in Table~\ref{tab:couplings}. The on-site interaction strength $U=0.75$\,eV is chosen based on the ambient pressure calculation (see Appendix~\ref{app:energymapping}). The Hund's rule coupling is fixed at $J_{\rm H}=0.72$\,eV~\cite{Mizokawa1996}. The data of Table~\ref{tab:couplings} are plotted in Figs.~\ref{fig:couplings}\,(a) and (c). The exchange paths are illustrated in Fig.~\ref{fig:couplings}\,(b). Based on the in-plane couplings $J_1$, $J_3$ and $J_6$, we can partially perform the sum in Eq.~\ref{eq:tcw} and obtain the in-plane contribution to the Curie-Weiss temperature; using the six interlayer couplings, we can obtain an interlayer contribution. These two parts of the Curie-Weiss temperature are shown in Fig.~\ref{fig:tcwparts}.

As mentioned above, magnetic order in van der Waals magnets occurs partly due to anisotropies. We add two anisotropic terms to the Hamiltonian $H$ of Eq.~\eqref{eq:hamiltonian}:
\begin{equation}
    H_{\rm anisotropic}=H+ \sum_{i<j} J^{zz} S^z_i S^z_j +K^z (S^z_i)^2\,,\label{eq:Haniso}
\end{equation}
where $J^{zz}$ is nearest neighbor exchange anisotropy and $K^z$ is the single ion anisotropy. Note that {\cgt} is isotropic in the $ab$ plane. The parameters are obtained from an energy mapping~\cite{Torelli2020} via
\begin{equation}
    \begin{split}
        K^z&=\frac{1}{2S^2}\big(E_{\rm FM}^z-E_{\rm FM}^{xy}+E_{\rm AFM}^z-E_{\rm AFM}^{xy}\big)\\
        J^{zz}&=\frac{1}{NS^2}\big(E_{\rm FM}^z-E_{\rm FM}^{xy}-E_{\rm AFM}^z+E_{\rm AFM}^{xy}\big)\,,\label{eq:KzJzz}
    \end{split}
\end{equation}
where $E_{\rm FM}^z$ and $E_{\rm FM}^{xy}$ are ferromagnetic energies with moments along $z$ and in the $xy$ plane, respectively. $E_{\rm AFM}^z$ and $E_{\rm AFM}^{xy}$ are the corresponding energies for the N{\'e}el state, and $N=3$ is the number of nearest neigbhors in the honeycomb lattice. Figure~\ref{fig:anisotropy} shows the evolution of $K^z$ and $J^{zz}$ with pressure, calculated with the same interaction parameters as the isotropic exchange in Fig.~\ref{fig:couplings}\,(a). The single ion anisotropy is substantial at ambient pressure and continously decreases with pressure. The exchange anisotropy initially decreases rapidly with pressure; in the metallic state, exchange becomes nearly isotropic. The quantity $\frac{3}{2}J^{zz}+K^z$, which determines the energy cost of spin orientation along the $z$ axis (easy axis) compared to within the $xy$ plane (easy plane), is very small at 0.4\,K to 0.6\,K for all pressures. It has been noted~\cite{Zeisner2019,Gong2017} that slightly larger $U$ values lead to a sign change of the single ion anisotropy, which brings DFT+U calculations into agreement with the small easy axis anisotropy observed experimentally.

\section{Additional DMFT results}

\begin{figure}[htb]
    \includegraphics[width=\columnwidth]{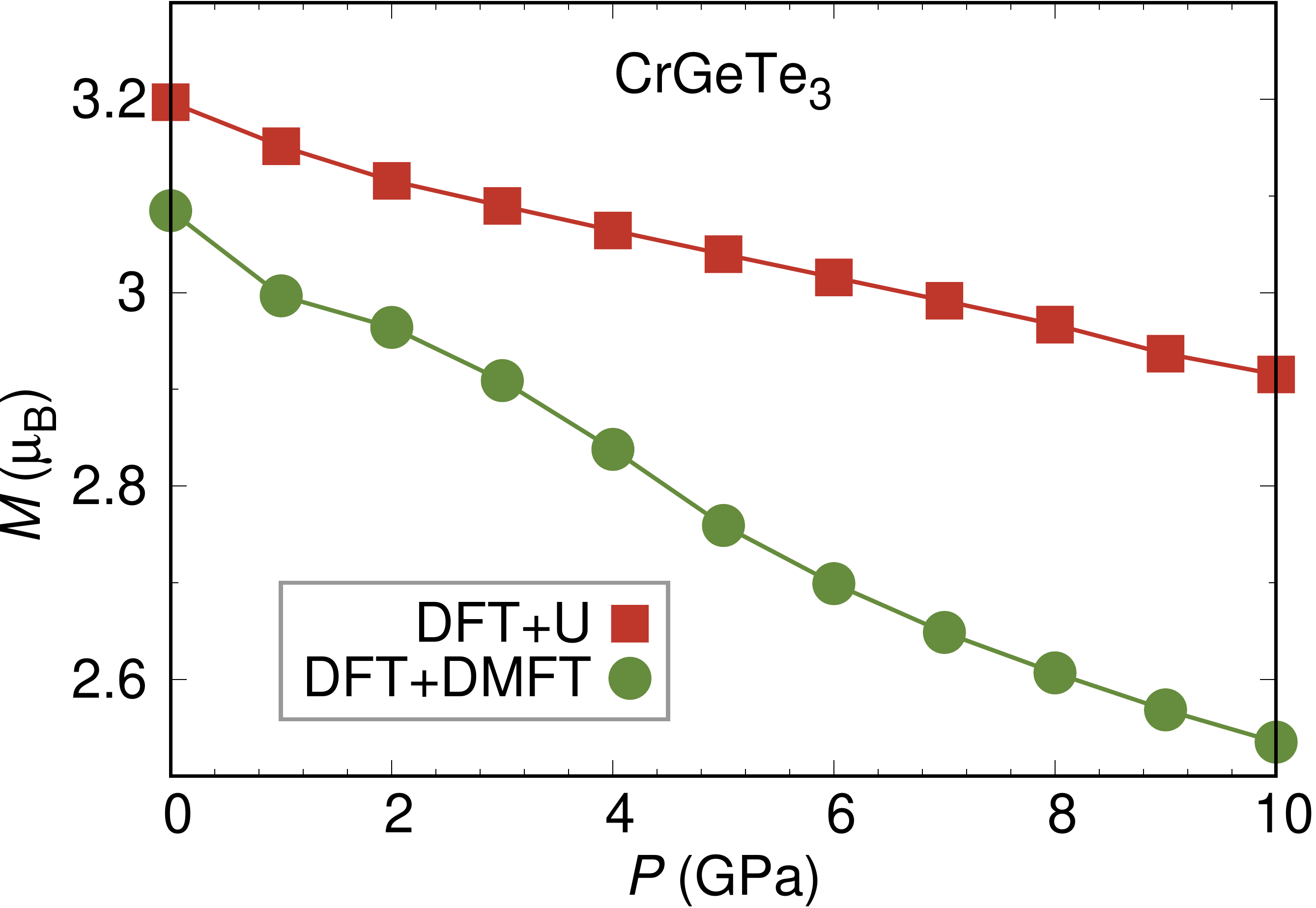}
    \caption{Cr$^{3+}$ magnetic moments as function of pressure. The DFT+U moments are averaged staggered moments from the 70 spin configurations used in the energy mapping, calculated at $U=0.75$\,eV.  }
    \label{fig:moments}
\end{figure}

\begin{figure}[htb]
    \includegraphics[width=\columnwidth]{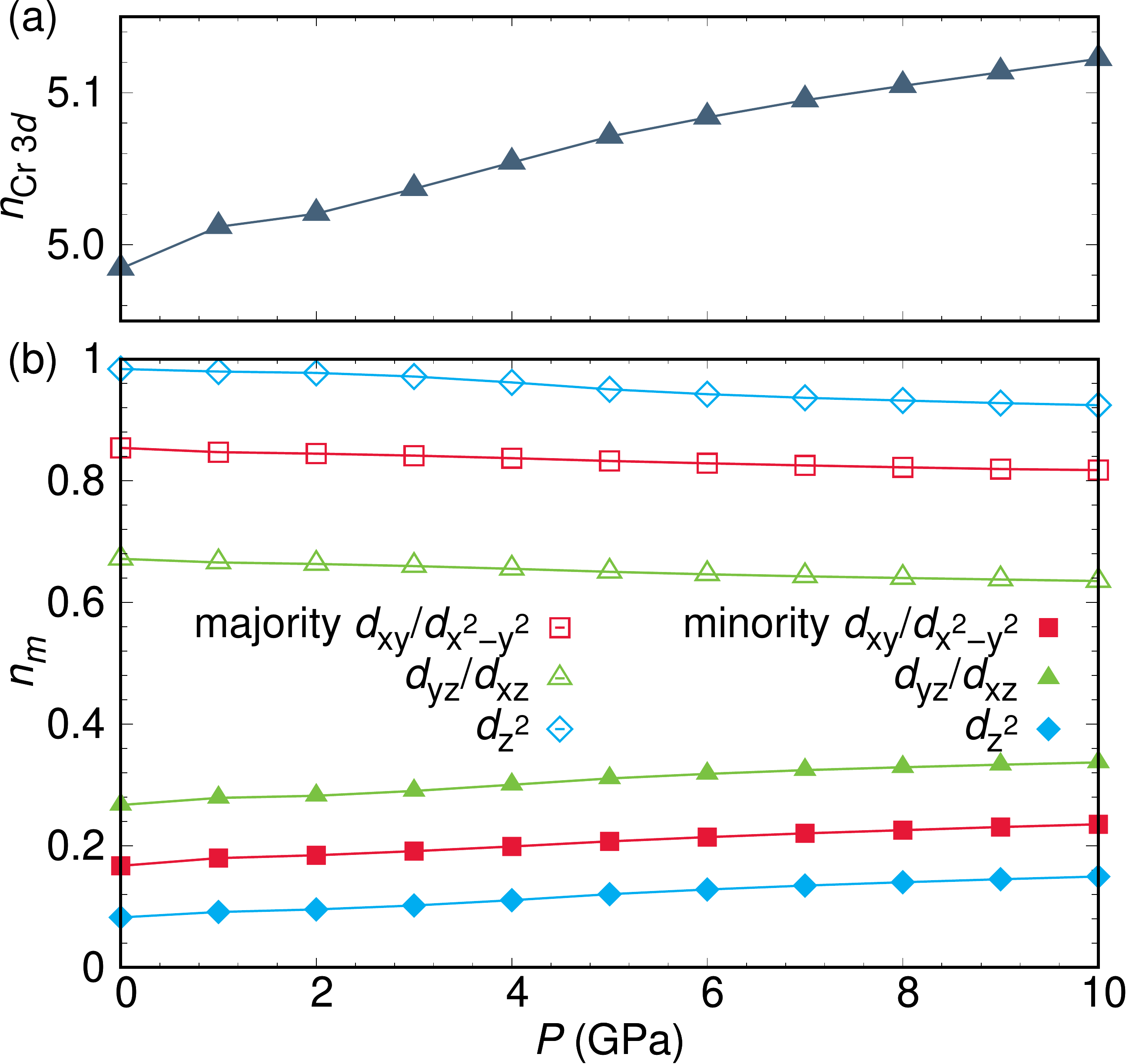}
    \caption{DFT+DMFT occupation of Cr $3d$ orbitals. (a) Total $3d$ occupation number. (b) Occupations $n_m$ resolved by orbital $m$. }
    \label{fig:occupations}
\end{figure}

In Fig.~\ref{fig:moments}, we show the evolution of the moments calculated within DFT+DMFT at $T=100$\,K in the ferromagnetic state with pressure. Moments are monotonously falling over the entire pressure range. The decrease of the moments appears to accelerate a bit at the insulator-to-metal transition but deviations from a constant rate of decrease are small. We compare the DFT+DMFT moments to the staggered moments obtained by averaging over the 70 DFT+U spin configurations per pressure value. DFT+U moments are initially slightly higher and are decreasing at a slightly lower rate compared to the DFT+DMFT moments. With increasing pressure, as Cr-Te distances decrease, covalency and Cr-Te hybridization increase; this leads to less localized magnetic moments and thus to a decrease in the size of the moment that can be ascribed to Cr. Note that we work with 64 antiferromagnetic spin configurations where total moment zero is unaffected by individual Cr moment size; the total moment $6\,\mu_{\rm B}$ of the other 6 configurations is precisely realized below 7~GPa and deviations are only 3\% even at 9~GPa. The DFT+DMFT moments decrease because increasing pressure leads to broadening of Cr $3d$ bands and thus partial occupation of minority $3d$ states and partial de-occupation of majority $3d$ states.

In Fig.~\ref{fig:occupations}, we show the occupation numbers from the DFT+DMFT calculations as function of pressure. In Fig.~\ref{fig:occupations}\,(a), a slow but steady increase of the total Cr $3d$ filling is found. The orbital resolved occupation numbers in Fig.~\ref{fig:occupations}\,(b) show a slow and nearly linear increase of minority occupation numbers and a similar decrease of majority occupation numbers.

\section{Details of the DMFT calculations}
\label{app:DMFT}
We study the effect of electronic correlations using dynamical mean-field theory (DMFT)~\cite{Georges1996,Kotliar2006}.
The Matsubara Green's function of the 36-band model is given by
\begin{equation}
\begin{split}
\hat{G}_{\sigma}(i\omega_n) = [ & i\omega_n + \mu - \hat{H}_\mathrm{DFT}(\bm{k}) \\ & + \hat{\Sigma}_\mathrm{dc} - \hat{\Sigma}_\mathrm{loc}(i\omega_n)]^{-1},  
\end{split}
\end{equation}
where quantities with hat are $(36 \times 36)$-matrices for the band indices.
$\hat{H}_\mathrm{DFT}(\bm{k})$ is the one-body Hamiltonian extracted from DFT calculations.
$\hat{\Sigma}_\mathrm{dc}$ is a double-counting correction. We apply the Hartree-Fock approximation to estimate $\hat{\Sigma}_\mathrm{dc}$ using the bare Green's function, $\hat{G}_{0, \sigma}(i\omega_n) = [i\omega_n + \mu - \hat{H}_\mathrm{DFT}(\bm{k})]^{-1}$.
$\hat{\Sigma}_\mathrm{loc}(i\omega_n)$ is the local (momentum-independent) self-energy to be computed by DMFT. As in the case of \ce{CeB6}~\cite{Otsuki2022}, we use a slight reduction of the double counting correction by 4\%.

We consider correlations only between electrons on the Cr $3d$ orbitals so that the local self-energy $\hat{\Sigma}_\mathrm{loc}(i\omega_n)$ has non-zero matrix elements only on the diagonals that represent the $3d$ orbitals on each Cr atom.
We represent these diagonal components of the self-energy as $\Sigma_{\alpha\sigma}^{i}(i\omega_n)$,
where the subscripts $\alpha=d_{z^2}, d_{xy}, d_{x^2-y^2}, d_{xy}, d_{yz}$ and $\sigma=\uparrow, \downarrow$ denote the orbital and spin components, respectively. The superscript $i=1,2$ is an index for two symmetry equivalent Cr atoms.
The local self-energy is computed by solving a single-impurity Anderson model.
We consider density-density terms in the Kanamori-type interaction, which is given by
\begin{equation}\begin{split}
    \mathcal{H}_\mathrm{int} = \, &U \sum_{\alpha} n_{\alpha \uparrow} n_{\alpha \downarrow}
    + U' \sum_{\sigma} n_{1\sigma} n_{2\bar{\sigma}}
    \\&+ (U'-J_{\rm H}) \sum_{\sigma} n_{1\sigma} n_{2\sigma},
\end{split}\end{equation}
where $\bar{\sigma}$ stands for the spin component opposite to $\sigma$.
The inter-orbital Coulomb interaction $U'$ is determined from $U$ and $J_H$ by $U'=U-2J_{\rm H}$.
The intra-orbital Coulomb interaction $U$ and the Hund's rule coupling $J_{\rm H}$ are fixed at $U=2$\,eV and $J_{\rm H}=0.72$\,eV. 
We performed the DMFT self-consistency calculations using DCore~\cite{Shinaoka2021} implemented on the TRIQS library~\cite{Aichhorn2016}.
The single-impurity problem was solved using an implementation~\cite{jo-cthyb} of the hybridization-expansion CT-QMC method~\cite{Werner2006,Gull2011} integrated into DCore.
Summations over $\bm{k}$ and $\omega_n$ are performed with $n_k=20^3$ points and $n_{iw}=3000$ points (for positive frequencies) at room temperature, $n_{iw}=9000$ at $T=100$\,K.
The momentum resolved spectral function for real frequencies $\omega$,  $A(\bm{k}, \omega) = -1/\pi \,\mathrm{Im}\, \mathrm{Tr}\, \hat{G}(\omega+i0)$
was computed by analytical continuation from Matsubara frequency to real frequency using the Pad\'e approximation~\cite{Vidberg1977}.

\section{Transforming FPLO Wannier functions to DCore input}
The projective Wannier function module of the FPLO code~\cite{Koepernik2023} provides hopping parameters, $t_{lm}(\Delta\bm{r})$, where $l$ and $m$ indices specify the lattice basis like orbital and sub-lattice, and $\Delta\bm{r}$ is a transfer vector of the hopping. On the other hand, DCore~\cite{Shinaoka2021} supports hopping parameters $t_{lm}(\Delta\bm{R})$ with transfer vectors $\Delta\bm{R}$ between Bravais lattice vectors. To relabel hopping parameters from $t_{lm}(\Delta\bm{r})$ to $t_{lm}(\Delta\bm{R})$, we use the relation
\begin{equation}
  \Delta{r} = \Delta{R} + \bm{\tau}_l - \bm{\tau}_m
\end{equation}
where $\bm{\tau}_l$ and $\bm{\tau}_m$ are basis positions in the lattice. While this is in princple enough to convert FPLO to Wannier90 format, we now give some details about the conversion of the Hamiltonian.
The FPLO hopping parameters $t_{lm}(\Delta\bm{r})$
assume a Hamiltonian,
\begin{equation}
  h_{lm}(\bm{k}) = \sum_{ \bm{R}, \bm{R}' } t_{lm} ( \bm{r}_i - \bm{r}_j ) e^{ -i \bm{k} \cdot ( \bm{r}_i - \bm{r}_j ) }
\end{equation}
while DCore supports a Hamiltonian of the form
\begin{equation}
  H_{lm}(\bm{k}) = \sum_{ \bm{R}, \bm{R}' } t_{lm} ( \bm{R} - \bm{R}' ) e^{ -i \bm{k} \cdot ( \bm{R} - \bm{R}' ) }
\end{equation}
where $l$ and $m$ represent the lattice basis, for example, orbital and sub-lattice, $\bm{R}$ and $\bm{R}'$ are Bravais lattice vectors, and $\bm{r}_i$ and $\bm{r}_j$ are positions of sites. These vectors are related by
\begin{equation}
\begin{split}
  \bm{r}_i &= \bm{R}  + \bm{\tau}_l \\
  \bm{r}_j &= \bm{R}' + \bm{\tau}_m
\end{split}
\label{eq:latticevectors}
\end{equation}
To transform $h_{lm}(\bm{k})$ to $H_{lm}(\bm{k})$, the following two steps are needed. First, the hopping parameters are relabeled from $t_{lm}(\bm{r}_i-\bm{r}_j)$ to $t_{lm}(\bm{R}-\bm{R}')$ with the relation of Eq.~\ref{eq:latticevectors}. Second, the phase factor is changed.
\begin{equation}
\begin{split}
  H_{lm}(\bm{k})
  &= \sum_{ \bm{R}, \bm{R}' } t_{lm} ( \bm{r} - \bm{r}' ) e^{ -i \bm{k} \cdot [ ( \bm{r}_i + \bm{\tau}_l ) - ( \bm{r}_j + \bm{\tau}_m ) ] } \\
  &= e^{ i\bm{k} \cdot ( \bm{\tau}_l - \bm{\tau}_m ) } h_{lm}(\bm{k})
\end{split}
\end{equation}

\bibliography{main}

\begin{thebibliography}{83}%
\makeatletter
\providecommand \@ifxundefined [1]{%
 \@ifx{#1\undefined}
}%
\providecommand \@ifnum [1]{%
 \ifnum #1\expandafter \@firstoftwo
 \else \expandafter \@secondoftwo
 \fi
}%
\providecommand \@ifx [1]{%
 \ifx #1\expandafter \@firstoftwo
 \else \expandafter \@secondoftwo
 \fi
}%
\providecommand \natexlab [1]{#1}%
\providecommand \enquote  [1]{``#1''}%
\providecommand \bibnamefont  [1]{#1}%
\providecommand \bibfnamefont [1]{#1}%
\providecommand \citenamefont [1]{#1}%
\providecommand \href@noop [0]{\@secondoftwo}%
\providecommand \href [0]{\begingroup \@sanitize@url \@href}%
\providecommand \@href[1]{\@@startlink{#1}\@@href}%
\providecommand \@@href[1]{\endgroup#1\@@endlink}%
\providecommand \@sanitize@url [0]{\catcode `\\12\catcode `\$12\catcode
  `\&12\catcode `\#12\catcode `\^12\catcode `\_12\catcode `\%12\relax}%
\providecommand \@@startlink[1]{}%
\providecommand \@@endlink[0]{}%
\providecommand \url  [0]{\begingroup\@sanitize@url \@url }%
\providecommand \@url [1]{\endgroup\@href {#1}{\urlprefix }}%
\providecommand \urlprefix  [0]{URL }%
\providecommand \Eprint [0]{\href }%
\providecommand \doibase [0]{https://doi.org/}%
\providecommand \selectlanguage [0]{\@gobble}%
\providecommand \bibinfo  [0]{\@secondoftwo}%
\providecommand \bibfield  [0]{\@secondoftwo}%
\providecommand \translation [1]{[#1]}%
\providecommand \BibitemOpen [0]{}%
\providecommand \bibitemStop [0]{}%
\providecommand \bibitemNoStop [0]{.\EOS\space}%
\providecommand \EOS [0]{\spacefactor3000\relax}%
\providecommand \BibitemShut  [1]{\csname bibitem#1\endcsname}%
\let\auto@bib@innerbib\@empty
%</preamble>
\bibitem [{\citenamefont {Ouvrard}\ \emph {et~al.}(1988)\citenamefont
  {Ouvrard}, \citenamefont {Sandre},\ and\ \citenamefont {Brec}}]{Ouvrard1988}%
  \BibitemOpen
  \bibfield  {author} {\bibinfo {author} {\bibfnamefont {G.}~\bibnamefont
  {Ouvrard}}, \bibinfo {author} {\bibfnamefont {E.}~\bibnamefont {Sandre}},\
  and\ \bibinfo {author} {\bibfnamefont {R.}~\bibnamefont {Brec}},\ }\bibfield
  {title} {\bibinfo {title} {Synthesis and crystal structure of a new layered
  phase: The chromium hexatellurosilicate \ce{Cr2Si2Te6}},\ }\href
  {https://doi.org/https://doi.org/10.1016/0022-4596(88)90049-7} {\bibfield
  {journal} {\bibinfo  {journal} {J. Solid State Chem.}\ }\textbf {\bibinfo
  {volume} {73}},\ \bibinfo {pages} {27} (\bibinfo {year} {1988})}\BibitemShut
  {NoStop}%
\bibitem [{\citenamefont {Marsh}(1988)}]{Marsh1988}%
  \BibitemOpen
  \bibfield  {author} {\bibinfo {author} {\bibfnamefont {R.~E.}\ \bibnamefont
  {Marsh}},\ }\bibfield  {title} {\bibinfo {title} {The crystal structure of
  \ce{Cr2Si2Te6}: Corrigendum},\ }\href
  {https://doi.org/https://doi.org/10.1016/0022-4596(88)90107-7} {\bibfield
  {journal} {\bibinfo  {journal} {J. Solid State Chem.}\ }\textbf {\bibinfo
  {volume} {77}},\ \bibinfo {pages} {190} (\bibinfo {year} {1988})}\BibitemShut
  {NoStop}%
\bibitem [{\citenamefont {Carteaux}\ \emph {et~al.}(1991)\citenamefont
  {Carteaux}, \citenamefont {Ouvrard}, \citenamefont {Grenier},\ and\
  \citenamefont {Laligant}}]{Carteaux1991}%
  \BibitemOpen
  \bibfield  {author} {\bibinfo {author} {\bibfnamefont {V.}~\bibnamefont
  {Carteaux}}, \bibinfo {author} {\bibfnamefont {G.}~\bibnamefont {Ouvrard}},
  \bibinfo {author} {\bibfnamefont {J.}~\bibnamefont {Grenier}},\ and\ \bibinfo
  {author} {\bibfnamefont {Y.}~\bibnamefont {Laligant}},\ }\bibfield  {title}
  {\bibinfo {title} {Magnetic structure of the new layered ferromagnetic
  chromium hexatellurosilicate \ce{Cr2Si2Te6}},\ }\href
  {https://doi.org/https://doi.org/10.1016/0304-8853(91)90121-P} {\bibfield
  {journal} {\bibinfo  {journal} {J. Mag. Mag. Mater.}\ }\textbf {\bibinfo
  {volume} {94}},\ \bibinfo {pages} {127} (\bibinfo {year} {1991})}\BibitemShut
  {NoStop}%
\bibitem [{\citenamefont {Carteaux}\ \emph
  {et~al.}(1995{\natexlab{a}})\citenamefont {Carteaux}, \citenamefont
  {Moussa},\ and\ \citenamefont {Spiesser}}]{Carteaux1995b}%
  \BibitemOpen
  \bibfield  {author} {\bibinfo {author} {\bibfnamefont {V.}~\bibnamefont
  {Carteaux}}, \bibinfo {author} {\bibfnamefont {F.}~\bibnamefont {Moussa}},\
  and\ \bibinfo {author} {\bibfnamefont {M.}~\bibnamefont {Spiesser}},\
  }\bibfield  {title} {\bibinfo {title} {2d ising-like ferromagnetic behaviour
  for the lamellar \ce{Cr2Si2Te6} compound: A neutron scattering
  investigation},\ }\href {https://doi.org/10.1209/0295-5075/29/3/011}
  {\bibfield  {journal} {\bibinfo  {journal} {Europhys. Lett.}\ }\textbf
  {\bibinfo {volume} {29}},\ \bibinfo {pages} {251} (\bibinfo {year}
  {1995}{\natexlab{a}})}\BibitemShut {NoStop}%
\bibitem [{\citenamefont {Carteaux}\ \emph
  {et~al.}(1995{\natexlab{b}})\citenamefont {Carteaux}, \citenamefont {Brunet},
  \citenamefont {Ouvrard},\ and\ \citenamefont {Andre}}]{Carteaux1995}%
  \BibitemOpen
  \bibfield  {author} {\bibinfo {author} {\bibfnamefont {V.}~\bibnamefont
  {Carteaux}}, \bibinfo {author} {\bibfnamefont {D.}~\bibnamefont {Brunet}},
  \bibinfo {author} {\bibfnamefont {G.}~\bibnamefont {Ouvrard}},\ and\ \bibinfo
  {author} {\bibfnamefont {G.}~\bibnamefont {Andre}},\ }\bibfield  {title}
  {\bibinfo {title} {Crystallographic, magnetic and electronic structures of a
  new layered ferromagnetic compound \ce{Cr2Ge2Te6}},\ }\href
  {https://doi.org/10.1088/0953-8984/7/1/008} {\bibfield  {journal} {\bibinfo
  {journal} {J. Phys.: Condens. Matter}\ }\textbf {\bibinfo {volume} {7}},\
  \bibinfo {pages} {69} (\bibinfo {year} {1995}{\natexlab{b}})}\BibitemShut
  {NoStop}%
\bibitem [{\citenamefont {Gong}\ \emph {et~al.}(2017)\citenamefont {Gong},
  \citenamefont {Li}, \citenamefont {Li}, \citenamefont {Ji}, \citenamefont
  {Stern}, \citenamefont {Xia}, \citenamefont {Cao}, \citenamefont {Bao},
  \citenamefont {Wang}, \citenamefont {Wang}, \citenamefont {Qiu},
  \citenamefont {Cava}, \citenamefont {Louie}, \citenamefont {Xia},\ and\
  \citenamefont {Zhang}}]{Gong2017}%
  \BibitemOpen
  \bibfield  {author} {\bibinfo {author} {\bibfnamefont {C.}~\bibnamefont
  {Gong}}, \bibinfo {author} {\bibfnamefont {L.}~\bibnamefont {Li}}, \bibinfo
  {author} {\bibfnamefont {Z.}~\bibnamefont {Li}}, \bibinfo {author}
  {\bibfnamefont {H.}~\bibnamefont {Ji}}, \bibinfo {author} {\bibfnamefont
  {A.}~\bibnamefont {Stern}}, \bibinfo {author} {\bibfnamefont
  {Y.}~\bibnamefont {Xia}}, \bibinfo {author} {\bibfnamefont {T.}~\bibnamefont
  {Cao}}, \bibinfo {author} {\bibfnamefont {W.}~\bibnamefont {Bao}}, \bibinfo
  {author} {\bibfnamefont {C.}~\bibnamefont {Wang}}, \bibinfo {author}
  {\bibfnamefont {Y.}~\bibnamefont {Wang}}, \bibinfo {author} {\bibfnamefont
  {Z.~Q.}\ \bibnamefont {Qiu}}, \bibinfo {author} {\bibfnamefont {R.~J.}\
  \bibnamefont {Cava}}, \bibinfo {author} {\bibfnamefont {S.~G.}\ \bibnamefont
  {Louie}}, \bibinfo {author} {\bibfnamefont {J.}~\bibnamefont {Xia}},\ and\
  \bibinfo {author} {\bibfnamefont {X.}~\bibnamefont {Zhang}},\ }\bibfield
  {title} {\bibinfo {title} {Discovery of intrinsic ferromagnetism in
  two-dimensional van der {W}aals crystals},\ }\href
  {https://doi.org/10.1038/nature22060} {\bibfield  {journal} {\bibinfo
  {journal} {Nature}\ }\textbf {\bibinfo {volume} {546}},\ \bibinfo {pages}
  {265} (\bibinfo {year} {2017})}\BibitemShut {NoStop}%
\bibitem [{\citenamefont {Li}\ and\ \citenamefont {Yang}(2014)}]{Li2014}%
  \BibitemOpen
  \bibfield  {author} {\bibinfo {author} {\bibfnamefont {X.}~\bibnamefont
  {Li}}\ and\ \bibinfo {author} {\bibfnamefont {J.}~\bibnamefont {Yang}},\
  }\bibfield  {title} {\bibinfo {title} {\ce{CrXTe3} ({X} = {S}i{,} {G}e)
  nanosheets: two dimensional intrinsic ferromagnetic semiconductors},\ }\href
  {https://doi.org/10.1039/C4TC01193G} {\bibfield  {journal} {\bibinfo
  {journal} {J. Mater. Chem. C}\ }\textbf {\bibinfo {volume} {2}},\ \bibinfo
  {pages} {7071} (\bibinfo {year} {2014})}\BibitemShut {NoStop}%
\bibitem [{\citenamefont {McGuire}\ \emph {et~al.}(2015)\citenamefont
  {McGuire}, \citenamefont {Dixit}, \citenamefont {Cooper},\ and\ \citenamefont
  {Sales}}]{McGuire2015}%
  \BibitemOpen
  \bibfield  {author} {\bibinfo {author} {\bibfnamefont {M.~A.}\ \bibnamefont
  {McGuire}}, \bibinfo {author} {\bibfnamefont {H.}~\bibnamefont {Dixit}},
  \bibinfo {author} {\bibfnamefont {V.~R.}\ \bibnamefont {Cooper}},\ and\
  \bibinfo {author} {\bibfnamefont {B.~C.}\ \bibnamefont {Sales}},\ }\bibfield
  {title} {\bibinfo {title} {Coupling of crystal structure and magnetism in the
  layered, ferromagnetic insulator \ce{CrI3}},\ }\href
  {https://doi.org/10.1021/cm504242t} {\bibfield  {journal} {\bibinfo
  {journal} {Chem. Mater.}\ }\textbf {\bibinfo {volume} {27}},\ \bibinfo
  {pages} {612} (\bibinfo {year} {2015})}\BibitemShut {NoStop}%
\bibitem [{\citenamefont {Huang}\ \emph {et~al.}(2017)\citenamefont {Huang},
  \citenamefont {Clark}, \citenamefont {Navarro-Moratalla}, \citenamefont
  {Klein}, \citenamefont {Cheng}, \citenamefont {Seyler}, \citenamefont
  {Zhong}, \citenamefont {Schmidgall}, \citenamefont {McGuire}, \citenamefont
  {Cobden}, \citenamefont {Yao}, \citenamefont {Xiao}, \citenamefont
  {Jarillo-Herrero},\ and\ \citenamefont {Xu}}]{Huang2017}%
  \BibitemOpen
  \bibfield  {author} {\bibinfo {author} {\bibfnamefont {B.}~\bibnamefont
  {Huang}}, \bibinfo {author} {\bibfnamefont {G.}~\bibnamefont {Clark}},
  \bibinfo {author} {\bibfnamefont {E.}~\bibnamefont {Navarro-Moratalla}},
  \bibinfo {author} {\bibfnamefont {D.~R.}\ \bibnamefont {Klein}}, \bibinfo
  {author} {\bibfnamefont {R.}~\bibnamefont {Cheng}}, \bibinfo {author}
  {\bibfnamefont {K.~L.}\ \bibnamefont {Seyler}}, \bibinfo {author}
  {\bibfnamefont {D.}~\bibnamefont {Zhong}}, \bibinfo {author} {\bibfnamefont
  {E.}~\bibnamefont {Schmidgall}}, \bibinfo {author} {\bibfnamefont {M.~A.}\
  \bibnamefont {McGuire}}, \bibinfo {author} {\bibfnamefont {D.~H.}\
  \bibnamefont {Cobden}}, \bibinfo {author} {\bibfnamefont {W.}~\bibnamefont
  {Yao}}, \bibinfo {author} {\bibfnamefont {D.}~\bibnamefont {Xiao}}, \bibinfo
  {author} {\bibfnamefont {P.}~\bibnamefont {Jarillo-Herrero}},\ and\ \bibinfo
  {author} {\bibfnamefont {X.}~\bibnamefont {Xu}},\ }\bibfield  {title}
  {\bibinfo {title} {Layer-dependent ferromagnetism in a van der {W}aals
  crystal down to the monolayer limit},\ }\href
  {https://doi.org/10.1038/nature22391} {\bibfield  {journal} {\bibinfo
  {journal} {Nature}\ }\textbf {\bibinfo {volume} {546}},\ \bibinfo {pages}
  {270} (\bibinfo {year} {2017})}\BibitemShut {NoStop}%
\bibitem [{\citenamefont {Huang}\ \emph {et~al.}(2018)\citenamefont {Huang},
  \citenamefont {Clark}, \citenamefont {Klein}, \citenamefont {MacNeill},
  \citenamefont {Navarro-Moratalla}, \citenamefont {Seyler}, \citenamefont
  {Wilson}, \citenamefont {McGuire}, \citenamefont {Cobden}, \citenamefont
  {Xiao}, \citenamefont {Yao}, \citenamefont {Jarillo-Herrero},\ and\
  \citenamefont {Xu}}]{Huang2018}%
  \BibitemOpen
  \bibfield  {author} {\bibinfo {author} {\bibfnamefont {B.}~\bibnamefont
  {Huang}}, \bibinfo {author} {\bibfnamefont {G.}~\bibnamefont {Clark}},
  \bibinfo {author} {\bibfnamefont {D.~R.}\ \bibnamefont {Klein}}, \bibinfo
  {author} {\bibfnamefont {D.}~\bibnamefont {MacNeill}}, \bibinfo {author}
  {\bibfnamefont {E.}~\bibnamefont {Navarro-Moratalla}}, \bibinfo {author}
  {\bibfnamefont {K.~L.}\ \bibnamefont {Seyler}}, \bibinfo {author}
  {\bibfnamefont {N.}~\bibnamefont {Wilson}}, \bibinfo {author} {\bibfnamefont
  {M.~A.}\ \bibnamefont {McGuire}}, \bibinfo {author} {\bibfnamefont {D.~H.}\
  \bibnamefont {Cobden}}, \bibinfo {author} {\bibfnamefont {D.}~\bibnamefont
  {Xiao}}, \bibinfo {author} {\bibfnamefont {W.}~\bibnamefont {Yao}}, \bibinfo
  {author} {\bibfnamefont {P.}~\bibnamefont {Jarillo-Herrero}},\ and\ \bibinfo
  {author} {\bibfnamefont {X.}~\bibnamefont {Xu}},\ }\bibfield  {title}
  {\bibinfo {title} {Electrical control of 2{D} magnetism in bilayer
  \ce{CrI3}},\ }\href {https://doi.org/10.1038/s41565-018-0121-3} {\bibfield
  {journal} {\bibinfo  {journal} {Nat. Nanotechnol.}\ }\textbf {\bibinfo
  {volume} {13}},\ \bibinfo {pages} {544} (\bibinfo {year} {2018})}\BibitemShut
  {NoStop}%
\bibitem [{\citenamefont {Jiang}\ \emph {et~al.}(2018)\citenamefont {Jiang},
  \citenamefont {Li}, \citenamefont {Wang}, \citenamefont {Mak},\ and\
  \citenamefont {Shan}}]{Jiang2018}%
  \BibitemOpen
  \bibfield  {author} {\bibinfo {author} {\bibfnamefont {S.}~\bibnamefont
  {Jiang}}, \bibinfo {author} {\bibfnamefont {L.}~\bibnamefont {Li}}, \bibinfo
  {author} {\bibfnamefont {Z.}~\bibnamefont {Wang}}, \bibinfo {author}
  {\bibfnamefont {K.~F.}\ \bibnamefont {Mak}},\ and\ \bibinfo {author}
  {\bibfnamefont {J.}~\bibnamefont {Shan}},\ }\bibfield  {title} {\bibinfo
  {title} {Controlling magnetism in 2{D} \ce{CrI3} by electrostatic doping},\
  }\href {https://doi.org/10.1038/s41565-018-0135-x} {\bibfield  {journal}
  {\bibinfo  {journal} {Nat. Nanotechnol.}\ }\textbf {\bibinfo {volume} {13}},\
  \bibinfo {pages} {549} (\bibinfo {year} {2018})}\BibitemShut {NoStop}%
\bibitem [{\citenamefont {Deng}\ \emph {et~al.}(2018)\citenamefont {Deng},
  \citenamefont {Yu}, \citenamefont {Song}, \citenamefont {Zhang},
  \citenamefont {Wang}, \citenamefont {Sun}, \citenamefont {Yi}, \citenamefont
  {Wu}, \citenamefont {Wu}, \citenamefont {Zhu}, \citenamefont {Wang},
  \citenamefont {Chen},\ and\ \citenamefont {Zhang}}]{Deng2018}%
  \BibitemOpen
  \bibfield  {author} {\bibinfo {author} {\bibfnamefont {Y.}~\bibnamefont
  {Deng}}, \bibinfo {author} {\bibfnamefont {Y.}~\bibnamefont {Yu}}, \bibinfo
  {author} {\bibfnamefont {Y.}~\bibnamefont {Song}}, \bibinfo {author}
  {\bibfnamefont {J.}~\bibnamefont {Zhang}}, \bibinfo {author} {\bibfnamefont
  {N.~Z.}\ \bibnamefont {Wang}}, \bibinfo {author} {\bibfnamefont
  {Z.}~\bibnamefont {Sun}}, \bibinfo {author} {\bibfnamefont {Y.}~\bibnamefont
  {Yi}}, \bibinfo {author} {\bibfnamefont {Y.~Z.}\ \bibnamefont {Wu}}, \bibinfo
  {author} {\bibfnamefont {S.}~\bibnamefont {Wu}}, \bibinfo {author}
  {\bibfnamefont {J.}~\bibnamefont {Zhu}}, \bibinfo {author} {\bibfnamefont
  {J.}~\bibnamefont {Wang}}, \bibinfo {author} {\bibfnamefont {X.~H.}\
  \bibnamefont {Chen}},\ and\ \bibinfo {author} {\bibfnamefont
  {Y.}~\bibnamefont {Zhang}},\ }\bibfield  {title} {\bibinfo {title}
  {Gate-tunable room-temperature ferromagnetism in two-dimensional
  \ce{Fe3GeTe2}},\ }\href {https://doi.org/10.1038/s41586-018-0626-9}
  {\bibfield  {journal} {\bibinfo  {journal} {Nature}\ }\textbf {\bibinfo
  {volume} {563}},\ \bibinfo {pages} {94} (\bibinfo {year} {2018})}\BibitemShut
  {NoStop}%
\bibitem [{\citenamefont {Burch}\ \emph {et~al.}(2018)\citenamefont {Burch},
  \citenamefont {Mandrus},\ and\ \citenamefont {Park}}]{Burch2018}%
  \BibitemOpen
  \bibfield  {author} {\bibinfo {author} {\bibfnamefont {K.~S.}\ \bibnamefont
  {Burch}}, \bibinfo {author} {\bibfnamefont {D.}~\bibnamefont {Mandrus}},\
  and\ \bibinfo {author} {\bibfnamefont {J.-G.}\ \bibnamefont {Park}},\
  }\bibfield  {title} {\bibinfo {title} {Magnetism in two-dimensional van der
  {W}aals materials},\ }\href {https://doi.org/10.1038/s41586-018-0631-z}
  {\bibfield  {journal} {\bibinfo  {journal} {Nature}\ }\textbf {\bibinfo
  {volume} {563}},\ \bibinfo {pages} {47} (\bibinfo {year} {2018})}\BibitemShut
  {NoStop}%
\bibitem [{\citenamefont {Gong}\ and\ \citenamefont {Zhang}(2019)}]{Gong2019}%
  \BibitemOpen
  \bibfield  {author} {\bibinfo {author} {\bibfnamefont {C.}~\bibnamefont
  {Gong}}\ and\ \bibinfo {author} {\bibfnamefont {X.}~\bibnamefont {Zhang}},\
  }\bibfield  {title} {\bibinfo {title} {Two-dimensional magnetic crystals and
  emergent heterostructure devices},\ }\href
  {https://doi.org/10.1126/science.aav4450} {\bibfield  {journal} {\bibinfo
  {journal} {Science}\ }\textbf {\bibinfo {volume} {363}},\ \bibinfo {pages}
  {eaav4450} (\bibinfo {year} {2019})}\BibitemShut {NoStop}%
\bibitem [{\citenamefont {Ji}\ \emph {et~al.}(2013)\citenamefont {Ji},
  \citenamefont {Stokes}, \citenamefont {Alegria}, \citenamefont {Blomberg},
  \citenamefont {Tanatar}, \citenamefont {Reijnders}, \citenamefont {Schoop},
  \citenamefont {Liang}, \citenamefont {Prozorov}, \citenamefont {Burch},
  \citenamefont {Ong}, \citenamefont {Petta},\ and\ \citenamefont
  {Cava}}]{Ji2013}%
  \BibitemOpen
  \bibfield  {author} {\bibinfo {author} {\bibfnamefont {H.}~\bibnamefont
  {Ji}}, \bibinfo {author} {\bibfnamefont {R.~A.}\ \bibnamefont {Stokes}},
  \bibinfo {author} {\bibfnamefont {L.~D.}\ \bibnamefont {Alegria}}, \bibinfo
  {author} {\bibfnamefont {E.~C.}\ \bibnamefont {Blomberg}}, \bibinfo {author}
  {\bibfnamefont {M.~A.}\ \bibnamefont {Tanatar}}, \bibinfo {author}
  {\bibfnamefont {A.}~\bibnamefont {Reijnders}}, \bibinfo {author}
  {\bibfnamefont {L.~M.}\ \bibnamefont {Schoop}}, \bibinfo {author}
  {\bibfnamefont {T.}~\bibnamefont {Liang}}, \bibinfo {author} {\bibfnamefont
  {R.}~\bibnamefont {Prozorov}}, \bibinfo {author} {\bibfnamefont {K.~S.}\
  \bibnamefont {Burch}}, \bibinfo {author} {\bibfnamefont {N.~P.}\ \bibnamefont
  {Ong}}, \bibinfo {author} {\bibfnamefont {J.~R.}\ \bibnamefont {Petta}},\
  and\ \bibinfo {author} {\bibfnamefont {R.~J.}\ \bibnamefont {Cava}},\
  }\bibfield  {title} {\bibinfo {title} {A ferromagnetic insulating substrate
  for the epitaxial growth of topological insulators},\ }\href
  {https://doi.org/10.1063/1.4822092} {\bibfield  {journal} {\bibinfo
  {journal} {J. Appl. Phys.}\ }\textbf {\bibinfo {volume} {114}},\ \bibinfo
  {pages} {114907} (\bibinfo {year} {2013})}\BibitemShut {NoStop}%
\bibitem [{\citenamefont {Mogi}\ \emph {et~al.}(2018)\citenamefont {Mogi},
  \citenamefont {Tsukazaki}, \citenamefont {Kaneko}, \citenamefont {Yoshimi},
  \citenamefont {Takahashi}, \citenamefont {Kawasaki},\ and\ \citenamefont
  {Tokura}}]{Mogi2018}%
  \BibitemOpen
  \bibfield  {author} {\bibinfo {author} {\bibfnamefont {M.}~\bibnamefont
  {Mogi}}, \bibinfo {author} {\bibfnamefont {A.}~\bibnamefont {Tsukazaki}},
  \bibinfo {author} {\bibfnamefont {Y.}~\bibnamefont {Kaneko}}, \bibinfo
  {author} {\bibfnamefont {R.}~\bibnamefont {Yoshimi}}, \bibinfo {author}
  {\bibfnamefont {K.~S.}\ \bibnamefont {Takahashi}}, \bibinfo {author}
  {\bibfnamefont {M.}~\bibnamefont {Kawasaki}},\ and\ \bibinfo {author}
  {\bibfnamefont {Y.}~\bibnamefont {Tokura}},\ }\bibfield  {title} {\bibinfo
  {title} {Ferromagnetic insulator \ce{Cr2Ge2Te6} thin films with perpendicular
  remanence},\ }\href {https://doi.org/10.1063/1.5046166} {\bibfield  {journal}
  {\bibinfo  {journal} {APL Mater.}\ }\textbf {\bibinfo {volume} {6}},\
  \bibinfo {pages} {091104} (\bibinfo {year} {2018})}\BibitemShut {NoStop}%
\bibitem [{\citenamefont {Hatayama}\ \emph {et~al.}(2018)\citenamefont
  {Hatayama}, \citenamefont {Sutou}, \citenamefont {Shindo}, \citenamefont
  {Saito}, \citenamefont {Song}, \citenamefont {Ando},\ and\ \citenamefont
  {Koike}}]{Hatayama2018}%
  \BibitemOpen
  \bibfield  {author} {\bibinfo {author} {\bibfnamefont {S.}~\bibnamefont
  {Hatayama}}, \bibinfo {author} {\bibfnamefont {Y.}~\bibnamefont {Sutou}},
  \bibinfo {author} {\bibfnamefont {S.}~\bibnamefont {Shindo}}, \bibinfo
  {author} {\bibfnamefont {Y.}~\bibnamefont {Saito}}, \bibinfo {author}
  {\bibfnamefont {Y.-H.}\ \bibnamefont {Song}}, \bibinfo {author}
  {\bibfnamefont {D.}~\bibnamefont {Ando}},\ and\ \bibinfo {author}
  {\bibfnamefont {J.}~\bibnamefont {Koike}},\ }\bibfield  {title} {\bibinfo
  {title} {Inverse resistance change \ce{Cr2Ge2Te6}-based {PCRAM} enabling
  ultralow-energy amorphization},\ }\href
  {https://doi.org/10.1021/acsami.7b16755} {\bibfield  {journal} {\bibinfo
  {journal} {ACS Appl. Mater. Interfaces}\ }\textbf {\bibinfo {volume} {10}},\
  \bibinfo {pages} {2725} (\bibinfo {year} {2018})}\BibitemShut {NoStop}%
\bibitem [{\citenamefont {Yang}\ \emph {et~al.}(2016)\citenamefont {Yang},
  \citenamefont {Yao}, \citenamefont {Chen}, \citenamefont {Peng},
  \citenamefont {Jiang}, \citenamefont {Lu}, \citenamefont {Uher},
  \citenamefont {Yang}, \citenamefont {Wang},\ and\ \citenamefont
  {Zhou}}]{Yang2016}%
  \BibitemOpen
  \bibfield  {author} {\bibinfo {author} {\bibfnamefont {D.}~\bibnamefont
  {Yang}}, \bibinfo {author} {\bibfnamefont {W.}~\bibnamefont {Yao}}, \bibinfo
  {author} {\bibfnamefont {Q.}~\bibnamefont {Chen}}, \bibinfo {author}
  {\bibfnamefont {K.}~\bibnamefont {Peng}}, \bibinfo {author} {\bibfnamefont
  {P.}~\bibnamefont {Jiang}}, \bibinfo {author} {\bibfnamefont
  {X.}~\bibnamefont {Lu}}, \bibinfo {author} {\bibfnamefont {C.}~\bibnamefont
  {Uher}}, \bibinfo {author} {\bibfnamefont {T.}~\bibnamefont {Yang}}, \bibinfo
  {author} {\bibfnamefont {G.}~\bibnamefont {Wang}},\ and\ \bibinfo {author}
  {\bibfnamefont {X.}~\bibnamefont {Zhou}},\ }\bibfield  {title} {\bibinfo
  {title} {\ce{Cr2Ge2Te6}: High thermoelectric performance from layered
  structure with high symmetry},\ }\href
  {https://doi.org/10.1021/acs.chemmater.5b04895} {\bibfield  {journal}
  {\bibinfo  {journal} {Chem. Mater.}\ }\textbf {\bibinfo {volume} {28}},\
  \bibinfo {pages} {1611} (\bibinfo {year} {2016})}\BibitemShut {NoStop}%
\bibitem [{\citenamefont {Lefèvre}\ \emph {et~al.}(2017)\citenamefont
  {Lefèvre}, \citenamefont {Berthebaud}, \citenamefont {Lebedev},
  \citenamefont {Pérez}, \citenamefont {Castro}, \citenamefont {Gascoin},
  \citenamefont {Chateigner},\ and\ \citenamefont {Gascoin}}]{Lefevre2017}%
  \BibitemOpen
  \bibfield  {author} {\bibinfo {author} {\bibfnamefont {R.}~\bibnamefont
  {Lefèvre}}, \bibinfo {author} {\bibfnamefont {D.}~\bibnamefont
  {Berthebaud}}, \bibinfo {author} {\bibfnamefont {O.}~\bibnamefont {Lebedev}},
  \bibinfo {author} {\bibfnamefont {O.}~\bibnamefont {Pérez}}, \bibinfo
  {author} {\bibfnamefont {C.}~\bibnamefont {Castro}}, \bibinfo {author}
  {\bibfnamefont {S.}~\bibnamefont {Gascoin}}, \bibinfo {author} {\bibfnamefont
  {D.}~\bibnamefont {Chateigner}},\ and\ \bibinfo {author} {\bibfnamefont
  {F.}~\bibnamefont {Gascoin}},\ }\bibfield  {title} {\bibinfo {title} {Layered
  tellurides: stacking faults induce low thermal conductivity in the new
  \ce{In2Ge2Te6} and thermoelectric properties of related compounds},\ }\href
  {https://doi.org/10.1039/C7TA04810F} {\bibfield  {journal} {\bibinfo
  {journal} {J. Mater. Chem. A}\ }\textbf {\bibinfo {volume} {5}},\ \bibinfo
  {pages} {19406} (\bibinfo {year} {2017})}\BibitemShut {NoStop}%
\bibitem [{\citenamefont {Lin}\ \emph {et~al.}(2016)\citenamefont {Lin},
  \citenamefont {Zhuang}, \citenamefont {Yan}, \citenamefont {Ward},
  \citenamefont {Puretzky}, \citenamefont {Rouleau}, \citenamefont {Gai},
  \citenamefont {Liang}, \citenamefont {Meunier}, \citenamefont {Sumpter},
  \citenamefont {Ganesh}, \citenamefont {Kent}, \citenamefont {Geohegan},
  \citenamefont {Mandrus},\ and\ \citenamefont {Xiao}}]{Lin2016}%
  \BibitemOpen
  \bibfield  {author} {\bibinfo {author} {\bibfnamefont {M.-W.}\ \bibnamefont
  {Lin}}, \bibinfo {author} {\bibfnamefont {H.~L.}\ \bibnamefont {Zhuang}},
  \bibinfo {author} {\bibfnamefont {J.}~\bibnamefont {Yan}}, \bibinfo {author}
  {\bibfnamefont {T.~Z.}\ \bibnamefont {Ward}}, \bibinfo {author}
  {\bibfnamefont {A.~A.}\ \bibnamefont {Puretzky}}, \bibinfo {author}
  {\bibfnamefont {C.~M.}\ \bibnamefont {Rouleau}}, \bibinfo {author}
  {\bibfnamefont {Z.}~\bibnamefont {Gai}}, \bibinfo {author} {\bibfnamefont
  {L.}~\bibnamefont {Liang}}, \bibinfo {author} {\bibfnamefont
  {V.}~\bibnamefont {Meunier}}, \bibinfo {author} {\bibfnamefont {B.~G.}\
  \bibnamefont {Sumpter}}, \bibinfo {author} {\bibfnamefont {P.}~\bibnamefont
  {Ganesh}}, \bibinfo {author} {\bibfnamefont {P.~R.~C.}\ \bibnamefont {Kent}},
  \bibinfo {author} {\bibfnamefont {D.~B.}\ \bibnamefont {Geohegan}}, \bibinfo
  {author} {\bibfnamefont {D.~G.}\ \bibnamefont {Mandrus}},\ and\ \bibinfo
  {author} {\bibfnamefont {K.}~\bibnamefont {Xiao}},\ }\bibfield  {title}
  {\bibinfo {title} {Ultrathin nanosheets of \ce{CrSiTe3}: a semiconducting
  two-dimensional ferromagnetic material},\ }\href
  {https://doi.org/10.1039/C5TC03463A} {\bibfield  {journal} {\bibinfo
  {journal} {J. Mater. Chem. C}\ }\textbf {\bibinfo {volume} {4}},\ \bibinfo
  {pages} {315} (\bibinfo {year} {2016})}\BibitemShut {NoStop}%
\bibitem [{\citenamefont {Casto}\ \emph {et~al.}(2015)\citenamefont {Casto},
  \citenamefont {Clune}, \citenamefont {Yokosuk}, \citenamefont {Musfeldt},
  \citenamefont {Williams}, \citenamefont {Zhuang}, \citenamefont {Lin},
  \citenamefont {Xiao}, \citenamefont {Hennig}, \citenamefont {Sales},
  \citenamefont {Yan},\ and\ \citenamefont {Mandrus}}]{Casto2015}%
  \BibitemOpen
  \bibfield  {author} {\bibinfo {author} {\bibfnamefont {L.~D.}\ \bibnamefont
  {Casto}}, \bibinfo {author} {\bibfnamefont {A.~J.}\ \bibnamefont {Clune}},
  \bibinfo {author} {\bibfnamefont {M.~O.}\ \bibnamefont {Yokosuk}}, \bibinfo
  {author} {\bibfnamefont {J.~L.}\ \bibnamefont {Musfeldt}}, \bibinfo {author}
  {\bibfnamefont {T.~J.}\ \bibnamefont {Williams}}, \bibinfo {author}
  {\bibfnamefont {H.~L.}\ \bibnamefont {Zhuang}}, \bibinfo {author}
  {\bibfnamefont {M.-W.}\ \bibnamefont {Lin}}, \bibinfo {author} {\bibfnamefont
  {K.}~\bibnamefont {Xiao}}, \bibinfo {author} {\bibfnamefont {R.~G.}\
  \bibnamefont {Hennig}}, \bibinfo {author} {\bibfnamefont {B.~C.}\
  \bibnamefont {Sales}}, \bibinfo {author} {\bibfnamefont {J.-Q.}\ \bibnamefont
  {Yan}},\ and\ \bibinfo {author} {\bibfnamefont {D.}~\bibnamefont {Mandrus}},\
  }\bibfield  {title} {\bibinfo {title} {Strong spin-lattice coupling in
  \ce{CrSiTe3}},\ }\href {https://doi.org/10.1063/1.4914134} {\bibfield
  {journal} {\bibinfo  {journal} {APL Mater.}\ }\textbf {\bibinfo {volume}
  {3}},\ \bibinfo {pages} {041515} (\bibinfo {year} {2015})}\BibitemShut
  {NoStop}%
\bibitem [{\citenamefont {Liu}\ \emph {et~al.}(2016)\citenamefont {Liu},
  \citenamefont {Zou}, \citenamefont {Zhang}, \citenamefont {Zhou},
  \citenamefont {Wang}, \citenamefont {Wang}, \citenamefont {Qu},\ and\
  \citenamefont {Zhang}}]{Liu2016}%
  \BibitemOpen
  \bibfield  {author} {\bibinfo {author} {\bibfnamefont {B.}~\bibnamefont
  {Liu}}, \bibinfo {author} {\bibfnamefont {Y.}~\bibnamefont {Zou}}, \bibinfo
  {author} {\bibfnamefont {L.}~\bibnamefont {Zhang}}, \bibinfo {author}
  {\bibfnamefont {S.}~\bibnamefont {Zhou}}, \bibinfo {author} {\bibfnamefont
  {Z.}~\bibnamefont {Wang}}, \bibinfo {author} {\bibfnamefont {W.}~\bibnamefont
  {Wang}}, \bibinfo {author} {\bibfnamefont {Z.}~\bibnamefont {Qu}},\ and\
  \bibinfo {author} {\bibfnamefont {Y.}~\bibnamefont {Zhang}},\ }\bibfield
  {title} {\bibinfo {title} {Critical behavior of the quasi-two-dimensional
  semiconducting ferromagnet \ce{CrSiTe3}},\ }\href
  {https://doi.org/10.1038/srep33873} {\bibfield  {journal} {\bibinfo
  {journal} {Sci. Rep.}\ }\textbf {\bibinfo {volume} {6}},\ \bibinfo {pages}
  {33873} (\bibinfo {year} {2016})}\BibitemShut {NoStop}%
\bibitem [{\citenamefont {Cai}\ \emph {et~al.}(2020)\citenamefont {Cai},
  \citenamefont {Sun}, \citenamefont {Xia}, \citenamefont {Wu}, \citenamefont
  {Liu}, \citenamefont {Liu}, \citenamefont {Gong}, \citenamefont {Yao},
  \citenamefont {Guo},\ and\ \citenamefont {Wang}}]{Cai2020}%
  \BibitemOpen
  \bibfield  {author} {\bibinfo {author} {\bibfnamefont {W.}~\bibnamefont
  {Cai}}, \bibinfo {author} {\bibfnamefont {H.}~\bibnamefont {Sun}}, \bibinfo
  {author} {\bibfnamefont {W.}~\bibnamefont {Xia}}, \bibinfo {author}
  {\bibfnamefont {C.}~\bibnamefont {Wu}}, \bibinfo {author} {\bibfnamefont
  {Y.}~\bibnamefont {Liu}}, \bibinfo {author} {\bibfnamefont {H.}~\bibnamefont
  {Liu}}, \bibinfo {author} {\bibfnamefont {Y.}~\bibnamefont {Gong}}, \bibinfo
  {author} {\bibfnamefont {D.-X.}\ \bibnamefont {Yao}}, \bibinfo {author}
  {\bibfnamefont {Y.}~\bibnamefont {Guo}},\ and\ \bibinfo {author}
  {\bibfnamefont {M.}~\bibnamefont {Wang}},\ }\bibfield  {title} {\bibinfo
  {title} {Pressure-induced superconductivity and structural transition in
  ferromagnetic \ce{CrSiTe3}},\ }\href
  {https://doi.org/10.1103/PhysRevB.102.144525} {\bibfield  {journal} {\bibinfo
   {journal} {Phys. Rev. B}\ }\textbf {\bibinfo {volume} {102}},\ \bibinfo
  {pages} {144525} (\bibinfo {year} {2020})}\BibitemShut {NoStop}%
\bibitem [{\citenamefont {Xu}\ \emph {et~al.}(2020)\citenamefont {Xu},
  \citenamefont {Yu}, \citenamefont {Xia}, \citenamefont {Xu}, \citenamefont
  {Mai}, \citenamefont {Wang}, \citenamefont {Guo}, \citenamefont {Miao},\ and\
  \citenamefont {Xu}}]{Xu2020}%
  \BibitemOpen
  \bibfield  {author} {\bibinfo {author} {\bibfnamefont {K.}~\bibnamefont
  {Xu}}, \bibinfo {author} {\bibfnamefont {Z.}~\bibnamefont {Yu}}, \bibinfo
  {author} {\bibfnamefont {W.}~\bibnamefont {Xia}}, \bibinfo {author}
  {\bibfnamefont {M.}~\bibnamefont {Xu}}, \bibinfo {author} {\bibfnamefont
  {X.}~\bibnamefont {Mai}}, \bibinfo {author} {\bibfnamefont {L.}~\bibnamefont
  {Wang}}, \bibinfo {author} {\bibfnamefont {Y.}~\bibnamefont {Guo}}, \bibinfo
  {author} {\bibfnamefont {X.}~\bibnamefont {Miao}},\ and\ \bibinfo {author}
  {\bibfnamefont {M.}~\bibnamefont {Xu}},\ }\bibfield  {title} {\bibinfo
  {title} {Unique {2D}–{3D} structure transformations in trichalcogenide
  \ce{CrSiTe3} under high pressure},\ }\href
  {https://doi.org/10.1021/acs.jpcc.0c03931} {\bibfield  {journal} {\bibinfo
  {journal} {J. Phys. Chem. C}\ }\textbf {\bibinfo {volume} {124}},\ \bibinfo
  {pages} {15600} (\bibinfo {year} {2020})}\BibitemShut {NoStop}%
\bibitem [{\citenamefont {Zhang}\ \emph {et~al.}(2021)\citenamefont {Zhang},
  \citenamefont {Gu}, \citenamefont {Wang}, \citenamefont {Huang},
  \citenamefont {Fu}, \citenamefont {Liu}, \citenamefont {Wang}, \citenamefont
  {Su}, \citenamefont {Mei}, \citenamefont {Zou},\ and\ \citenamefont
  {Dai}}]{Zhang2021}%
  \BibitemOpen
  \bibfield  {author} {\bibinfo {author} {\bibfnamefont {C.}~\bibnamefont
  {Zhang}}, \bibinfo {author} {\bibfnamefont {Y.}~\bibnamefont {Gu}}, \bibinfo
  {author} {\bibfnamefont {L.}~\bibnamefont {Wang}}, \bibinfo {author}
  {\bibfnamefont {L.-L.}\ \bibnamefont {Huang}}, \bibinfo {author}
  {\bibfnamefont {Y.}~\bibnamefont {Fu}}, \bibinfo {author} {\bibfnamefont
  {C.}~\bibnamefont {Liu}}, \bibinfo {author} {\bibfnamefont {S.}~\bibnamefont
  {Wang}}, \bibinfo {author} {\bibfnamefont {H.}~\bibnamefont {Su}}, \bibinfo
  {author} {\bibfnamefont {J.-W.}\ \bibnamefont {Mei}}, \bibinfo {author}
  {\bibfnamefont {X.}~\bibnamefont {Zou}},\ and\ \bibinfo {author}
  {\bibfnamefont {J.-F.}\ \bibnamefont {Dai}},\ }\bibfield  {title} {\bibinfo
  {title} {Pressure-enhanced ferromagnetism in layered \ce{CrSiTe3} flakes},\
  }\href {https://doi.org/10.1021/acs.nanolett.1c01994} {\bibfield  {journal}
  {\bibinfo  {journal} {Nano Lett.}\ }\textbf {\bibinfo {volume} {21}},\
  \bibinfo {pages} {7946} (\bibinfo {year} {2021})}\BibitemShut {NoStop}%
\bibitem [{\citenamefont {Williams}\ \emph {et~al.}(2015)\citenamefont
  {Williams}, \citenamefont {Aczel}, \citenamefont {Lumsden}, \citenamefont
  {Nagler}, \citenamefont {Stone}, \citenamefont {Yan},\ and\ \citenamefont
  {Mandrus}}]{Williams2015}%
  \BibitemOpen
  \bibfield  {author} {\bibinfo {author} {\bibfnamefont {T.~J.}\ \bibnamefont
  {Williams}}, \bibinfo {author} {\bibfnamefont {A.~A.}\ \bibnamefont {Aczel}},
  \bibinfo {author} {\bibfnamefont {M.~D.}\ \bibnamefont {Lumsden}}, \bibinfo
  {author} {\bibfnamefont {S.~E.}\ \bibnamefont {Nagler}}, \bibinfo {author}
  {\bibfnamefont {M.~B.}\ \bibnamefont {Stone}}, \bibinfo {author}
  {\bibfnamefont {J.-Q.}\ \bibnamefont {Yan}},\ and\ \bibinfo {author}
  {\bibfnamefont {D.}~\bibnamefont {Mandrus}},\ }\bibfield  {title} {\bibinfo
  {title} {Magnetic correlations in the quasi-two-dimensional semiconducting
  ferromagnet \ce{CrSiTe3}},\ }\href
  {https://doi.org/10.1103/PhysRevB.92.144404} {\bibfield  {journal} {\bibinfo
  {journal} {Phys. Rev. B}\ }\textbf {\bibinfo {volume} {92}},\ \bibinfo
  {pages} {144404} (\bibinfo {year} {2015})}\BibitemShut {NoStop}%
\bibitem [{\citenamefont {Achinuq}\ \emph {et~al.}()\citenamefont {Achinuq},
  \citenamefont {Fujita}, \citenamefont {Xia}, \citenamefont {Guo},
  \citenamefont {Bencok}, \citenamefont {van~der Laan},\ and\ \citenamefont
  {Hesjedal}}]{Achinuq2021}%
  \BibitemOpen
  \bibfield  {author} {\bibinfo {author} {\bibfnamefont {B.}~\bibnamefont
  {Achinuq}}, \bibinfo {author} {\bibfnamefont {R.}~\bibnamefont {Fujita}},
  \bibinfo {author} {\bibfnamefont {W.}~\bibnamefont {Xia}}, \bibinfo {author}
  {\bibfnamefont {Y.}~\bibnamefont {Guo}}, \bibinfo {author} {\bibfnamefont
  {P.}~\bibnamefont {Bencok}}, \bibinfo {author} {\bibfnamefont
  {G.}~\bibnamefont {van~der Laan}},\ and\ \bibinfo {author} {\bibfnamefont
  {T.}~\bibnamefont {Hesjedal}},\ }\bibfield  {title} {\bibinfo {title}
  {Covalent mixing in the {2D} ferromagnet \ce{CrSiTe3} evidenced by magnetic
  {X}-ray circular dichroism},\ }\href
  {https://doi.org/https://doi.org/10.1002/pssr.202100566} {\bibfield
  {journal} {\bibinfo  {journal} {Phys. Stat. Sol. RRL}\ }\textbf {\bibinfo
  {volume} {2021}},\ \bibinfo {pages} {2100566}}\BibitemShut {NoStop}%
\bibitem [{\citenamefont {Niu}\ \emph {et~al.}(2021)\citenamefont {Niu},
  \citenamefont {Zhang}, \citenamefont {Wang}, \citenamefont {Sun},
  \citenamefont {Xu}, \citenamefont {He}, \citenamefont {Liu},\ and\
  \citenamefont {Pu}}]{Niu2021}%
  \BibitemOpen
  \bibfield  {author} {\bibinfo {author} {\bibfnamefont {W.}~\bibnamefont
  {Niu}}, \bibinfo {author} {\bibfnamefont {X.}~\bibnamefont {Zhang}}, \bibinfo
  {author} {\bibfnamefont {W.}~\bibnamefont {Wang}}, \bibinfo {author}
  {\bibfnamefont {J.}~\bibnamefont {Sun}}, \bibinfo {author} {\bibfnamefont
  {Y.}~\bibnamefont {Xu}}, \bibinfo {author} {\bibfnamefont {L.}~\bibnamefont
  {He}}, \bibinfo {author} {\bibfnamefont {W.}~\bibnamefont {Liu}},\ and\
  \bibinfo {author} {\bibfnamefont {Y.}~\bibnamefont {Pu}},\ }\bibfield
  {title} {\bibinfo {title} {Probing the atomic-scale ferromagnetism in van der
  {W}aals magnet \ce{CrSiTe3}},\ }\href {https://doi.org/10.1063/5.0069885}
  {\bibfield  {journal} {\bibinfo  {journal} {Appl. Phys. Lett.}\ }\textbf
  {\bibinfo {volume} {119}},\ \bibinfo {pages} {172402} (\bibinfo {year}
  {2021})}\BibitemShut {NoStop}%
\bibitem [{\citenamefont {Zhang}\ \emph {et~al.}(2019)\citenamefont {Zhang},
  \citenamefont {Cai}, \citenamefont {Xia}, \citenamefont {Liang},
  \citenamefont {Huang}, \citenamefont {Wang}, \citenamefont {Yang},
  \citenamefont {Yuan}, \citenamefont {Chen}, \citenamefont {Zhang},
  \citenamefont {Guo}, \citenamefont {Liu},\ and\ \citenamefont
  {Li}}]{Zhang2019}%
  \BibitemOpen
  \bibfield  {author} {\bibinfo {author} {\bibfnamefont {J.}~\bibnamefont
  {Zhang}}, \bibinfo {author} {\bibfnamefont {X.}~\bibnamefont {Cai}}, \bibinfo
  {author} {\bibfnamefont {W.}~\bibnamefont {Xia}}, \bibinfo {author}
  {\bibfnamefont {A.}~\bibnamefont {Liang}}, \bibinfo {author} {\bibfnamefont
  {J.}~\bibnamefont {Huang}}, \bibinfo {author} {\bibfnamefont
  {C.}~\bibnamefont {Wang}}, \bibinfo {author} {\bibfnamefont {L.}~\bibnamefont
  {Yang}}, \bibinfo {author} {\bibfnamefont {H.}~\bibnamefont {Yuan}}, \bibinfo
  {author} {\bibfnamefont {Y.}~\bibnamefont {Chen}}, \bibinfo {author}
  {\bibfnamefont {S.}~\bibnamefont {Zhang}}, \bibinfo {author} {\bibfnamefont
  {Y.}~\bibnamefont {Guo}}, \bibinfo {author} {\bibfnamefont {Z.}~\bibnamefont
  {Liu}},\ and\ \bibinfo {author} {\bibfnamefont {G.}~\bibnamefont {Li}},\
  }\bibfield  {title} {\bibinfo {title} {Unveiling electronic correlation and
  the ferromagnetic superexchange mechanism in the van der {W}aals crystal
  \ce{CrSiTe3}},\ }\href {https://doi.org/10.1103/PhysRevLett.123.047203}
  {\bibfield  {journal} {\bibinfo  {journal} {Phys. Rev. Lett.}\ }\textbf
  {\bibinfo {volume} {123}},\ \bibinfo {pages} {047203} (\bibinfo {year}
  {2019})}\BibitemShut {NoStop}%
\bibitem [{\citenamefont {Milosavljevi\ifmmode~\acute{c}\else \'{c}\fi{}}\
  \emph {et~al.}(2018)\citenamefont {Milosavljevi\ifmmode~\acute{c}\else
  \'{c}\fi{}}, \citenamefont {\ifmmode \check{S}\else
  \v{S}\fi{}olaji\ifmmode~\acute{c}\else \'{c}\fi{}}, \citenamefont {Pe\ifmmode
  \check{s}\else \v{s}\fi{}i\ifmmode~\acute{c}\else \'{c}\fi{}}, \citenamefont
  {Liu}, \citenamefont {Petrovic}, \citenamefont
  {Lazarevi\ifmmode~\acute{c}\else \'{c}\fi{}},\ and\ \citenamefont
  {Popovi\ifmmode~\acute{c}\else \'{c}\fi{}}}]{Milosavljevic2018}%
  \BibitemOpen
  \bibfield  {author} {\bibinfo {author} {\bibfnamefont {A.}~\bibnamefont
  {Milosavljevi\ifmmode~\acute{c}\else \'{c}\fi{}}}, \bibinfo {author}
  {\bibfnamefont {A.}~\bibnamefont {\ifmmode \check{S}\else
  \v{S}\fi{}olaji\ifmmode~\acute{c}\else \'{c}\fi{}}}, \bibinfo {author}
  {\bibfnamefont {J.}~\bibnamefont {Pe\ifmmode \check{s}\else
  \v{s}\fi{}i\ifmmode~\acute{c}\else \'{c}\fi{}}}, \bibinfo {author}
  {\bibfnamefont {Y.}~\bibnamefont {Liu}}, \bibinfo {author} {\bibfnamefont
  {C.}~\bibnamefont {Petrovic}}, \bibinfo {author} {\bibfnamefont
  {N.}~\bibnamefont {Lazarevi\ifmmode~\acute{c}\else \'{c}\fi{}}},\ and\
  \bibinfo {author} {\bibfnamefont {Z.~V.}\ \bibnamefont
  {Popovi\ifmmode~\acute{c}\else \'{c}\fi{}}},\ }\bibfield  {title} {\bibinfo
  {title} {Evidence of spin-phonon coupling in \ce{CrSiTe3}},\ }\href
  {https://doi.org/10.1103/PhysRevB.98.104306} {\bibfield  {journal} {\bibinfo
  {journal} {Phys. Rev. B}\ }\textbf {\bibinfo {volume} {98}},\ \bibinfo
  {pages} {104306} (\bibinfo {year} {2018})}\BibitemShut {NoStop}%
\bibitem [{\citenamefont {Sivadas}\ \emph {et~al.}(2015)\citenamefont
  {Sivadas}, \citenamefont {Daniels}, \citenamefont {Swendsen}, \citenamefont
  {Okamoto},\ and\ \citenamefont {Xiao}}]{Sivadas2015}%
  \BibitemOpen
  \bibfield  {author} {\bibinfo {author} {\bibfnamefont {N.}~\bibnamefont
  {Sivadas}}, \bibinfo {author} {\bibfnamefont {M.~W.}\ \bibnamefont
  {Daniels}}, \bibinfo {author} {\bibfnamefont {R.~H.}\ \bibnamefont
  {Swendsen}}, \bibinfo {author} {\bibfnamefont {S.}~\bibnamefont {Okamoto}},\
  and\ \bibinfo {author} {\bibfnamefont {D.}~\bibnamefont {Xiao}},\ }\bibfield
  {title} {\bibinfo {title} {Magnetic ground state of semiconducting
  transition-metal trichalcogenide monolayers},\ }\href
  {https://doi.org/10.1103/PhysRevB.91.235425} {\bibfield  {journal} {\bibinfo
  {journal} {Phys. Rev. B}\ }\textbf {\bibinfo {volume} {91}},\ \bibinfo
  {pages} {235425} (\bibinfo {year} {2015})}\BibitemShut {NoStop}%
\bibitem [{\citenamefont {Zhuang}\ \emph {et~al.}(2015)\citenamefont {Zhuang},
  \citenamefont {Xie}, \citenamefont {Kent},\ and\ \citenamefont
  {Ganesh}}]{Zhuang2015}%
  \BibitemOpen
  \bibfield  {author} {\bibinfo {author} {\bibfnamefont {H.~L.}\ \bibnamefont
  {Zhuang}}, \bibinfo {author} {\bibfnamefont {Y.}~\bibnamefont {Xie}},
  \bibinfo {author} {\bibfnamefont {P.~R.~C.}\ \bibnamefont {Kent}},\ and\
  \bibinfo {author} {\bibfnamefont {P.}~\bibnamefont {Ganesh}},\ }\bibfield
  {title} {\bibinfo {title} {Computational discovery of ferromagnetic
  semiconducting single-layer \ce{CrSnTe3}},\ }\href
  {https://doi.org/10.1103/PhysRevB.92.035407} {\bibfield  {journal} {\bibinfo
  {journal} {Phys. Rev. B}\ }\textbf {\bibinfo {volume} {92}},\ \bibinfo
  {pages} {035407} (\bibinfo {year} {2015})}\BibitemShut {NoStop}%
\bibitem [{\citenamefont {Kang}\ \emph {et~al.}(2019)\citenamefont {Kang},
  \citenamefont {Kang},\ and\ \citenamefont {Yu}}]{Kang2019}%
  \BibitemOpen
  \bibfield  {author} {\bibinfo {author} {\bibfnamefont {S.}~\bibnamefont
  {Kang}}, \bibinfo {author} {\bibfnamefont {S.}~\bibnamefont {Kang}},\ and\
  \bibinfo {author} {\bibfnamefont {J.}~\bibnamefont {Yu}},\ }\bibfield
  {title} {\bibinfo {title} {Effect of {C}oulomb interactions on the electronic
  and magnetic properties of two-dimensional \ce{CrSiTe3} and \ce{CrGeTe3}
  materials},\ }\href {https://doi.org/10.1007/s11664-018-6601-2} {\bibfield
  {journal} {\bibinfo  {journal} {J. Electr. Mater.}\ }\textbf {\bibinfo
  {volume} {48}},\ \bibinfo {pages} {1441} (\bibinfo {year}
  {2019})}\BibitemShut {NoStop}%
\bibitem [{\citenamefont {Chen}\ \emph {et~al.}(2020)\citenamefont {Chen},
  \citenamefont {Long},\ and\ \citenamefont {Wang}}]{Chen2020}%
  \BibitemOpen
  \bibfield  {author} {\bibinfo {author} {\bibfnamefont {X.-Y.}\ \bibnamefont
  {Chen}}, \bibinfo {author} {\bibfnamefont {M.-Q.}\ \bibnamefont {Long}},\
  and\ \bibinfo {author} {\bibfnamefont {Y.-P.}\ \bibnamefont {Wang}},\
  }\bibfield  {title} {\bibinfo {title} {Paramagnetic phases of two-dimensional
  magnetic materials},\ }\href {https://doi.org/10.1103/PhysRevB.102.214417}
  {\bibfield  {journal} {\bibinfo  {journal} {Phys. Rev. B}\ }\textbf {\bibinfo
  {volume} {102}},\ \bibinfo {pages} {214417} (\bibinfo {year}
  {2020})}\BibitemShut {NoStop}%
\bibitem [{\citenamefont {Bhoi}\ \emph {et~al.}(2021)\citenamefont {Bhoi},
  \citenamefont {Gouchi}, \citenamefont {Hiraoka}, \citenamefont {Zhang},
  \citenamefont {Ogita}, \citenamefont {Hasegawa}, \citenamefont {Kitagawa},
  \citenamefont {Takagi}, \citenamefont {Kim},\ and\ \citenamefont
  {Uwatoko}}]{Bhoi2021}%
  \BibitemOpen
  \bibfield  {author} {\bibinfo {author} {\bibfnamefont {D.}~\bibnamefont
  {Bhoi}}, \bibinfo {author} {\bibfnamefont {J.}~\bibnamefont {Gouchi}},
  \bibinfo {author} {\bibfnamefont {N.}~\bibnamefont {Hiraoka}}, \bibinfo
  {author} {\bibfnamefont {Y.}~\bibnamefont {Zhang}}, \bibinfo {author}
  {\bibfnamefont {N.}~\bibnamefont {Ogita}}, \bibinfo {author} {\bibfnamefont
  {T.}~\bibnamefont {Hasegawa}}, \bibinfo {author} {\bibfnamefont
  {K.}~\bibnamefont {Kitagawa}}, \bibinfo {author} {\bibfnamefont
  {H.}~\bibnamefont {Takagi}}, \bibinfo {author} {\bibfnamefont {K.~H.}\
  \bibnamefont {Kim}},\ and\ \bibinfo {author} {\bibfnamefont {Y.}~\bibnamefont
  {Uwatoko}},\ }\bibfield  {title} {\bibinfo {title} {Nearly room-temperature
  ferromagnetism in a pressure-induced correlated metallic state of the van der
  {W}aals insulator \ce{CrGeTe3}},\ }\href
  {https://doi.org/10.1103/PhysRevLett.127.217203} {\bibfield  {journal}
  {\bibinfo  {journal} {Phys. Rev. Lett.}\ }\textbf {\bibinfo {volume} {127}},\
  \bibinfo {pages} {217203} (\bibinfo {year} {2021})}\BibitemShut {NoStop}%
\bibitem [{\citenamefont {Yu}\ \emph {et~al.}(2019)\citenamefont {Yu},
  \citenamefont {Xia}, \citenamefont {Xu}, \citenamefont {Xu}, \citenamefont
  {Wang}, \citenamefont {Wang}, \citenamefont {Yu}, \citenamefont {Zou},
  \citenamefont {Zhao}, \citenamefont {Wang}, \citenamefont {Miao},\ and\
  \citenamefont {Guo}}]{Yu2019}%
  \BibitemOpen
  \bibfield  {author} {\bibinfo {author} {\bibfnamefont {Z.}~\bibnamefont
  {Yu}}, \bibinfo {author} {\bibfnamefont {W.}~\bibnamefont {Xia}}, \bibinfo
  {author} {\bibfnamefont {K.}~\bibnamefont {Xu}}, \bibinfo {author}
  {\bibfnamefont {M.}~\bibnamefont {Xu}}, \bibinfo {author} {\bibfnamefont
  {H.}~\bibnamefont {Wang}}, \bibinfo {author} {\bibfnamefont {X.}~\bibnamefont
  {Wang}}, \bibinfo {author} {\bibfnamefont {N.}~\bibnamefont {Yu}}, \bibinfo
  {author} {\bibfnamefont {Z.}~\bibnamefont {Zou}}, \bibinfo {author}
  {\bibfnamefont {J.}~\bibnamefont {Zhao}}, \bibinfo {author} {\bibfnamefont
  {L.}~\bibnamefont {Wang}}, \bibinfo {author} {\bibfnamefont {X.}~\bibnamefont
  {Miao}},\ and\ \bibinfo {author} {\bibfnamefont {Y.}~\bibnamefont {Guo}},\
  }\bibfield  {title} {\bibinfo {title} {Pressure-induced structural phase
  transition and a special amorphization phase of two-dimensional ferromagnetic
  semiconductor \ce{Cr2Ge2Te6}},\ }\href
  {https://doi.org/10.1021/acs.jpcc.9b02415} {\bibfield  {journal} {\bibinfo
  {journal} {J. Phys. Chem. C}\ }\textbf {\bibinfo {volume} {123}},\ \bibinfo
  {pages} {13885} (\bibinfo {year} {2019})}\BibitemShut {NoStop}%
\bibitem [{\citenamefont {Wang}\ \emph {et~al.}(2019)\citenamefont {Wang},
  \citenamefont {Tang}, \citenamefont {Shi}, \citenamefont {Zhang},
  \citenamefont {Zhuo}, \citenamefont {Liu}, \citenamefont {Meng},
  \citenamefont {Ma}, \citenamefont {Ying}, \citenamefont {Zou}, \citenamefont
  {Sun},\ and\ \citenamefont {Chen}}]{Wang2019}%
  \BibitemOpen
  \bibfield  {author} {\bibinfo {author} {\bibfnamefont {N.}~\bibnamefont
  {Wang}}, \bibinfo {author} {\bibfnamefont {H.}~\bibnamefont {Tang}}, \bibinfo
  {author} {\bibfnamefont {M.}~\bibnamefont {Shi}}, \bibinfo {author}
  {\bibfnamefont {H.}~\bibnamefont {Zhang}}, \bibinfo {author} {\bibfnamefont
  {W.}~\bibnamefont {Zhuo}}, \bibinfo {author} {\bibfnamefont {D.}~\bibnamefont
  {Liu}}, \bibinfo {author} {\bibfnamefont {F.}~\bibnamefont {Meng}}, \bibinfo
  {author} {\bibfnamefont {L.}~\bibnamefont {Ma}}, \bibinfo {author}
  {\bibfnamefont {J.}~\bibnamefont {Ying}}, \bibinfo {author} {\bibfnamefont
  {L.}~\bibnamefont {Zou}}, \bibinfo {author} {\bibfnamefont {Z.}~\bibnamefont
  {Sun}},\ and\ \bibinfo {author} {\bibfnamefont {X.}~\bibnamefont {Chen}},\
  }\bibfield  {title} {\bibinfo {title} {Transition from ferromagnetic
  semiconductor to ferromagnetic metal with enhanced {C}urie temperature in
  \ce{Cr2Ge2Te6} via organic ion intercalation},\ }\href
  {https://doi.org/10.1021/jacs.9b06929} {\bibfield  {journal} {\bibinfo
  {journal} {J. Am. Chem. Soc.}\ }\textbf {\bibinfo {volume} {141}},\ \bibinfo
  {pages} {17166} (\bibinfo {year} {2019})}\BibitemShut {NoStop}%
\bibitem [{\citenamefont {Wang}\ \emph {et~al.}(2018)\citenamefont {Wang},
  \citenamefont {Zhang}, \citenamefont {Ding}, \citenamefont {Dong},
  \citenamefont {Li}, \citenamefont {Chen}, \citenamefont {Li}, \citenamefont
  {Huang}, \citenamefont {Wang}, \citenamefont {Zhao}, \citenamefont {Li},
  \citenamefont {Li}, \citenamefont {Jia}, \citenamefont {Sun}, \citenamefont
  {Guo}, \citenamefont {Ye}, \citenamefont {Sun}, \citenamefont {Chen},
  \citenamefont {Yang}, \citenamefont {Zhang}, \citenamefont {Ono},
  \citenamefont {Han},\ and\ \citenamefont {Zhang}}]{Wang2018}%
  \BibitemOpen
  \bibfield  {author} {\bibinfo {author} {\bibfnamefont {Z.}~\bibnamefont
  {Wang}}, \bibinfo {author} {\bibfnamefont {T.}~\bibnamefont {Zhang}},
  \bibinfo {author} {\bibfnamefont {M.}~\bibnamefont {Ding}}, \bibinfo {author}
  {\bibfnamefont {B.}~\bibnamefont {Dong}}, \bibinfo {author} {\bibfnamefont
  {Y.}~\bibnamefont {Li}}, \bibinfo {author} {\bibfnamefont {M.}~\bibnamefont
  {Chen}}, \bibinfo {author} {\bibfnamefont {X.}~\bibnamefont {Li}}, \bibinfo
  {author} {\bibfnamefont {J.}~\bibnamefont {Huang}}, \bibinfo {author}
  {\bibfnamefont {H.}~\bibnamefont {Wang}}, \bibinfo {author} {\bibfnamefont
  {X.}~\bibnamefont {Zhao}}, \bibinfo {author} {\bibfnamefont {Y.}~\bibnamefont
  {Li}}, \bibinfo {author} {\bibfnamefont {D.}~\bibnamefont {Li}}, \bibinfo
  {author} {\bibfnamefont {C.}~\bibnamefont {Jia}}, \bibinfo {author}
  {\bibfnamefont {L.}~\bibnamefont {Sun}}, \bibinfo {author} {\bibfnamefont
  {H.}~\bibnamefont {Guo}}, \bibinfo {author} {\bibfnamefont {Y.}~\bibnamefont
  {Ye}}, \bibinfo {author} {\bibfnamefont {D.}~\bibnamefont {Sun}}, \bibinfo
  {author} {\bibfnamefont {Y.}~\bibnamefont {Chen}}, \bibinfo {author}
  {\bibfnamefont {T.}~\bibnamefont {Yang}}, \bibinfo {author} {\bibfnamefont
  {J.}~\bibnamefont {Zhang}}, \bibinfo {author} {\bibfnamefont
  {S.}~\bibnamefont {Ono}}, \bibinfo {author} {\bibfnamefont {Z.}~\bibnamefont
  {Han}},\ and\ \bibinfo {author} {\bibfnamefont {Z.}~\bibnamefont {Zhang}},\
  }\bibfield  {title} {\bibinfo {title} {Electric-field control of magnetism in
  a few-layered van der waals ferromagnetic semiconductor},\ }\href
  {https://doi.org/10.1038/s41565-018-0186-z} {\bibfield  {journal} {\bibinfo
  {journal} {Nat. Nanotechnol.}\ }\textbf {\bibinfo {volume} {13}},\ \bibinfo
  {pages} {554} (\bibinfo {year} {2018})}\BibitemShut {NoStop}%
\bibitem [{\citenamefont {Verzhbitskiy}\ \emph {et~al.}(2020)\citenamefont
  {Verzhbitskiy}, \citenamefont {Kurebayashi}, \citenamefont {Cheng},
  \citenamefont {Zhou}, \citenamefont {Khan}, \citenamefont {Feng},\ and\
  \citenamefont {Eda}}]{Verzhbitskiy2020}%
  \BibitemOpen
  \bibfield  {author} {\bibinfo {author} {\bibfnamefont {I.~A.}\ \bibnamefont
  {Verzhbitskiy}}, \bibinfo {author} {\bibfnamefont {H.}~\bibnamefont
  {Kurebayashi}}, \bibinfo {author} {\bibfnamefont {H.}~\bibnamefont {Cheng}},
  \bibinfo {author} {\bibfnamefont {J.}~\bibnamefont {Zhou}}, \bibinfo {author}
  {\bibfnamefont {S.}~\bibnamefont {Khan}}, \bibinfo {author} {\bibfnamefont
  {Y.~P.}\ \bibnamefont {Feng}},\ and\ \bibinfo {author} {\bibfnamefont
  {G.}~\bibnamefont {Eda}},\ }\bibfield  {title} {\bibinfo {title} {Controlling
  the magnetic anisotropy in \ce{Cr2Ge2Te6} by electrostatic gating},\ }\href
  {https://doi.org/10.1038/s41928-020-0427-7} {\bibfield  {journal} {\bibinfo
  {journal} {Nat. Electron.}\ }\textbf {\bibinfo {volume} {3}},\ \bibinfo
  {pages} {460} (\bibinfo {year} {2020})}\BibitemShut {NoStop}%
\bibitem [{\citenamefont {Watson}\ \emph {et~al.}(2020)\citenamefont {Watson},
  \citenamefont {Markovi\ifmmode~\acute{c}\else \'{c}\fi{}}, \citenamefont
  {Mazzola}, \citenamefont {Rajan}, \citenamefont {Morales}, \citenamefont
  {Burn}, \citenamefont {Hesjedal}, \citenamefont {van~der Laan}, \citenamefont
  {Mukherjee}, \citenamefont {Kim}, \citenamefont {Bigi}, \citenamefont
  {Vobornik}, \citenamefont {Ciomaga~Hatnean}, \citenamefont {Balakrishnan},\
  and\ \citenamefont {King}}]{Watson2020}%
  \BibitemOpen
  \bibfield  {author} {\bibinfo {author} {\bibfnamefont {M.~D.}\ \bibnamefont
  {Watson}}, \bibinfo {author} {\bibfnamefont {I.}~\bibnamefont
  {Markovi\ifmmode~\acute{c}\else \'{c}\fi{}}}, \bibinfo {author}
  {\bibfnamefont {F.}~\bibnamefont {Mazzola}}, \bibinfo {author} {\bibfnamefont
  {A.}~\bibnamefont {Rajan}}, \bibinfo {author} {\bibfnamefont {E.~A.}\
  \bibnamefont {Morales}}, \bibinfo {author} {\bibfnamefont {D.~M.}\
  \bibnamefont {Burn}}, \bibinfo {author} {\bibfnamefont {T.}~\bibnamefont
  {Hesjedal}}, \bibinfo {author} {\bibfnamefont {G.}~\bibnamefont {van~der
  Laan}}, \bibinfo {author} {\bibfnamefont {S.}~\bibnamefont {Mukherjee}},
  \bibinfo {author} {\bibfnamefont {T.~K.}\ \bibnamefont {Kim}}, \bibinfo
  {author} {\bibfnamefont {C.}~\bibnamefont {Bigi}}, \bibinfo {author}
  {\bibfnamefont {I.}~\bibnamefont {Vobornik}}, \bibinfo {author}
  {\bibfnamefont {M.}~\bibnamefont {Ciomaga~Hatnean}}, \bibinfo {author}
  {\bibfnamefont {G.}~\bibnamefont {Balakrishnan}},\ and\ \bibinfo {author}
  {\bibfnamefont {P.~D.~C.}\ \bibnamefont {King}},\ }\bibfield  {title}
  {\bibinfo {title} {Direct observation of the energy gain underpinning
  ferromagnetic superexchange in the electronic structure of \ce{CrGeTe3}},\
  }\href {https://doi.org/10.1103/PhysRevB.101.205125} {\bibfield  {journal}
  {\bibinfo  {journal} {Phys. Rev. B}\ }\textbf {\bibinfo {volume} {101}},\
  \bibinfo {pages} {205125} (\bibinfo {year} {2020})}\BibitemShut {NoStop}%
\bibitem [{\citenamefont {Sakurai}\ \emph {et~al.}(2021)\citenamefont
  {Sakurai}, \citenamefont {Rubrecht}, \citenamefont {Corredor}, \citenamefont
  {Takehara}, \citenamefont {Yasutani}, \citenamefont {Zeisner}, \citenamefont
  {Alfonsov}, \citenamefont {Selter}, \citenamefont {Aswartham}, \citenamefont
  {Wolter}, \citenamefont {B\"uchner}, \citenamefont {Ohta},\ and\
  \citenamefont {Kataev}}]{Sakurai2021}%
  \BibitemOpen
  \bibfield  {author} {\bibinfo {author} {\bibfnamefont {T.}~\bibnamefont
  {Sakurai}}, \bibinfo {author} {\bibfnamefont {B.}~\bibnamefont {Rubrecht}},
  \bibinfo {author} {\bibfnamefont {L.~T.}\ \bibnamefont {Corredor}}, \bibinfo
  {author} {\bibfnamefont {R.}~\bibnamefont {Takehara}}, \bibinfo {author}
  {\bibfnamefont {M.}~\bibnamefont {Yasutani}}, \bibinfo {author}
  {\bibfnamefont {J.}~\bibnamefont {Zeisner}}, \bibinfo {author} {\bibfnamefont
  {A.}~\bibnamefont {Alfonsov}}, \bibinfo {author} {\bibfnamefont
  {S.}~\bibnamefont {Selter}}, \bibinfo {author} {\bibfnamefont
  {S.}~\bibnamefont {Aswartham}}, \bibinfo {author} {\bibfnamefont {A.~U.~B.}\
  \bibnamefont {Wolter}}, \bibinfo {author} {\bibfnamefont {B.}~\bibnamefont
  {B\"uchner}}, \bibinfo {author} {\bibfnamefont {H.}~\bibnamefont {Ohta}},\
  and\ \bibinfo {author} {\bibfnamefont {V.}~\bibnamefont {Kataev}},\
  }\bibfield  {title} {\bibinfo {title} {Pressure control of the magnetic
  anisotropy of the quasi-two-dimensional van der {W}aals ferromagnet
  \ce{Cr2Ge2Te6}},\ }\href {https://doi.org/10.1103/PhysRevB.103.024404}
  {\bibfield  {journal} {\bibinfo  {journal} {Phys. Rev. B}\ }\textbf {\bibinfo
  {volume} {103}},\ \bibinfo {pages} {024404} (\bibinfo {year}
  {2021})}\BibitemShut {NoStop}%
\bibitem [{\citenamefont {Saito}\ \emph {et~al.}(2021)\citenamefont {Saito},
  \citenamefont {Hatayama}, \citenamefont {Shuang}, \citenamefont {Fons},
  \citenamefont {Kolobov},\ and\ \citenamefont {Sutou}}]{Saito2021}%
  \BibitemOpen
  \bibfield  {author} {\bibinfo {author} {\bibfnamefont {Y.}~\bibnamefont
  {Saito}}, \bibinfo {author} {\bibfnamefont {S.}~\bibnamefont {Hatayama}},
  \bibinfo {author} {\bibfnamefont {Y.}~\bibnamefont {Shuang}}, \bibinfo
  {author} {\bibfnamefont {P.}~\bibnamefont {Fons}}, \bibinfo {author}
  {\bibfnamefont {A.~V.}\ \bibnamefont {Kolobov}},\ and\ \bibinfo {author}
  {\bibfnamefont {Y.}~\bibnamefont {Sutou}},\ }\bibfield  {title} {\bibinfo
  {title} {Dimensional transformation of chemical bonding during
  crystallization in a layered chalcogenide material},\ }\href
  {https://doi.org/10.1038/s41598-020-80301-5} {\bibfield  {journal} {\bibinfo
  {journal} {Sci. Rep.}\ }\textbf {\bibinfo {volume} {11}},\ \bibinfo {pages}
  {4782} (\bibinfo {year} {2021})}\BibitemShut {NoStop}%
\bibitem [{\citenamefont {Liu}\ and\ \citenamefont {Petrovic}(2017)}]{Liu2017}%
  \BibitemOpen
  \bibfield  {author} {\bibinfo {author} {\bibfnamefont {Y.}~\bibnamefont
  {Liu}}\ and\ \bibinfo {author} {\bibfnamefont {C.}~\bibnamefont {Petrovic}},\
  }\bibfield  {title} {\bibinfo {title} {Critical behavior of
  quasi-two-dimensional semiconducting ferromagnet \ce{Cr2Ge2Te6}},\ }\href
  {https://doi.org/10.1103/PhysRevB.96.054406} {\bibfield  {journal} {\bibinfo
  {journal} {Phys. Rev. B}\ }\textbf {\bibinfo {volume} {96}},\ \bibinfo
  {pages} {054406} (\bibinfo {year} {2017})}\BibitemShut {NoStop}%
\bibitem [{\citenamefont {Lin}\ \emph {et~al.}(2017)\citenamefont {Lin},
  \citenamefont {Zhuang}, \citenamefont {Luo}, \citenamefont {Liu},
  \citenamefont {Chen}, \citenamefont {Yan}, \citenamefont {Sun}, \citenamefont
  {Zhou}, \citenamefont {Lu}, \citenamefont {Tong}, \citenamefont {Sheng},
  \citenamefont {Qu}, \citenamefont {Song}, \citenamefont {Zhu},\ and\
  \citenamefont {Sun}}]{Lin2017}%
  \BibitemOpen
  \bibfield  {author} {\bibinfo {author} {\bibfnamefont {G.~T.}\ \bibnamefont
  {Lin}}, \bibinfo {author} {\bibfnamefont {H.~L.}\ \bibnamefont {Zhuang}},
  \bibinfo {author} {\bibfnamefont {X.}~\bibnamefont {Luo}}, \bibinfo {author}
  {\bibfnamefont {B.~J.}\ \bibnamefont {Liu}}, \bibinfo {author} {\bibfnamefont
  {F.~C.}\ \bibnamefont {Chen}}, \bibinfo {author} {\bibfnamefont
  {J.}~\bibnamefont {Yan}}, \bibinfo {author} {\bibfnamefont {Y.}~\bibnamefont
  {Sun}}, \bibinfo {author} {\bibfnamefont {J.}~\bibnamefont {Zhou}}, \bibinfo
  {author} {\bibfnamefont {W.~J.}\ \bibnamefont {Lu}}, \bibinfo {author}
  {\bibfnamefont {P.}~\bibnamefont {Tong}}, \bibinfo {author} {\bibfnamefont
  {Z.~G.}\ \bibnamefont {Sheng}}, \bibinfo {author} {\bibfnamefont
  {Z.}~\bibnamefont {Qu}}, \bibinfo {author} {\bibfnamefont {W.~H.}\
  \bibnamefont {Song}}, \bibinfo {author} {\bibfnamefont {X.~B.}\ \bibnamefont
  {Zhu}},\ and\ \bibinfo {author} {\bibfnamefont {Y.~P.}\ \bibnamefont {Sun}},\
  }\bibfield  {title} {\bibinfo {title} {Tricritical behavior of the
  two-dimensional intrinsically ferromagnetic semiconductor \ce{CrGeTe3}},\
  }\href {https://doi.org/10.1103/PhysRevB.95.245212} {\bibfield  {journal}
  {\bibinfo  {journal} {Phys. Rev. B}\ }\textbf {\bibinfo {volume} {95}},\
  \bibinfo {pages} {245212} (\bibinfo {year} {2017})}\BibitemShut {NoStop}%
\bibitem [{\citenamefont {Zeisner}\ \emph {et~al.}(2019)\citenamefont
  {Zeisner}, \citenamefont {Alfonsov}, \citenamefont {Selter}, \citenamefont
  {Aswartham}, \citenamefont {Ghimire}, \citenamefont {Richter}, \citenamefont
  {van~den Brink}, \citenamefont {B\"uchner},\ and\ \citenamefont
  {Kataev}}]{Zeisner2019}%
  \BibitemOpen
  \bibfield  {author} {\bibinfo {author} {\bibfnamefont {J.}~\bibnamefont
  {Zeisner}}, \bibinfo {author} {\bibfnamefont {A.}~\bibnamefont {Alfonsov}},
  \bibinfo {author} {\bibfnamefont {S.}~\bibnamefont {Selter}}, \bibinfo
  {author} {\bibfnamefont {S.}~\bibnamefont {Aswartham}}, \bibinfo {author}
  {\bibfnamefont {M.~P.}\ \bibnamefont {Ghimire}}, \bibinfo {author}
  {\bibfnamefont {M.}~\bibnamefont {Richter}}, \bibinfo {author} {\bibfnamefont
  {J.}~\bibnamefont {van~den Brink}}, \bibinfo {author} {\bibfnamefont
  {B.}~\bibnamefont {B\"uchner}},\ and\ \bibinfo {author} {\bibfnamefont
  {V.}~\bibnamefont {Kataev}},\ }\bibfield  {title} {\bibinfo {title} {Magnetic
  anisotropy and spin-polarized two-dimensional electron gas in the van der
  {W}aals ferromagnet \ce{Cr2Ge2Te6}},\ }\href
  {https://doi.org/10.1103/PhysRevB.99.165109} {\bibfield  {journal} {\bibinfo
  {journal} {Phys. Rev. B}\ }\textbf {\bibinfo {volume} {99}},\ \bibinfo
  {pages} {165109} (\bibinfo {year} {2019})}\BibitemShut {NoStop}%
\bibitem [{\citenamefont {Suzuki}\ \emph {et~al.}(2019)\citenamefont {Suzuki},
  \citenamefont {Gao}, \citenamefont {Koshiishi}, \citenamefont {Nakata},
  \citenamefont {Hagiwara}, \citenamefont {Lin}, \citenamefont {Wan},
  \citenamefont {Kumigashira}, \citenamefont {Ono}, \citenamefont {Kang},
  \citenamefont {Kang}, \citenamefont {Yu}, \citenamefont {Kobayashi},
  \citenamefont {Cheong},\ and\ \citenamefont {Fujimori}}]{Suzuki2019}%
  \BibitemOpen
  \bibfield  {author} {\bibinfo {author} {\bibfnamefont {M.}~\bibnamefont
  {Suzuki}}, \bibinfo {author} {\bibfnamefont {B.}~\bibnamefont {Gao}},
  \bibinfo {author} {\bibfnamefont {K.}~\bibnamefont {Koshiishi}}, \bibinfo
  {author} {\bibfnamefont {S.}~\bibnamefont {Nakata}}, \bibinfo {author}
  {\bibfnamefont {K.}~\bibnamefont {Hagiwara}}, \bibinfo {author}
  {\bibfnamefont {C.}~\bibnamefont {Lin}}, \bibinfo {author} {\bibfnamefont
  {Y.~X.}\ \bibnamefont {Wan}}, \bibinfo {author} {\bibfnamefont
  {H.}~\bibnamefont {Kumigashira}}, \bibinfo {author} {\bibfnamefont
  {K.}~\bibnamefont {Ono}}, \bibinfo {author} {\bibfnamefont {S.}~\bibnamefont
  {Kang}}, \bibinfo {author} {\bibfnamefont {S.}~\bibnamefont {Kang}}, \bibinfo
  {author} {\bibfnamefont {J.}~\bibnamefont {Yu}}, \bibinfo {author}
  {\bibfnamefont {M.}~\bibnamefont {Kobayashi}}, \bibinfo {author}
  {\bibfnamefont {S.-W.}\ \bibnamefont {Cheong}},\ and\ \bibinfo {author}
  {\bibfnamefont {A.}~\bibnamefont {Fujimori}},\ }\bibfield  {title} {\bibinfo
  {title} {Coulomb-interaction effect on the two-dimensional electronic
  structure of the van der {W}aals ferromagnet
  $\mathrm{C}{\mathrm{r}}_{2}\mathrm{G}{\mathrm{e}}_{2}\mathrm{T}{\mathrm{e}}_{6}$},\
  }\href {https://doi.org/10.1103/PhysRevB.99.161401} {\bibfield  {journal}
  {\bibinfo  {journal} {Phys. Rev. B}\ }\textbf {\bibinfo {volume} {99}},\
  \bibinfo {pages} {161401(R)} (\bibinfo {year} {2019})}\BibitemShut {NoStop}%
\bibitem [{\citenamefont {Fang}\ \emph {et~al.}(2018)\citenamefont {Fang},
  \citenamefont {Wu}, \citenamefont {Zhu},\ and\ \citenamefont
  {Guo}}]{Fang2018}%
  \BibitemOpen
  \bibfield  {author} {\bibinfo {author} {\bibfnamefont {Y.}~\bibnamefont
  {Fang}}, \bibinfo {author} {\bibfnamefont {S.}~\bibnamefont {Wu}}, \bibinfo
  {author} {\bibfnamefont {Z.-Z.}\ \bibnamefont {Zhu}},\ and\ \bibinfo {author}
  {\bibfnamefont {G.-Y.}\ \bibnamefont {Guo}},\ }\bibfield  {title} {\bibinfo
  {title} {Large magneto-optical effects and magnetic anisotropy energy in
  two-dimensional \ce{Cr2Ge2Te6}},\ }\href
  {https://doi.org/10.1103/PhysRevB.98.125416} {\bibfield  {journal} {\bibinfo
  {journal} {Phys. Rev. B}\ }\textbf {\bibinfo {volume} {98}},\ \bibinfo
  {pages} {125416} (\bibinfo {year} {2018})}\BibitemShut {NoStop}%
\bibitem [{\citenamefont {Lee}\ \emph {et~al.}(2020)\citenamefont {Lee},
  \citenamefont {Kotani},\ and\ \citenamefont {Ke}}]{Lee2020}%
  \BibitemOpen
  \bibfield  {author} {\bibinfo {author} {\bibfnamefont {Y.}~\bibnamefont
  {Lee}}, \bibinfo {author} {\bibfnamefont {T.}~\bibnamefont {Kotani}},\ and\
  \bibinfo {author} {\bibfnamefont {L.}~\bibnamefont {Ke}},\ }\bibfield
  {title} {\bibinfo {title} {Role of nonlocality in exchange correlation for
  magnetic two-dimensional van der {W}aals materials},\ }\href
  {https://doi.org/10.1103/PhysRevB.101.241409} {\bibfield  {journal} {\bibinfo
   {journal} {Phys. Rev. B}\ }\textbf {\bibinfo {volume} {101}},\ \bibinfo
  {pages} {241409(R)} (\bibinfo {year} {2020})}\BibitemShut {NoStop}%
\bibitem [{\citenamefont {O.~Fumega}\ \emph {et~al.}(2020)\citenamefont
  {O.~Fumega}, \citenamefont {Blanco-Canosa}, \citenamefont {Babu-Vasili},
  \citenamefont {Gargiani}, \citenamefont {Li}, \citenamefont {Zhou},
  \citenamefont {Rivadulla},\ and\ \citenamefont {Pardo}}]{Fumega2020}%
  \BibitemOpen
  \bibfield  {author} {\bibinfo {author} {\bibfnamefont {A.}~\bibnamefont
  {O.~Fumega}}, \bibinfo {author} {\bibfnamefont {S.}~\bibnamefont
  {Blanco-Canosa}}, \bibinfo {author} {\bibfnamefont {H.}~\bibnamefont
  {Babu-Vasili}}, \bibinfo {author} {\bibfnamefont {P.}~\bibnamefont
  {Gargiani}}, \bibinfo {author} {\bibfnamefont {H.}~\bibnamefont {Li}},
  \bibinfo {author} {\bibfnamefont {J.-S.}\ \bibnamefont {Zhou}}, \bibinfo
  {author} {\bibfnamefont {F.}~\bibnamefont {Rivadulla}},\ and\ \bibinfo
  {author} {\bibfnamefont {V.}~\bibnamefont {Pardo}},\ }\bibfield  {title}
  {\bibinfo {title} {{Electronic structure and magnetic exchange interactions
  of {C}r-based van der {W}aals ferromagnets. A comparative study between
  \ce{CrBr3} and \ce{Cr2Ge2Te6}}},\ }\href {https://doi.org/10.1039/D0TC02003F}
  {\bibfield  {journal} {\bibinfo  {journal} {J. Mater. Chem. C}\ }\textbf
  {\bibinfo {volume} {8}},\ \bibinfo {pages} {13582} (\bibinfo {year}
  {2020})}\BibitemShut {NoStop}%
\bibitem [{\citenamefont {Tiwari}\ \emph
  {et~al.}(2021{\natexlab{a}})\citenamefont {Tiwari}, \citenamefont {Van~de
  Put}, \citenamefont {Sor\'ee},\ and\ \citenamefont
  {Vandenberghe}}]{Tiwari2021}%
  \BibitemOpen
  \bibfield  {author} {\bibinfo {author} {\bibfnamefont {S.}~\bibnamefont
  {Tiwari}}, \bibinfo {author} {\bibfnamefont {M.~L.}\ \bibnamefont {Van~de
  Put}}, \bibinfo {author} {\bibfnamefont {B.}~\bibnamefont {Sor\'ee}},\ and\
  \bibinfo {author} {\bibfnamefont {W.~G.}\ \bibnamefont {Vandenberghe}},\
  }\bibfield  {title} {\bibinfo {title} {Critical behavior of the ferromagnets
  \ce{CrI3}, \ce{CrBr3}, and \ce{CrGeTe3} and the antiferromagnet \ce{FeCl2}: A
  detailed first-principles study},\ }\href
  {https://doi.org/10.1103/PhysRevB.103.014432} {\bibfield  {journal} {\bibinfo
   {journal} {Phys. Rev. B}\ }\textbf {\bibinfo {volume} {103}},\ \bibinfo
  {pages} {014432} (\bibinfo {year} {2021}{\natexlab{a}})}\BibitemShut
  {NoStop}%
\bibitem [{\citenamefont {Koepernik}\ and\ \citenamefont
  {Eschrig}(1999)}]{Koepernik1999}%
  \BibitemOpen
  \bibfield  {author} {\bibinfo {author} {\bibfnamefont {K.}~\bibnamefont
  {Koepernik}}\ and\ \bibinfo {author} {\bibfnamefont {H.}~\bibnamefont
  {Eschrig}},\ }\bibfield  {title} {\bibinfo {title} {Full-potential
  nonorthogonal local-orbital minimum-basis band-structure scheme},\ }\href
  {https://doi.org/10.1103/PhysRevB.59.1743} {\bibfield  {journal} {\bibinfo
  {journal} {Phys. Rev. B}\ }\textbf {\bibinfo {volume} {59}},\ \bibinfo
  {pages} {1743} (\bibinfo {year} {1999})}\BibitemShut {NoStop}%
\bibitem [{\citenamefont {Perdew}\ \emph {et~al.}(1996)\citenamefont {Perdew},
  \citenamefont {Burke},\ and\ \citenamefont {Ernzerhof}}]{Perdew1996}%
  \BibitemOpen
  \bibfield  {author} {\bibinfo {author} {\bibfnamefont {J.~P.}\ \bibnamefont
  {Perdew}}, \bibinfo {author} {\bibfnamefont {K.}~\bibnamefont {Burke}},\ and\
  \bibinfo {author} {\bibfnamefont {M.}~\bibnamefont {Ernzerhof}},\ }\bibfield
  {title} {\bibinfo {title} {Generalized gradient approximation made simple},\
  }\href {https://doi.org/10.1103/PhysRevLett.77.3865} {\bibfield  {journal}
  {\bibinfo  {journal} {Phys. Rev. Lett.}\ }\textbf {\bibinfo {volume} {77}},\
  \bibinfo {pages} {3865} (\bibinfo {year} {1996})}\BibitemShut {NoStop}%
\bibitem [{\citenamefont {Eschrig}\ and\ \citenamefont
  {Koepernik}(2009)}]{Eschrig2009}%
  \BibitemOpen
  \bibfield  {author} {\bibinfo {author} {\bibfnamefont {H.}~\bibnamefont
  {Eschrig}}\ and\ \bibinfo {author} {\bibfnamefont {K.}~\bibnamefont
  {Koepernik}},\ }\bibfield  {title} {\bibinfo {title} {Tight-binding models
  for the iron-based superconductors},\ }\href
  {https://doi.org/10.1103/PhysRevB.80.104503} {\bibfield  {journal} {\bibinfo
  {journal} {Phys. Rev. B}\ }\textbf {\bibinfo {volume} {80}},\ \bibinfo
  {pages} {104503} (\bibinfo {year} {2009})}\BibitemShut {NoStop}%
\bibitem [{\citenamefont {Koepernik}\ \emph {et~al.}(2023)\citenamefont
  {Koepernik}, \citenamefont {Janson}, \citenamefont {Sun},\ and\ \citenamefont
  {van~den Brink}}]{Koepernik2023}%
  \BibitemOpen
  \bibfield  {author} {\bibinfo {author} {\bibfnamefont {K.}~\bibnamefont
  {Koepernik}}, \bibinfo {author} {\bibfnamefont {O.}~\bibnamefont {Janson}},
  \bibinfo {author} {\bibfnamefont {Y.}~\bibnamefont {Sun}},\ and\ \bibinfo
  {author} {\bibfnamefont {J.}~\bibnamefont {van~den Brink}},\ }\bibfield
  {title} {\bibinfo {title} {Symmetry-conserving maximally projected {W}annier
  functions},\ }\href {https://doi.org/10.1103/PhysRevB.107.235135} {\bibfield
  {journal} {\bibinfo  {journal} {Phys. Rev. B}\ }\textbf {\bibinfo {volume}
  {107}},\ \bibinfo {pages} {235135} (\bibinfo {year} {2023})}\BibitemShut
  {NoStop}%
\bibitem [{\citenamefont {Jeschke}\ \emph {et~al.}(2013)\citenamefont
  {Jeschke}, \citenamefont {Salvat-Pujol},\ and\ \citenamefont
  {Valent\'{\i}}}]{Jeschke2013}%
  \BibitemOpen
  \bibfield  {author} {\bibinfo {author} {\bibfnamefont {H.~O.}\ \bibnamefont
  {Jeschke}}, \bibinfo {author} {\bibfnamefont {F.}~\bibnamefont
  {Salvat-Pujol}},\ and\ \bibinfo {author} {\bibfnamefont {R.}~\bibnamefont
  {Valent\'{\i}}},\ }\bibfield  {title} {\bibinfo {title} {First-principles
  determination of {H}eisenberg hamiltonian parameters for the
  spin-$\frac{1}{2}$ kagome antiferromagnet \ce{ZnCu3(OH)6Cl2}},\ }\href
  {https://doi.org/10.1103/PhysRevB.88.075106} {\bibfield  {journal} {\bibinfo
  {journal} {Phys. Rev. B}\ }\textbf {\bibinfo {volume} {88}},\ \bibinfo
  {pages} {075106} (\bibinfo {year} {2013})}\BibitemShut {NoStop}%
\bibitem [{\citenamefont {Iqbal}\ \emph {et~al.}(2017)\citenamefont {Iqbal},
  \citenamefont {M\"uller}, \citenamefont {Riedl}, \citenamefont {Reuther},
  \citenamefont {Rachel}, \citenamefont {Valent\'{\i}}, \citenamefont
  {Gingras}, \citenamefont {Thomale},\ and\ \citenamefont
  {Jeschke}}]{Iqbal2017}%
  \BibitemOpen
  \bibfield  {author} {\bibinfo {author} {\bibfnamefont {Y.}~\bibnamefont
  {Iqbal}}, \bibinfo {author} {\bibfnamefont {T.}~\bibnamefont {M\"uller}},
  \bibinfo {author} {\bibfnamefont {K.}~\bibnamefont {Riedl}}, \bibinfo
  {author} {\bibfnamefont {J.}~\bibnamefont {Reuther}}, \bibinfo {author}
  {\bibfnamefont {S.}~\bibnamefont {Rachel}}, \bibinfo {author} {\bibfnamefont
  {R.}~\bibnamefont {Valent\'{\i}}}, \bibinfo {author} {\bibfnamefont
  {M.~J.~P.}\ \bibnamefont {Gingras}}, \bibinfo {author} {\bibfnamefont
  {R.}~\bibnamefont {Thomale}},\ and\ \bibinfo {author} {\bibfnamefont {H.~O.}\
  \bibnamefont {Jeschke}},\ }\bibfield  {title} {\bibinfo {title} {Signatures
  of a gearwheel quantum spin liquid in a spin-$\frac{1}{2}$ pyrochlore
  molybdate {H}eisenberg antiferromagnet},\ }\href
  {https://doi.org/10.1103/PhysRevMaterials.1.071201} {\bibfield  {journal}
  {\bibinfo  {journal} {Phys. Rev. Mater.}\ }\textbf {\bibinfo {volume} {1}},\
  \bibinfo {pages} {071201(R)} (\bibinfo {year} {2017})}\BibitemShut {NoStop}%
\bibitem [{\citenamefont {Guterding}\ \emph {et~al.}(2016)\citenamefont
  {Guterding}, \citenamefont {Valent\'{\i}},\ and\ \citenamefont
  {Jeschke}}]{Guterding2016}%
  \BibitemOpen
  \bibfield  {author} {\bibinfo {author} {\bibfnamefont {D.}~\bibnamefont
  {Guterding}}, \bibinfo {author} {\bibfnamefont {R.}~\bibnamefont
  {Valent\'{\i}}},\ and\ \bibinfo {author} {\bibfnamefont {H.~O.}\ \bibnamefont
  {Jeschke}},\ }\bibfield  {title} {\bibinfo {title} {Reduction of magnetic
  interlayer coupling in barlowite through isoelectronic substitution},\ }\href
  {https://doi.org/10.1103/PhysRevB.94.125136} {\bibfield  {journal} {\bibinfo
  {journal} {Phys. Rev. B}\ }\textbf {\bibinfo {volume} {94}},\ \bibinfo
  {pages} {125136} (\bibinfo {year} {2016})}\BibitemShut {NoStop}%
\bibitem [{\citenamefont {Liechtenstein}\ \emph {et~al.}(1995)\citenamefont
  {Liechtenstein}, \citenamefont {Anisimov},\ and\ \citenamefont
  {Zaanen}}]{Liechtenstein1995}%
  \BibitemOpen
  \bibfield  {author} {\bibinfo {author} {\bibfnamefont {A.~I.}\ \bibnamefont
  {Liechtenstein}}, \bibinfo {author} {\bibfnamefont {V.~I.}\ \bibnamefont
  {Anisimov}},\ and\ \bibinfo {author} {\bibfnamefont {J.}~\bibnamefont
  {Zaanen}},\ }\bibfield  {title} {\bibinfo {title} {Density-functional theory
  and strong interactions: Orbital ordering in {M}ott-{H}ubbard insulators},\
  }\href {https://doi.org/10.1103/PhysRevB.52.R5467} {\bibfield  {journal}
  {\bibinfo  {journal} {Phys. Rev. B}\ }\textbf {\bibinfo {volume} {52}},\
  \bibinfo {pages} {R5467} (\bibinfo {year} {1995})}\BibitemShut {NoStop}%
\bibitem [{\citenamefont {Mizokawa}\ and\ \citenamefont
  {Fujimori}(1996)}]{Mizokawa1996}%
  \BibitemOpen
  \bibfield  {author} {\bibinfo {author} {\bibfnamefont {T.}~\bibnamefont
  {Mizokawa}}\ and\ \bibinfo {author} {\bibfnamefont {A.}~\bibnamefont
  {Fujimori}},\ }\bibfield  {title} {\bibinfo {title} {Electronic structure and
  orbital ordering in perovskite-type 3d transition-metal oxides studied by
  {H}artree-{F}ock band-structure calculations},\ }\href
  {https://doi.org/10.1103/PhysRevB.54.5368} {\bibfield  {journal} {\bibinfo
  {journal} {Phys. Rev. B}\ }\textbf {\bibinfo {volume} {54}},\ \bibinfo
  {pages} {5368} (\bibinfo {year} {1996})}\BibitemShut {NoStop}%
\bibitem [{\citenamefont {Shinaoka}\ \emph {et~al.}(2021)\citenamefont
  {Shinaoka}, \citenamefont {Otsuki}, \citenamefont {Kawamura}, \citenamefont
  {Takemori},\ and\ \citenamefont {Yoshimi}}]{Shinaoka2021}%
  \BibitemOpen
  \bibfield  {author} {\bibinfo {author} {\bibfnamefont {H.}~\bibnamefont
  {Shinaoka}}, \bibinfo {author} {\bibfnamefont {J.}~\bibnamefont {Otsuki}},
  \bibinfo {author} {\bibfnamefont {M.}~\bibnamefont {Kawamura}}, \bibinfo
  {author} {\bibfnamefont {N.}~\bibnamefont {Takemori}},\ and\ \bibinfo
  {author} {\bibfnamefont {K.}~\bibnamefont {Yoshimi}},\ }\bibfield  {title}
  {\bibinfo {title} {{DCore: Integrated DMFT software for correlated
  electrons}},\ }\href {https://doi.org/10.21468/SciPostPhys.10.5.117}
  {\bibfield  {journal} {\bibinfo  {journal} {SciPost Phys.}\ }\textbf
  {\bibinfo {volume} {10}},\ \bibinfo {pages} {117} (\bibinfo {year}
  {2021})}\BibitemShut {NoStop}%
\bibitem [{\citenamefont {Werner}\ \emph {et~al.}(2006)\citenamefont {Werner},
  \citenamefont {Comanac}, \citenamefont {de' Medici}, \citenamefont {Troyer},\
  and\ \citenamefont {Millis}}]{Werner2006}%
  \BibitemOpen
  \bibfield  {author} {\bibinfo {author} {\bibfnamefont {P.}~\bibnamefont
  {Werner}}, \bibinfo {author} {\bibfnamefont {A.}~\bibnamefont {Comanac}},
  \bibinfo {author} {\bibfnamefont {L.}~\bibnamefont {de' Medici}}, \bibinfo
  {author} {\bibfnamefont {M.}~\bibnamefont {Troyer}},\ and\ \bibinfo {author}
  {\bibfnamefont {A.~J.}\ \bibnamefont {Millis}},\ }\bibfield  {title}
  {\bibinfo {title} {Continuous-time solver for quantum impurity models},\
  }\href {https://doi.org/10.1103/PhysRevLett.97.076405} {\bibfield  {journal}
  {\bibinfo  {journal} {Phys. Rev. Lett.}\ }\textbf {\bibinfo {volume} {97}},\
  \bibinfo {pages} {076405} (\bibinfo {year} {2006})}\BibitemShut {NoStop}%
\bibitem [{\citenamefont {Gull}\ \emph {et~al.}(2011)\citenamefont {Gull},
  \citenamefont {Millis}, \citenamefont {Lichtenstein}, \citenamefont
  {Rubtsov}, \citenamefont {Troyer},\ and\ \citenamefont {Werner}}]{Gull2011}%
  \BibitemOpen
  \bibfield  {author} {\bibinfo {author} {\bibfnamefont {E.}~\bibnamefont
  {Gull}}, \bibinfo {author} {\bibfnamefont {A.~J.}\ \bibnamefont {Millis}},
  \bibinfo {author} {\bibfnamefont {A.~I.}\ \bibnamefont {Lichtenstein}},
  \bibinfo {author} {\bibfnamefont {A.~N.}\ \bibnamefont {Rubtsov}}, \bibinfo
  {author} {\bibfnamefont {M.}~\bibnamefont {Troyer}},\ and\ \bibinfo {author}
  {\bibfnamefont {P.}~\bibnamefont {Werner}},\ }\bibfield  {title} {\bibinfo
  {title} {Continuous-time {M}onte {C}arlo methods for quantum impurity
  models},\ }\href {https://doi.org/10.1103/RevModPhys.83.349} {\bibfield
  {journal} {\bibinfo  {journal} {Rev. Mod. Phys.}\ }\textbf {\bibinfo {volume}
  {83}},\ \bibinfo {pages} {349} (\bibinfo {year} {2011})}\BibitemShut
  {NoStop}%
\bibitem [{\citenamefont {Anisimov}\ \emph {et~al.}(1993)\citenamefont
  {Anisimov}, \citenamefont {Solovyev}, \citenamefont {Korotin}, \citenamefont
  {Czy\ifmmode~\dot{z}\else \.{z}\fi{}yk},\ and\ \citenamefont
  {Sawatzky}}]{Anisimov1993}%
  \BibitemOpen
  \bibfield  {author} {\bibinfo {author} {\bibfnamefont {V.~I.}\ \bibnamefont
  {Anisimov}}, \bibinfo {author} {\bibfnamefont {I.~V.}\ \bibnamefont
  {Solovyev}}, \bibinfo {author} {\bibfnamefont {M.~A.}\ \bibnamefont
  {Korotin}}, \bibinfo {author} {\bibfnamefont {M.~T.}\ \bibnamefont
  {Czy\ifmmode~\dot{z}\else \.{z}\fi{}yk}},\ and\ \bibinfo {author}
  {\bibfnamefont {G.~A.}\ \bibnamefont {Sawatzky}},\ }\bibfield  {title}
  {\bibinfo {title} {Density-functional theory and \ce{NiO} photoemission
  spectra},\ }\href {https://doi.org/10.1103/PhysRevB.48.16929} {\bibfield
  {journal} {\bibinfo  {journal} {Phys. Rev. B}\ }\textbf {\bibinfo {volume}
  {48}},\ \bibinfo {pages} {16929} (\bibinfo {year} {1993})}\BibitemShut
  {NoStop}%
\bibitem [{\citenamefont {Ghosh}\ \emph {et~al.}(2019)\citenamefont {Ghosh},
  \citenamefont {Iqbal}, \citenamefont {M\"{u}ller}, \citenamefont
  {Ponnaganti}, \citenamefont {Thomale}, \citenamefont {Narayanan},
  \citenamefont {Reuther}, \citenamefont {Gingras},\ and\ \citenamefont
  {Jeschke}}]{Ghosh2019}%
  \BibitemOpen
  \bibfield  {author} {\bibinfo {author} {\bibfnamefont {P.}~\bibnamefont
  {Ghosh}}, \bibinfo {author} {\bibfnamefont {Y.}~\bibnamefont {Iqbal}},
  \bibinfo {author} {\bibfnamefont {T.}~\bibnamefont {M\"{u}ller}}, \bibinfo
  {author} {\bibfnamefont {R.~T.}\ \bibnamefont {Ponnaganti}}, \bibinfo
  {author} {\bibfnamefont {R.}~\bibnamefont {Thomale}}, \bibinfo {author}
  {\bibfnamefont {R.}~\bibnamefont {Narayanan}}, \bibinfo {author}
  {\bibfnamefont {J.}~\bibnamefont {Reuther}}, \bibinfo {author} {\bibfnamefont
  {M.~J.~P.}\ \bibnamefont {Gingras}},\ and\ \bibinfo {author} {\bibfnamefont
  {H.~O.}\ \bibnamefont {Jeschke}},\ }\bibfield  {title} {\bibinfo {title}
  {Breathing chromium spinels: a showcase for a variety of pyrochlore
  {H}eisenberg hamiltonians},\ }\href
  {https://doi.org/10.1038/s41535-019-0202-z} {\bibfield  {journal} {\bibinfo
  {journal} {npj Quantum Mater.}\ }\textbf {\bibinfo {volume} {4}},\ \bibinfo
  {pages} {63} (\bibinfo {year} {2019})}\BibitemShut {NoStop}%
\bibitem [{\citenamefont {Yasuda}\ \emph {et~al.}(2005)\citenamefont {Yasuda},
  \citenamefont {Todo}, \citenamefont {Hukushima}, \citenamefont {Alet},
  \citenamefont {Keller}, \citenamefont {Troyer},\ and\ \citenamefont
  {Takayama}}]{Yasuda2005}%
  \BibitemOpen
  \bibfield  {author} {\bibinfo {author} {\bibfnamefont {C.}~\bibnamefont
  {Yasuda}}, \bibinfo {author} {\bibfnamefont {S.}~\bibnamefont {Todo}},
  \bibinfo {author} {\bibfnamefont {K.}~\bibnamefont {Hukushima}}, \bibinfo
  {author} {\bibfnamefont {F.}~\bibnamefont {Alet}}, \bibinfo {author}
  {\bibfnamefont {M.}~\bibnamefont {Keller}}, \bibinfo {author} {\bibfnamefont
  {M.}~\bibnamefont {Troyer}},\ and\ \bibinfo {author} {\bibfnamefont
  {H.}~\bibnamefont {Takayama}},\ }\bibfield  {title} {\bibinfo {title} {N\'eel
  temperature of quasi-low-dimensional {H}eisenberg antiferromagnets},\ }\href
  {https://doi.org/10.1103/PhysRevLett.94.217201} {\bibfield  {journal}
  {\bibinfo  {journal} {Phys. Rev. Lett.}\ }\textbf {\bibinfo {volume} {94}},\
  \bibinfo {pages} {217201} (\bibinfo {year} {2005})}\BibitemShut {NoStop}%
\bibitem [{\citenamefont {Lado}\ and\ \citenamefont
  {Fernández-Rossier}(2017)}]{Lado2017}%
  \BibitemOpen
  \bibfield  {author} {\bibinfo {author} {\bibfnamefont {J.~L.}\ \bibnamefont
  {Lado}}\ and\ \bibinfo {author} {\bibfnamefont {J.}~\bibnamefont
  {Fernández-Rossier}},\ }\bibfield  {title} {\bibinfo {title} {On the origin
  of magnetic anisotropy in two dimensional \ce{CrI3}},\ }\href
  {https://doi.org/10.1088/2053-1583/aa75ed} {\bibfield  {journal} {\bibinfo
  {journal} {2D Mater.}\ }\textbf {\bibinfo {volume} {4}},\ \bibinfo {pages}
  {035002} (\bibinfo {year} {2017})}\BibitemShut {NoStop}%
\bibitem [{\citenamefont {Tiwari}\ \emph
  {et~al.}(2021{\natexlab{b}})\citenamefont {Tiwari}, \citenamefont {Vanherck},
  \citenamefont {Van~de Put}, \citenamefont {Vandenberghe},\ and\ \citenamefont
  {Sor\'ee}}]{Tiwari2021b}%
  \BibitemOpen
  \bibfield  {author} {\bibinfo {author} {\bibfnamefont {S.}~\bibnamefont
  {Tiwari}}, \bibinfo {author} {\bibfnamefont {J.}~\bibnamefont {Vanherck}},
  \bibinfo {author} {\bibfnamefont {M.~L.}\ \bibnamefont {Van~de Put}},
  \bibinfo {author} {\bibfnamefont {W.~G.}\ \bibnamefont {Vandenberghe}},\ and\
  \bibinfo {author} {\bibfnamefont {B.}~\bibnamefont {Sor\'ee}},\ }\bibfield
  {title} {\bibinfo {title} {Computing curie temperature of two-dimensional
  ferromagnets in the presence of exchange anisotropy},\ }\href
  {https://doi.org/10.1103/PhysRevResearch.3.043024} {\bibfield  {journal}
  {\bibinfo  {journal} {Phys. Rev. Res.}\ }\textbf {\bibinfo {volume} {3}},\
  \bibinfo {pages} {043024} (\bibinfo {year} {2021}{\natexlab{b}})}\BibitemShut
  {NoStop}%
\bibitem [{\citenamefont {Kim}\ \emph {et~al.}(2017)\citenamefont {Kim},
  \citenamefont {Khmelevskyi}, \citenamefont {Mazin}, \citenamefont
  {Agterberg},\ and\ \citenamefont {Franchini}}]{Kim2017}%
  \BibitemOpen
  \bibfield  {author} {\bibinfo {author} {\bibfnamefont {B.}~\bibnamefont
  {Kim}}, \bibinfo {author} {\bibfnamefont {S.}~\bibnamefont {Khmelevskyi}},
  \bibinfo {author} {\bibfnamefont {I.~I.}\ \bibnamefont {Mazin}}, \bibinfo
  {author} {\bibfnamefont {D.~F.}\ \bibnamefont {Agterberg}},\ and\ \bibinfo
  {author} {\bibfnamefont {C.}~\bibnamefont {Franchini}},\ }\bibfield  {title}
  {\bibinfo {title} {Anisotropy of magnetic interactions and symmetry of the
  order parameter in unconventional superconductor \ce{Sr2RuO4}},\ }\href
  {https://doi.org/10.1038/s41535-017-0041-8} {\bibfield  {journal} {\bibinfo
  {journal} {npj Quant. Mater.}\ }\textbf {\bibinfo {volume} {2}},\ \bibinfo
  {pages} {37} (\bibinfo {year} {2017})}\BibitemShut {NoStop}%
\bibitem [{\citenamefont {Huang}\ \emph {et~al.}(2023)\citenamefont {Huang},
  \citenamefont {Jeschke},\ and\ \citenamefont {Mazin}}]{Huang2023}%
  \BibitemOpen
  \bibfield  {author} {\bibinfo {author} {\bibfnamefont {Y.~N.}\ \bibnamefont
  {Huang}}, \bibinfo {author} {\bibfnamefont {H.~O.}\ \bibnamefont {Jeschke}},\
  and\ \bibinfo {author} {\bibfnamefont {I.~I.}\ \bibnamefont {Mazin}},\
  }\bibfield  {title} {\bibinfo {title} {Cr{R}h{A}s: a member of a large family
  of metallic kagome antiferromagnets},\ }\href
  {https://doi.org/10.1038/s41535-023-00562-x} {\bibfield  {journal} {\bibinfo
  {journal} {npj Quant. Mater.}\ }\textbf {\bibinfo {volume} {8}},\ \bibinfo
  {pages} {32} (\bibinfo {year} {2023})}\BibitemShut {NoStop}%
\bibitem [{\citenamefont {Irkhin}\ \emph {et~al.}(1999)\citenamefont {Irkhin},
  \citenamefont {Katanin},\ and\ \citenamefont {Katsnelson}}]{Irkhin1999}%
  \BibitemOpen
  \bibfield  {author} {\bibinfo {author} {\bibfnamefont {V.~Y.}\ \bibnamefont
  {Irkhin}}, \bibinfo {author} {\bibfnamefont {A.~A.}\ \bibnamefont
  {Katanin}},\ and\ \bibinfo {author} {\bibfnamefont {M.~I.}\ \bibnamefont
  {Katsnelson}},\ }\bibfield  {title} {\bibinfo {title} {Self-consistent
  spin-wave theory of layered {H}eisenberg magnets},\ }\href
  {https://doi.org/10.1103/PhysRevB.60.1082} {\bibfield  {journal} {\bibinfo
  {journal} {Phys. Rev. B}\ }\textbf {\bibinfo {volume} {60}},\ \bibinfo
  {pages} {1082} (\bibinfo {year} {1999})}\BibitemShut {NoStop}%
\bibitem [{\citenamefont {Costa}\ and\ \citenamefont
  {Pires}(2003)}]{Costa2003}%
  \BibitemOpen
  \bibfield  {author} {\bibinfo {author} {\bibfnamefont {B.}~\bibnamefont
  {Costa}}\ and\ \bibinfo {author} {\bibfnamefont {A.}~\bibnamefont {Pires}},\
  }\bibfield  {title} {\bibinfo {title} {Phase diagrams of a two-dimensional
  {H}eisenberg antiferromagnet with single-ion anisotropy},\ }\href
  {https://doi.org/10.1016/S0304-8853(02)01527-5} {\bibfield  {journal}
  {\bibinfo  {journal} {J. Mag. Mag. Mater.}\ }\textbf {\bibinfo {volume}
  {262}},\ \bibinfo {pages} {316} (\bibinfo {year} {2003})}\BibitemShut
  {NoStop}%
\bibitem [{\citenamefont {Setyawan}\ and\ \citenamefont
  {Curtarolo}(2010)}]{Setyawan2010}%
  \BibitemOpen
  \bibfield  {author} {\bibinfo {author} {\bibfnamefont {W.}~\bibnamefont
  {Setyawan}}\ and\ \bibinfo {author} {\bibfnamefont {S.}~\bibnamefont
  {Curtarolo}},\ }\bibfield  {title} {\bibinfo {title} {High-throughput
  electronic band structure calculations: Challenges and tools},\ }\href
  {https://doi.org/10.1016/j.commatsci.2010.05.010} {\bibfield  {journal}
  {\bibinfo  {journal} {Comput. Mater. Sci.}\ }\textbf {\bibinfo {volume}
  {49}},\ \bibinfo {pages} {299} (\bibinfo {year} {2010})}\BibitemShut
  {NoStop}%
\bibitem [{\citenamefont {Arsenault}\ \emph {et~al.}(2012)\citenamefont
  {Arsenault}, \citenamefont {S\'emon},\ and\ \citenamefont
  {Tremblay}}]{Arsenault2012}%
  \BibitemOpen
  \bibfield  {author} {\bibinfo {author} {\bibfnamefont {L.-F.}\ \bibnamefont
  {Arsenault}}, \bibinfo {author} {\bibfnamefont {P.}~\bibnamefont {S\'emon}},\
  and\ \bibinfo {author} {\bibfnamefont {A.-M.~S.}\ \bibnamefont {Tremblay}},\
  }\bibfield  {title} {\bibinfo {title} {Benchmark of a modified iterated
  perturbation theory approach on the fcc lattice at strong coupling},\ }\href
  {https://doi.org/10.1103/PhysRevB.86.085133} {\bibfield  {journal} {\bibinfo
  {journal} {Phys. Rev. B}\ }\textbf {\bibinfo {volume} {86}},\ \bibinfo
  {pages} {085133} (\bibinfo {year} {2012})}\BibitemShut {NoStop}%
\bibitem [{\citenamefont {Kvashnin}\ \emph {et~al.}(2022)\citenamefont
  {Kvashnin}, \citenamefont {Rudenko}, \citenamefont {Thunstr\"om},
  \citenamefont {R\"osner},\ and\ \citenamefont {Katsnelson}}]{Kvashnin2022}%
  \BibitemOpen
  \bibfield  {author} {\bibinfo {author} {\bibfnamefont {Y.~O.}\ \bibnamefont
  {Kvashnin}}, \bibinfo {author} {\bibfnamefont {A.~N.}\ \bibnamefont
  {Rudenko}}, \bibinfo {author} {\bibfnamefont {P.}~\bibnamefont
  {Thunstr\"om}}, \bibinfo {author} {\bibfnamefont {M.}~\bibnamefont
  {R\"osner}},\ and\ \bibinfo {author} {\bibfnamefont {M.~I.}\ \bibnamefont
  {Katsnelson}},\ }\bibfield  {title} {\bibinfo {title} {Dynamical correlations
  in single-layer \ce{CrI3}},\ }\href
  {https://doi.org/10.1103/PhysRevB.105.205124} {\bibfield  {journal} {\bibinfo
   {journal} {Phys. Rev. B}\ }\textbf {\bibinfo {volume} {105}},\ \bibinfo
  {pages} {205124} (\bibinfo {year} {2022})}\BibitemShut {NoStop}%
\bibitem [{\citenamefont {Anderson}\ and\ \citenamefont
  {Hasegawa}(1955)}]{Anderson1955}%
  \BibitemOpen
  \bibfield  {author} {\bibinfo {author} {\bibfnamefont {P.~W.}\ \bibnamefont
  {Anderson}}\ and\ \bibinfo {author} {\bibfnamefont {H.}~\bibnamefont
  {Hasegawa}},\ }\bibfield  {title} {\bibinfo {title} {Considerations on double
  exchange},\ }\href {https://doi.org/10.1103/PhysRev.100.675} {\bibfield
  {journal} {\bibinfo  {journal} {Phys. Rev.}\ }\textbf {\bibinfo {volume}
  {100}},\ \bibinfo {pages} {675} (\bibinfo {year} {1955})}\BibitemShut
  {NoStop}%
\bibitem [{\citenamefont {Fujiwara}\ \emph {et~al.}(2022)\citenamefont
  {Fujiwara}, \citenamefont {Terashima}, \citenamefont {Otsuki}, \citenamefont
  {Takemori}, \citenamefont {Jeschke}, \citenamefont {Wakita}, \citenamefont
  {Yano}, \citenamefont {Hosoda}, \citenamefont {Kataoka}, \citenamefont
  {Teruya}, \citenamefont {Kakihana}, \citenamefont {Hedo}, \citenamefont
  {Nakama}, \citenamefont {Onuki}, \citenamefont {Yaji}, \citenamefont
  {Harasawa}, \citenamefont {Kuroda}, \citenamefont {Shin}, \citenamefont
  {Horiba}, \citenamefont {Kumigashira}, \citenamefont {Muraoka},\ and\
  \citenamefont {Yokoya}}]{Fujiwara2022}%
  \BibitemOpen
  \bibfield  {author} {\bibinfo {author} {\bibfnamefont {H.}~\bibnamefont
  {Fujiwara}}, \bibinfo {author} {\bibfnamefont {K.}~\bibnamefont {Terashima}},
  \bibinfo {author} {\bibfnamefont {J.}~\bibnamefont {Otsuki}}, \bibinfo
  {author} {\bibfnamefont {N.}~\bibnamefont {Takemori}}, \bibinfo {author}
  {\bibfnamefont {H.~O.}\ \bibnamefont {Jeschke}}, \bibinfo {author}
  {\bibfnamefont {T.}~\bibnamefont {Wakita}}, \bibinfo {author} {\bibfnamefont
  {Y.}~\bibnamefont {Yano}}, \bibinfo {author} {\bibfnamefont {W.}~\bibnamefont
  {Hosoda}}, \bibinfo {author} {\bibfnamefont {N.}~\bibnamefont {Kataoka}},
  \bibinfo {author} {\bibfnamefont {A.}~\bibnamefont {Teruya}}, \bibinfo
  {author} {\bibfnamefont {M.}~\bibnamefont {Kakihana}}, \bibinfo {author}
  {\bibfnamefont {M.}~\bibnamefont {Hedo}}, \bibinfo {author} {\bibfnamefont
  {T.}~\bibnamefont {Nakama}}, \bibinfo {author} {\bibfnamefont
  {Y.}~\bibnamefont {Onuki}}, \bibinfo {author} {\bibfnamefont
  {K.}~\bibnamefont {Yaji}}, \bibinfo {author} {\bibfnamefont {A.}~\bibnamefont
  {Harasawa}}, \bibinfo {author} {\bibfnamefont {K.}~\bibnamefont {Kuroda}},
  \bibinfo {author} {\bibfnamefont {S.}~\bibnamefont {Shin}}, \bibinfo {author}
  {\bibfnamefont {K.}~\bibnamefont {Horiba}}, \bibinfo {author} {\bibfnamefont
  {H.}~\bibnamefont {Kumigashira}}, \bibinfo {author} {\bibfnamefont
  {Y.}~\bibnamefont {Muraoka}},\ and\ \bibinfo {author} {\bibfnamefont
  {T.}~\bibnamefont {Yokoya}},\ }\bibfield  {title} {\bibinfo {title}
  {Anomalously large spin-dependent electron correlation in the nearly
  half-metallic ferromagnet \ce{CoS2}},\ }\href
  {https://doi.org/10.1103/PhysRevB.106.085114} {\bibfield  {journal} {\bibinfo
   {journal} {Phys. Rev. B}\ }\textbf {\bibinfo {volume} {106}},\ \bibinfo
  {pages} {085114} (\bibinfo {year} {2022})}\BibitemShut {NoStop}%
\bibitem [{\citenamefont {Torelli}\ \emph {et~al.}(2020)\citenamefont
  {Torelli}, \citenamefont {Moustafa}, \citenamefont {Jacobsen},\ and\
  \citenamefont {Olsen}}]{Torelli2020}%
  \BibitemOpen
  \bibfield  {author} {\bibinfo {author} {\bibfnamefont {D.}~\bibnamefont
  {Torelli}}, \bibinfo {author} {\bibfnamefont {H.}~\bibnamefont {Moustafa}},
  \bibinfo {author} {\bibfnamefont {K.~W.}\ \bibnamefont {Jacobsen}},\ and\
  \bibinfo {author} {\bibfnamefont {T.}~\bibnamefont {Olsen}},\ }\bibfield
  {title} {\bibinfo {title} {High-throughput computational screening for
  two-dimensional magnetic materials based on experimental databases of
  three-dimensional compounds},\ }\href
  {https://doi.org/10.1038/s41524-020-00428-x} {\bibfield  {journal} {\bibinfo
  {journal} {npj Comput. Mater.}\ }\textbf {\bibinfo {volume} {6}},\ \bibinfo
  {pages} {158} (\bibinfo {year} {2020})}\BibitemShut {NoStop}%
\bibitem [{\citenamefont {Georges}\ \emph {et~al.}(1996)\citenamefont
  {Georges}, \citenamefont {Kotliar}, \citenamefont {Krauth},\ and\
  \citenamefont {Rozenberg}}]{Georges1996}%
  \BibitemOpen
  \bibfield  {author} {\bibinfo {author} {\bibfnamefont {A.}~\bibnamefont
  {Georges}}, \bibinfo {author} {\bibfnamefont {G.}~\bibnamefont {Kotliar}},
  \bibinfo {author} {\bibfnamefont {W.}~\bibnamefont {Krauth}},\ and\ \bibinfo
  {author} {\bibfnamefont {M.~J.}\ \bibnamefont {Rozenberg}},\ }\bibfield
  {title} {\bibinfo {title} {{Dynamical mean-field theory of strongly
  correlated fermion systems and the limit of infinite dimensions}},\ }\href
  {https://doi.org/10.1103/RevModPhys.68.13} {\bibfield  {journal} {\bibinfo
  {journal} {Rev. Mod. Phys.}\ }\textbf {\bibinfo {volume} {68}},\ \bibinfo
  {pages} {13} (\bibinfo {year} {1996})}\BibitemShut {NoStop}%
\bibitem [{\citenamefont {Kotliar}\ \emph {et~al.}(2006)\citenamefont
  {Kotliar}, \citenamefont {Savrasov}, \citenamefont {Haule}, \citenamefont
  {Oudovenko}, \citenamefont {Parcollet},\ and\ \citenamefont
  {Marianetti}}]{Kotliar2006}%
  \BibitemOpen
  \bibfield  {author} {\bibinfo {author} {\bibfnamefont {G.}~\bibnamefont
  {Kotliar}}, \bibinfo {author} {\bibfnamefont {S.~Y.}\ \bibnamefont
  {Savrasov}}, \bibinfo {author} {\bibfnamefont {K.}~\bibnamefont {Haule}},
  \bibinfo {author} {\bibfnamefont {V.~S.}\ \bibnamefont {Oudovenko}}, \bibinfo
  {author} {\bibfnamefont {O.}~\bibnamefont {Parcollet}},\ and\ \bibinfo
  {author} {\bibfnamefont {C.~A.}\ \bibnamefont {Marianetti}},\ }\bibfield
  {title} {\bibinfo {title} {{Electronic structure calculations with dynamical
  mean-field theory}},\ }\href {https://doi.org/10.1103/RevModPhys.78.865}
  {\bibfield  {journal} {\bibinfo  {journal} {Rev. Mod. Phys.}\ }\textbf
  {\bibinfo {volume} {78}},\ \bibinfo {pages} {865} (\bibinfo {year}
  {2006})}\BibitemShut {NoStop}%
\bibitem [{\citenamefont {Otsuki}\ \emph {et~al.}(2022)\citenamefont {Otsuki},
  \citenamefont {Yoshimi}, \citenamefont {Shinaoka},\ and\ \citenamefont
  {Jeschke}}]{Otsuki2022}%
  \BibitemOpen
  \bibfield  {author} {\bibinfo {author} {\bibfnamefont {J.}~\bibnamefont
  {Otsuki}}, \bibinfo {author} {\bibfnamefont {K.}~\bibnamefont {Yoshimi}},
  \bibinfo {author} {\bibfnamefont {H.}~\bibnamefont {Shinaoka}},\ and\
  \bibinfo {author} {\bibfnamefont {H.~O.}\ \bibnamefont {Jeschke}},\
  }\href@noop {} {\bibinfo {title} {Multipolar ordering from dynamical mean
  field theory with application to \ce{CeB6}}} (\bibinfo {year} {2022}),\
  \Eprint {https://arxiv.org/abs/2209.10429} {arXiv:2209.10429
  [cond-mat.str-el]} \BibitemShut {NoStop}%
\bibitem [{\citenamefont {Aichhorn}\ \emph {et~al.}(2016)\citenamefont
  {Aichhorn}, \citenamefont {Pourovskii}, \citenamefont {Seth}, \citenamefont
  {Vildosola}, \citenamefont {Zingl}, \citenamefont {Peil}, \citenamefont
  {Deng}, \citenamefont {Mravlje}, \citenamefont {Kraberger}, \citenamefont
  {Martins}, \citenamefont {Ferrero},\ and\ \citenamefont
  {Parcollet}}]{Aichhorn2016}%
  \BibitemOpen
  \bibfield  {author} {\bibinfo {author} {\bibfnamefont {M.}~\bibnamefont
  {Aichhorn}}, \bibinfo {author} {\bibfnamefont {L.}~\bibnamefont
  {Pourovskii}}, \bibinfo {author} {\bibfnamefont {P.}~\bibnamefont {Seth}},
  \bibinfo {author} {\bibfnamefont {V.}~\bibnamefont {Vildosola}}, \bibinfo
  {author} {\bibfnamefont {M.}~\bibnamefont {Zingl}}, \bibinfo {author}
  {\bibfnamefont {O.~E.}\ \bibnamefont {Peil}}, \bibinfo {author}
  {\bibfnamefont {X.}~\bibnamefont {Deng}}, \bibinfo {author} {\bibfnamefont
  {J.}~\bibnamefont {Mravlje}}, \bibinfo {author} {\bibfnamefont {G.~J.}\
  \bibnamefont {Kraberger}}, \bibinfo {author} {\bibfnamefont {C.}~\bibnamefont
  {Martins}}, \bibinfo {author} {\bibfnamefont {M.}~\bibnamefont {Ferrero}},\
  and\ \bibinfo {author} {\bibfnamefont {O.}~\bibnamefont {Parcollet}},\
  }\bibfield  {title} {\bibinfo {title} {{TRIQS}/{DFTT}ools: A {TRIQS}
  application for ab initio calculations of correlated materials},\ }\href
  {https://doi.org/10.1016/j.cpc.2016.03.014} {\bibfield  {journal} {\bibinfo
  {journal} {Comput. Phys. Commun.}\ }\textbf {\bibinfo {volume} {204}},\
  \bibinfo {pages} {200} (\bibinfo {year} {2016})}\BibitemShut {NoStop}%
\bibitem [{jo-()}]{jo-cthyb}%
  \BibitemOpen
  \href@noop {} {}\bibinfo {note}
  {\url{https://github.com/j-otsuki/cthyb}}\BibitemShut {NoStop}%
\bibitem [{\citenamefont {Vidberg}\ and\ \citenamefont
  {Serene}(1977)}]{Vidberg1977}%
  \BibitemOpen
  \bibfield  {author} {\bibinfo {author} {\bibfnamefont {H.~J.}\ \bibnamefont
  {Vidberg}}\ and\ \bibinfo {author} {\bibfnamefont {J.~W.}\ \bibnamefont
  {Serene}},\ }\bibfield  {title} {\bibinfo {title} {Solving the {E}liashberg
  equations by means of {N}-point {P}ad{\'{e}} approximants},\ }\href
  {https://doi.org/10.1007/BF00655090} {\bibfield  {journal} {\bibinfo
  {journal} {J. Low Temp. Phys.}\ }\textbf {\bibinfo {volume} {29}},\ \bibinfo
  {pages} {179} (\bibinfo {year} {1977})}\BibitemShut {NoStop}%
\end{thebibliography}%

\end{document}